\documentclass[12pt, twocolumnfalse]{IEEEtran}

\usepackage{epsfig,graphics,amssymb,color,rotating}
\usepackage{amscd}
\usepackage{amsmath}
\usepackage{enumerate,multirow}
\usepackage{theorem}
\usepackage{epsfig}
\usepackage{verbatim}
\definecolor{blau}{rgb}{.15,.4,.8}
\definecolor{gruen}{rgb}{.25,.8,.25}
\definecolor{rot}{rgb}{.8,0,0}
\definecolor{gelb}{rgb}{.8,.8,0}
\definecolor{lila}{rgb}{.8,0,.8}
\definecolor{flieder}{rgb}{.77,.79,.91}
\definecolor{beige}{rgb}{.96,.79,.48}
\definecolor{orange}{rgb}{.8,.4,.1}
\definecolor{lightgreen}{rgb}{.73,.95,.76}
\definecolor{lightred}{rgb}{1,.71,.67}
\definecolor{lightblue}{rgb}{.52,.83,.98}

\definecolor{white}{rgb}{1,1,1}
\definecolor{blue}{rgb}{.15,.4,.8}
\definecolor{green}{rgb}{0,.8,.2}
\definecolor{red}{rgb}{.8,0,0}
\definecolor{yellow}{rgb}{.8,.8,0}
\definecolor{purple}{rgb}{.8,0,.8}

\newcommand{\be}{\begin{equation}}
\newcommand{\ee}{\end{equation}}
\newcommand{\ba}{\begin{array}}
\newcommand{\ea}{\end{array}}
\newcommand{\bea}{\begin{eqnarray}}
\newcommand{\eea}{\end{eqnarray}}

\def\mathlette#1#2{{\mathchoice{\mbox{#1$\displaystyle #2$}}%
                               {\mbox{#1$\textstyle #2$}}%
                               {\mbox{#1$\scriptstyle #2$}}%
                               {\mbox{#1$\scriptscriptstyle #2$}}}}

\newcommand{\Mat}[1]{\ensuremath{\mathlette{\boldmath}{#1}}}
\renewcommand{\Vec}[1]{\ensuremath{\mathlette{\boldmath}{#1}}}

\providecommand{\K}{\ensuremath{K}}

\providecommand{\N}{\ensuremath{N}}

\providecommand{\matS}{\ensuremath{{\Mat{S}}}}
\providecommand{\matT}{\ensuremath{{\Mat{T}}}}
\providecommand{\matR}{\ensuremath{{\Mat{R}}}}
\providecommand{\matH}{\ensuremath{{\Mat{H}}}}

\providecommand{\K}{\ensuremath{K}}

\providecommand{\N}{\ensuremath{N}}

\providecommand{\matS}{\ensuremath{{\Mat{S}}}}

\providecommand{\matT}{\ensuremath{{\Mat{T}}}}

\providecommand{\matR}{\ensuremath{{\Mat{R}}}}

\providecommand{\matH}{\ensuremath{{\Mat{H}}}}

\providecommand{\matA}{\ensuremath{{\Mat{A}}}}

\providecommand{\vecx}{\ensuremath{{\Vec{x}}}}

\providecommand{\stackH}{\ensuremath{{\boldsymbol{\mathcal{H}}}}}

\providecommand{\stacky}{\ensuremath{{\boldsymbol{\mathcal{Y}}}}}
\providecommand{\stackb}{\ensuremath{{\boldsymbol{\mathcal{B}}}}}
\providecommand{\stackh}{\ensuremath{{\boldsymbol{\mathcal{U}}}}}

\providecommand{\stackW}{\ensuremath{{\boldsymbol{\mathcal{W}}}}}
\providecommand{\stackM}{\ensuremath{{\boldsymbol{\mathcal{M}}}}}
\providecommand{\stackT}{\ensuremath{{\boldsymbol{\mathcal{T}}}}}
\providecommand{\stackR}{\ensuremath{{\boldsymbol{\mathcal{R}}}}}

\providecommand{\vecb}{\ensuremath{{\Vec{b}}}}

\providecommand{\vecy}{\ensuremath{{\Vec{y}}}}

\providecommand{\vecs}{\ensuremath{{\Vec{s}}}}
\providecommand{\vech}{\ensuremath{{\Vec{h}}}}

\providecommand{\vecw}{\ensuremath{{\Vec{w}}}}
\newcommand{\I}{\Mat{I}}
\newcommand{\E}{\mathrm{E}}

\newtheorem{theor}{Theorem}

\newtheorem{corollary}{Corollary}
\newtheorem{algorithm}{Algorithm}

\newtheorem{definition}{Definition}

\makeatletter
\def\ifundefined{\@ifundefined}
\makeatother

\renewcommand{\baselinestretch}{1.66}
\begin{document}

\title{Asynchronous
CDMA Systems with Random Spreading--Part II: Design Criteria}

\author{Laura Cottatellucci, Ralf R. M\"uller, and M\'erouane Debbah
\thanks{This work was presented in part at the IEEE Information Theory Workshop (ITW 2006), Punta de l'Este, Uruguay, March 2006 and  at the IEEE Wireless Communications and Networking Conference, Hong Kong, March 2007. It partly appears in Laura
Cottatellucci, ``Low Complexity Multistage Detectors for Randomly
Spread CDMA Systems'', Ph.D. thesis, Vienna University of
Technology, March 2006. }
\thanks{This work was supported in part by the  French ANR "Masses de Données" project SESAME  and by the Research Council of Norway under grant 171133/V30. }
\thanks{Laura Cottatellucci is with Eurecom, Sophia Antipolis, France (e-mail:
laura.cottatellucci@eurecom.fr). She was with Institute of
Telecommunications Research, University of South Australia,
Adelaide, SA, Australia. Ralf M\"uller is with Norwegian
University of Science and Technology, Trondheim, Norway, (e-mail:
mueller@iet.ntnu.no). M\'erouane Debbah was with Eurecom, Sophia Antipolis, France. He is currently with SUPELEC,  91192 Gif-sur-Yvette, France (e-mail: merouane.debbah@supelec.fr).} }

\ifundefined{IEEEtransversionmajor}{%
   
   \newlength{\IEEEilabelindent}
   \newlength{\IEEEilabelindentA}
   \newlength{\IEEEilabelindentB}
   \newlength{\IEEEelabelindent}
   \newlength{\IEEEdlabelindent}
   \newlength{\labelindent}
   \newlength{\IEEEiednormlabelsep}
   \newlength{\IEEEiedmathlabelsep}
   \newlength{\IEEEiedtopsep}

   \providecommand{\IEEElabelindentfactori}{1.0}
   \providecommand{\IEEElabelindentfactorii}{0.75}
   \providecommand{\IEEElabelindentfactoriii}{0.0}
   \providecommand{\IEEElabelindentfactoriv}{0.0}
   \providecommand{\IEEElabelindentfactorv}{0.0}
   \providecommand{\IEEElabelindentfactorvi}{0.0}
   \providecommand{\labelindentfactor}{1.0}

   \providecommand{\iedlistdecl}{\relax}
   \providecommand{\calcleftmargin}[1]{
                   \setlength{\leftmargin}{#1}
                   \addtolength{\leftmargin}{\labelwidth}
                   \addtolength{\leftmargin}{\labelsep}}
   \providecommand{\setlabelwidth}[1]{
                   \settowidth{\labelwidth}{#1}}
   \providecommand{\usemathlabelsep}{\relax}
   \providecommand{\iedlabeljustifyl}{\relax}
   \providecommand{\iedlabeljustifyc}{\relax}
   \providecommand{\iedlabeljustifyr}{\relax}

   \newif\ifnocalcleftmargin
   \nocalcleftmarginfalse

   \newif\ifnolabelindentfactor
   \nolabelindentfactorfalse

   \newif\ifcenterfigcaptions
   \centerfigcaptionsfalse

   \let\OLDitemize\itemize
   \let\OLDenumerate\enumerate
   \let\OLDdescription\description

   \renewcommand{\itemize}[1][\relax]{\OLDitemize}
   \renewcommand{\enumerate}[1][\relax]{\OLDenumerate}
   \renewcommand{\description}[1][\relax]{\OLDdescription}

   \providecommand{\pubid}[1]{\relax}
   \providecommand{\pubidadjcol}{\relax}
   \providecommand{\specialpapernotice}[1]{\relax}
   \providecommand{\overrideIEEEmargins}{\relax}

   \let\CMPARstart\PARstart

   \let\OLDappendix\appendix
   \renewcommand{\appendix}[1][\relax]{\OLDappendix}

   \newif\ifuseRomanappendices
   \useRomanappendicestrue

   \let\OLDbiography\biography
   \let\OLDendbiography\endbiography
   \renewcommand{\biography}[2][\relax]{\OLDbiography{#2}}
   \renewcommand{\endbiography}{\OLDendbiography}

   \markboth{A Test for IEEEtran.cls--- {\tiny \bfseries
   [Running Older Class]}}{Shell: A Test for IEEEtran.cls}}{

   \markboth{}%
   {}}


\renewcommand{\baselinestretch}{1.66}

\maketitle

\begin{abstract}
Totally asynchronous code-division multiple-access
(CDMA) systems are addressed. In Part I, the fundamental limits of asynchronous
CDMA systems are analyzed in terms of spectral efficiency and SINR at
the output of the optimum linear detector. The focus of Part II is
the design of low-complexity implementations  of linear multiuser
detectors in systems  with many users that  admit a multistage representation, e.g.\ reduced rank multistage Wiener filters, polynomial expansion detectors, weighted linear parallel interference cancellers.

The effects of excess bandwidth, chip-pulse shaping, and time delay distribution on CDMA
with suboptimum linear receiver structures are investigated.
Recursive expressions for universal weight design are given.
The performance in terms of SINR is derived in the large-system limit and the performance improvement over synchronous systems is quantified.
The considerations distinguish between two ways of forming discrete-time statistics: chip-matched filtering and oversampling.

 \vspace{10mm} \emph{Index Terms} - Asynchronous code-division multiple-access (CDMA), channel capacity, effective interference, minimum mean-square error (MMSE) detector, multistage detector, multiuser detection, random matrix theory, random spreading sequences.
\end{abstract}

\renewcommand{\baselinestretch}{2}
\vspace{0mm}

\section{Introduction}\label{chap:async_sec:introduction}
In Part I of this  paper \cite{cottatellucci:07b}, we analyzed
asynchronous CDMA systems with random spreading sequences in terms
of spectral efficiency constrained to a given chip pulse waveform and in terms of
SINR at the output of an optimum linear multiuser detector. The
analysis showed that under realistic conditions, chip-asynchronous CDMA systems significantly outperform chip-synchronous CDMA systems.
In order to utilize the benefits from chip-asynchronous\footnote{As already shown in Part I of this  paper \cite{cottatellucci:07b}, asynchronism is beneficial when the relative delays between users are \emph{not} integer multiples of a chip interval. To emphasize this requirement we use the term chip-asynchronism instead of asynchronism.} CDMA, we need efficient algorithms to cope with multiuser detection for chip-asynchronous users. Therefore,
in part II of this work, we focus on the generalization of known design rules for low-complexity multiuser detectors to chip-asynchronous CDMA.

A unified framework for the design and analysis of multiuser detectors that admit a multistage representation for synchronous users was given in \cite{cottatellucci:04}.
The
class of multiuser detectors that admit a multistage
representation is large and includes popular linear
multiuser detectors like linear MMSE detectors (e.g.
\cite{verdu:98}), reduced rank multistage Wiener filters
\cite{goldstein:97,goldstein:98}, polynomial expansion detectors
\cite{moshavi:96b} or conjugate gradient methods (e.g.
\cite{golub:96}), linear parallel interference cancellers (PIC,
e.g. \cite{divsalar:94,divsalar:98}), eventually weighted (e.g.
\cite{trichard:02}), and the single-user matched filters. Multistage detectors
are constructed around the matched filter concept. They consist of
a projection of the signal into a subspace
of the whole signal space by successive matched filtering and
re-spreading followed by a linear filter in the subspace.

Multistage detectors based on
universal weights have been proposed in
\cite{mueller:00,mueller:01a} for CDMA systems in AWGN channels
and extended to more realistic scenarios in
\cite{hachem:04,li:04,cottatellucci:04}.
These references make use of the self-averaging properties of large random matrices to find {\it universal} weighting coefficients for the linear filter in the subspace.
More
specifically, the universal weights are obtained by approximating
the precise weights designed according to some optimality
criterion with asymptotically optimum weights, i.e.\ the optimum
weights for a CDMA system whose number of users and spreading
factor tend to infinity with constant ratio. Thanks to the
properties of random matrices, asymptotically, these weights become
independent of the users' spreading sequences and depend only on
few macroscopic system parameters, as the system load or number of
transmitted symbols per chip, the variance of the noise, and the
distribution of the fading.
In this way, the weight design for long-code CDMA
simplifies considerably, its complexity becomes independent of both the
number of users in the system and the spreading factor.
Moreover, the weights need updating only when the macroscopic
system parameters change.


The fact that users are not received in a time-synchronized manner at the receiver causes two main problems from a signal processing perspective:
(i) the need for an infinite observation window to implement a linear MMSE detector and (ii) the potential need for oversampling to form sufficient discrete-time statistics.
The need for an infinite observation window is primarily related to asynchronism on the symbol-level, not the chip-level.
This aspect was addressed in \cite{cottatellucci:04b,cottatellucci:04g} where it was found that multistage detectors need not have infinite observation windows and can be efficiently implemented without windowing at all.
A detailed overview of the state of art about statistics, sufficient or not,  for multiuser CDMA systems and  how to form them was addressed in Part I of this paper \cite{cottatellucci:07b}. In part I we presented general results with the only constraint that the sampled noise at the output of the front-end was white. For the sake of clarity  and to get insights into systems of practical interests, in this part II we focus on two groups of statistics implementable in practical systems:
\begin{enumerate}[(A)]
    \item \label{sufficient_statistics} Sufficient statistics
    obtained by filtering the received signal by a lowpass filter
    with bandwidth $B_{\mathrm{LOW}}$ larger than the chip-pulse
    bandwidth and subsequent sampling at rate $2B_{\mathrm{LOW}}$.
    \item \label{suboptimal_statistics} Statistics
    obtained by sampling the output of a
    filter matched to the chip waveform at the chip rate (\textit{chip rate sampling}). In this case, the sampling instants need to be synchronized with the time delay of each user of interest. Thus, different statistics for each user are required. Additionally, the chip pulses at the output of matched filter need to satisfy the Nyquist criterion. In the following we refer to them as root Nyquist
    chip-pulse waveforms.
\end{enumerate}
General results for the design of linear multistage
detectors with both kind of statistics are provided in this work.
The chip pulse waveforms are assumed to be identical for all
users.

For asynchronous CDMA, low-complexity detectors
with universal weights are conveniently designed for statistics
(\ref{sufficient_statistics}). In fact, these observables enable a
joint processing of all users without loss of information.
Multistage detectors with universal weights and statistics (A)
have a complexity order per bit equal to $\mathcal{O}(rK)$ if the
sampling rate is $\frac{r}{T_c}.$ On the contrary, discretization scheme
(\ref{suboptimal_statistics}) provides different observables for
each user and does not allow for simultaneous joint detection of all users. An
implementation of multistage detectors with universal weights
using such statistics implies a complexity order \emph{per bit}
equal to $\mathcal{O}(K^2).$ This approach is still interesting
from a complexity point of view if detection of a single user is
required. However, it suffers from a performance degradation due
to the sub-optimality of the statistics.

This work is organized in six additional sections. Section
\ref{sec:notation} and \ref{sec:system_model} introduce the notation
and the system model for asynchronous CDMA, respectively.
In Section \ref{IVa},
multistage detectors for asynchronous CDMA are reviewed and a implementation which does not suffer from truncation effects is given. The design of universal weighting is addressed in Section \ref{IVb}.
Finally, the analytical results are applied to gain
further insight into the system in Section
\ref{chap:async_sec:chip_async_symb_sync_subsec:pulse_shape} where methods for pulse-shaping, forming sufficient statistics and synchronization are compared. Conclusions are summed up in Section
\ref{chap:async_sec:conclusions}.

\section{Notation and Some Useful Definitions}\label{sec:notation}
Throughout Part II  we adopt the same notation and definitions
already introduced in Part I of this work
\cite{cottatellucci:07b}. In order to make Part II self-contained
we repeat here definitions useful in this part. Upper and lower
boldface symbols are used respectively for matrices and vectors
corresponding to signals spanning a specific symbol interval $m.$
Matrices and vectors describing signals spanning more than a
symbol interval are denoted by upper boldface calligraphic
letters.

In the following, we utilize \emph{unitary} Fourier transforms
both in the continuous time and in the discrete time domain.
The unitary Fourier transform of a function
$f(t)$ in the continuous time domain is given by $F(\omega)=\frac{1}{\sqrt{2 \pi}} \int f(t) \mathrm{e}^{-j \omega t}
\mathrm{d}t $. The unitary Fourier transform of a sequence $\{
\ldots, c_{-1}, c_{0}, c_{1}, \ldots \}$ in the discrete time
domain is given by $c(\Omega)= \frac{1}{\sqrt{2
\pi}} \sum_{n=-\infty}^{+\infty} c_n \mathrm{e}^{-j \Omega n}$.
We will refer to them shortly as Fourier transform.
We denote the argument of a Fourier transform of a continuous
function by $\omega$ and the argument of a Fourier transform of a
sequence by $\Omega.$ { They are the angular frequency and the normalized angular frequency, respectively. A function in $\Omega$ is periodic with respect to integer multiples of $2\pi$.}

For further studies it is convenient to define  the concept of
\emph{$r$-block-wise circulant matrices of order $N$}.
\begin{definition} Let $r$ and $\N$ be  positive integers. An
$r$-block-wise circulant matrix of order $\N$  is an $r \N \times
\N$ matrix of the form
\begin{equation}
\Mat{C}=\left( \begin{array}{cccc}
 \Mat{B}_0 & \Mat{B}_1 & \cdots & \Mat{B}_{\N-1} \\
  \Mat{B}_{\N-1} & \Mat{B}_0 & \cdots & \Mat{B}_{\N-2} \\
  \vdots & \vdots &  &\vdots \\
 \Mat{B}_{1} & \Mat{B}_{2} & \cdots & \Mat{B}_{0} \\
\end{array} \right)
\end{equation}
with $\Mat{B}_i=(c_{1,i},c_{2,i}, \ldots, c_{r,i})^T.$
\end{definition}
In the matrix $\Mat{C}$ an $r \times N$ block row is obtained by
circularly right shift of the previous block. Since the matrix
$\Mat{C}$ is univocally defined by the unitary Fourier transforms
of the sequences $\{ c_{s,0}, c_{s,1}, \ldots c_{s,N-1}\}$, for
$s=1...r,$
\begin{equation*} c_s(\Omega)=\frac{1}{\sqrt{2 \pi}} \sum_{k=0}^{\N-1}
c_{sk} \mathrm{e}^{-j \Omega k } \qquad s=1, \ldots, r,
\end{equation*}
there exists a bijection $\mathfrak F$ from  the frequency dependent vector $\Mat c(\Omega)=[c_1(\Omega), c_2(\Omega),\dots,c_r(\Omega)]$ to $\Mat C$. Thus,
\begin{equation}\label{bijection}
\Mat C = \mathfrak F\{\Mat c(\Omega)\}.
\end{equation}

Furthermore, the superscripts $\cdot^T,$ $\cdot^H,$ and  $\cdot^*,$ denote the
transpose, the conjugate transpose, and the conjugate of the matrix argument,
respectively. $\I_{n}$ is the identity matrix of size $n \times n$
and $\mathbb{C}$, $\mathbb{Z}$, $\mathbb{Z}^{+},$ $\mathbb{N},$ and
$\mathbb{R}$ are the fields of complex, integer, nonnegative
integers, natural, and real numbers, respectively. $\mathrm{tr}(\cdot)$ is
the trace of the matrix argument and
$\mathrm{span}(\Mat{v}_1,\Mat{v}_2, \ldots, \Mat{v}_s )$ denotes
the vector space spanned by the $s$ vectors $\Mat{v}_1, \Mat{v}_2,
\ldots \Mat{v}_s. $ $\mathrm{diag}(\ldots): \mathbb{C}^{n}
\rightarrow \mathbb{C}^{n \times n}$ transforms an $n$-dimensional
vector $\Vec{v}$ into a diagonal matrix of size $n$ having as
diagonal elements the components of $\Vec{v}$ in the same order.
$\E\{\cdot\}$ and $\mathrm{Pr}\{\cdot\}$ are the expectation and
probability operators, respectively. $\delta_{ij}$ is the
Kronecker symbol and $\delta(\lambda)$ is the Dirac's delta
function. $\mathrm{mod}$ denotes the modulus and $\lfloor \cdot
\rfloor$ is the operator that yields the maximum integer not
greater than its argument.

\section{System Model}\label{sec:system_model}
In this section we recall briefly the system model for
asynchronous CDMA introduced in Section IV and VII of Part I of this
work \cite{cottatellucci:07b}. The reader interested in the
details of the derivation can refer to \cite{cottatellucci:07b}.

Let us consider an asynchronous CDMA system with $K$ active users
in the uplink  channel with spreading factor $N$. Each user and the
base station are equipped with a single antenna. The channel is
flat fading and impaired by additive white Gaussian noise with power spectral density $N_0$. The symbol interval is denoted
with $T_s$ and $T_c=\frac{T_s}{N}$ is the chip interval. The
modulation of all users is based on the same chip pulse waveform
$\psi(t)$ bandlimited with bandwidth $B,$ unitary Fourier transform $\Psi(\omega),$ and energy $E_{\psi}=\int_{-\infty}^{\infty} |\psi(t)|^2 \mathrm{d}t.$

The time delays of the $K$ users are denoted with $\tau_k,$
$k=1,\ldots,K.$ Without loss of generality we can assume (i) user
1 as reference user so that $\tau_1=0$, (ii) the users ordered
according to increasing time delay with respect to the reference
user, i.e. $\tau_1 \leq \tau_2 \leq \ldots \leq \tau_K$; (iii) the
time delay to be, at most, one symbol interval so that $\tau_k \in
[0,T_s)$.\footnote{For a thorough discussion on this assumption
the reader can refer to \cite{verdu:98}. }

As for the results presented in Part I, the mathematical results presented in this second part hold for any front-end that keeps the sampled noise white at its output. However, in order to get better insights into the physical system  we focus on two front-ends of practical and theoretical interest. Both of them satisfy the more general assumption underlying the results in Part I. We refer to them as Front-end Type A and Front-end Type B\footnote{For the sake of compactness of some of the results, we adopt a different normalization from the one in Part I. Here, the signal energy at the output of the front-end is equal to one. In Part I, the energy of the analog filter's impulse response is normalized to unity. The variance of the sampled noise at the front-end output changes accordingly.}.

\noindent\textbf{\emph{Front-end Type A}} consists of
\begin{itemize}
    \item An ideal lowpass filter with cut-off frequency $\omega=
    \frac{\pi r}{Tc}$ where $r \in \mathbb{Z}^{+}$ satisfies the
    constraint $B \leq \frac{r}{2 T_c} $ such that the sampling theorem applies. The
    filter is normalized to obtain a unit overall amplification factor, i.e.\ the transfer function is
    \begin{equation}
    G(\omega)= \begin{cases}
    \frac{1}{\sqrt{E_{\psi}}} & |\omega| \leq \frac{\pi r}{ T_c} \\
    0 & |\omega| > \frac{\pi r}{ T_c}. \\
     \end{cases}
    \end{equation}
    \item A subsequent continuous-discrete time conversion by
    sampling at rate $\frac{r}{T_c}.$
\end{itemize}
This front-end satisfies the conditions of the sampling
theorem and, thus, provides sufficient discrete-time statistics.
For convenience, the sampling rate is an integer multiple of the chip rate.
Additionally, the discrete-time noise process is white with zero mean and variance
$\sigma^2=\frac{N_0 r}{E_{\psi} T_c}.$

\noindent\textbf{\emph{Front-end Type B}} consists of
\begin{itemize}
\item A filter $G(\omega)$ matched to the chip pulse and
normalized to the chip pulse energy, i.e.\ $G(\omega)=
{\Psi^{*}(\omega)}{{E_{\psi}^{-\frac12}}};$
\item Subsequent sampling at the chip rate.
\end{itemize}
When used
with root Nyquist chip pulses, the
discrete time noise process $\{w[p]\}$ is white with variance
$\frac{N_0}{E_{\psi} T_c}$. For a synchronous systems with square
root Nyquist chip pulses, this front end provides
sufficient statistics whereas the observables are not sufficient
if the system is asynchronous.

{The chip waveform at the filter output is denoted by $\phi(t)$ and its unitary Fourier transform by $\Phi(\omega).$ The well-known relations $\phi(t)=\psi(t) \ast g(t)$ and $\Phi(\omega)=\Psi(\omega) G(\omega)$ hold.  The unitary Fourier transform of the chip pulse waveform $\phi(t)$ sampled at rate $\frac{1}{T_c}$ and delay $\tau$ is given by
\begin{equation}\label{discrete_Fourier_phi} {\phi}(\Omega,\tau)\overset{\triangle}{=} \frac{1}{T_c}
\sum_{s=-\infty}^{+\infty} \mathrm{e}^{  j  \frac{\tau}{T_c}(\Omega+2 \pi s)} \Phi^{*} \left(\tfrac{j(\Omega+2 \pi s)}{T_c} \right).
\end{equation}  }

{Sufficient statistics for asynchronous CDMA require an infinite observation window. In the following, we introduce a matrix system model corresponding to an infinite observation window.}

Let us denote with $\vecb^{(m)}$ and $\vecy^{(m)}$ the vectors of
transmitted and received signals at time instants $m \in
\mathbb{Z}.$ The baseband discrete-time asynchronous system is
given by
\begin{equation}\label{matrix_model}
\boldsymbol{\mathcal{Y}}=
\boldsymbol{\mathcal{H}}\boldsymbol{\mathcal{B}}+\boldsymbol{\mathcal{W}}
\end{equation}
where $\stacky=[\ldots,\vecy^{(m-1)T},\vecy^{(m)T},\vecy^{(m+1)T}
\ldots]^T$ and
$\stackb=[\ldots,\vecb^{(m-1)T},\vecb^{(m)T},\vecb^{(m+1)T} \ldots]^T$
are infinite-dimensional vectors of received and transmitted
symbols respectively; $\stackW$ is an infinite-dimensional noise
vector; and $\stackH$ is a bi-diagonal block matrix of infinite size
given by
\begin{equation}\label{unlimited_stack_H} \stackH = \left[ \begin{array}{ccccccc} \ddots & \ddots
&\ddots &\ddots &\ddots &\ddots &\ddots
 \\
   \ldots & \mathbf{0} & \matH_d^{(m-1)} & \matH_u^{(m)} & \mathbf{0} &
  \ldots & \ldots  \\  \ldots & \ldots  & \mathbf{0} & \matH_d^{(m)} & \matH_u^{(m+1)} & \mathbf{0} &
  \ldots   \\ \ddots & \ddots &\ddots
&\ddots &\ddots &\ddots &\ddots
\\
\end{array} \right].
\end{equation}
Here, $\matH_u^{(m)}$ and $\matH_d^{(m)}$ are matrices of size $rN
\times K $ obtained by the decomposition of the $2 r \N \times \K$
matrix $\matH^{(m)}$ into two parts such that $\matH^{(m)}=
[\matH_u^{(m)T},\matH_d^{(m)T}]^T.$ For $\matH^{(m)}$ the relation
\begin{equation}
\matH^{(m)}=\matS^{(m)} \matA
\end{equation}
holds where $\matA$ is the $\K \times \K$ diagonal matrix of the
received amplitudes $a_{k}$ and $\matS^{(m)}$ is the $2 r \N \times \K$
matrix whose $k$-th column accounts for the spreading of the symbol transmitted by user $k$ in the symbol interval $m$ and due to the actual spreading sequence, the channel delay, and filtering and sampling at the front-end. We refer to it as the matrix of virtual spreading. More specifically, the  matrix of virtual spreading is given by
\begin{equation}
\matS^{(m)}=\left(\Mat{\Phi}_1 \vecs_{1}^{(m)}, \Mat{\Phi}_2 \vecs_{2}^{(m)},
\ldots \Mat{\Phi}_K \vecs_{K}^{(m)} \right)
\end{equation}
where  $\vecs_{k}^{(m)}$ is the $N$-dimensional column vector of the
spreading sequence of user $k$ for the transmitted symbol $m$ and
$\Mat{\Phi}_k$ is the $2 r N \times N$ matrix taking into account
the effects of the chip pulse shape and the time delay $\tau_k$ of user $k.$ Let us decompose $\tau_k$ in $\overline{\tau}_k= \left \lfloor \frac{\tau_k}{T_c} \right \rfloor $ and $\widetilde{\tau}_k=\tau_k-T_c\overline{\tau}_k= \tau_k \mod T_c,$ the integer number of chips the signal is delayed and its delay within a chip, respectively.
 The matrix $\Mat{\Phi}_{k}$ is of the form
\begin{equation}\label{matrice_Phi}
\Mat{\Phi}_{k} = \left[ \begin{array}{c}
  \Mat{0}_{\overline{\tau}_k} \\
  \widetilde{\Mat{\Phi}}_k \\
  \Mat{0}_{N-\overline{\tau}_k}
\end{array} \right]
\end{equation}
where $\Mat{0}_{\overline{\tau}_k}$ and $\Mat{0}_{N-\overline{\tau}_k}$ are zero matrices of dimensions ${\overline{\tau}_k} \times N$ and
$\left(N-\overline{\tau}_k \right)\times N $,
respectively; $\widetilde{\Mat{\Phi}}_k$ is an
$r$-block-wise circulant matrix of order $N$ as in (\ref{bijection})
\begin{equation}\label{circulant_C_phi_r}
\widetilde{\Mat{\Phi}}_k = \mathfrak{F}(\Vec{c}(\widetilde{\tau}_k)),
\end{equation}
with
\begin{equation*}
    \Vec{c}(\widetilde{\tau}_k)= \left[ {\phi}(\Omega,\widetilde{\tau}_k) {\phi}(\Omega,\widetilde{\tau}_k -\tfrac{T_c}{r}),\ldots,
{\phi}(\Omega, \widetilde{\tau}_k -\tfrac{(r-1)T_c}{r}) \right].
\end{equation*}
Thus, the virtual spreading sequences are the samples of the delayed continuous-time spreading waveforms at sampling rate $r/T_c$.

Throughout this work we assume that the transmitted symbols are
uncorrelated and identically distributed random variables with
unitary variance and zero mean, i.e.
$\E(\stackb)=\boldsymbol{\mathcal{O}}$ and
$\E(\stackb\stackb^H)=\boldsymbol{\mathcal{I}}$ being
$\boldsymbol{\mathcal{O}}$ and $\boldsymbol{\mathcal{I}}$ the
unlimited zero vector and the unlimited identity matrix,
respectively. The elements of the spreading sequences
$\vecs_{k}^{(m)}$ are assumed to be zero mean  i.i.d. Gaussian random variables
over all the users, chips, and symbols with
$\mathrm{E}\{\vecs_{k}^{(m)}\vecs_{k}^{(m)H}\}=\frac{1}{N}\Mat{I}_{N}$.
Finally, $\stackh_{k}^{(m)}$ denotes that column of the matrix
$\stackH$ containing the $k^{\mathrm{th}}$ column of the matrix
$\Mat{H}^{(m)}.$ We define the correlation
matrices $\stackT=\stackH\stackH^H$ and
$\stackR=\stackH^H \stackH.$ The system load $\beta=\frac{K}{N}$
is the number of transmitted symbols per chip.


\section{Multistage Structures for Asynchronous CDMA}
\label{IVa}

We consider the large class of linear multistage detectors for
asynchronous CDMA.  Let $\chi_{L,k}^{(m)}(\stackH)$ be the
Krylov subspace \cite{vorst:03} of rank $L \in \mathbb{Z}^{+}$ given by
\begin{equation}
\chi_{L,k}^{(m)}(\stackH)= \mathrm{span} (\stackT^{\ell}
\stackh_{k}^{(m)})|_{\ell=0}^{L-1}.
\end{equation}
A multistage detector of rank $L \in \mathbb{Z}^{+}$ for user $k$
is given by
\begin{equation}\label{gen_multistage_detector}
\widehat{b}_k =\sum_{\ell=0}^{L-1} (\vecw_{k}^{(m)})_{\ell}
\stackh_{k}^{(m)H} \stackT^{\ell}\stacky
\end{equation}
where $\vecw_{k}^{(m)}$ is the $L$-dimensional vector of weight
coefficients.

\begin{figure}
\begin{center}
\centerline{\epsfig{file=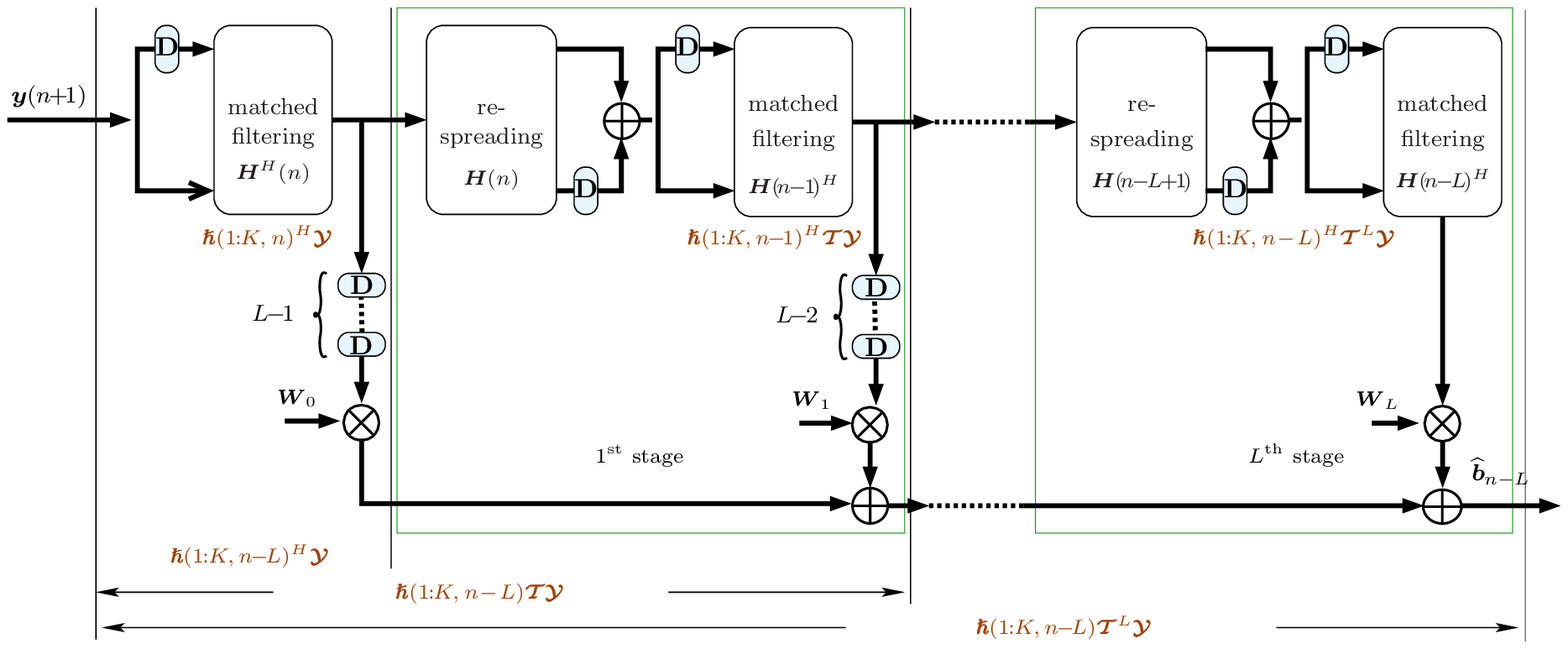,width=\textwidth}}
 \end{center}
\vspace*{-15mm} \caption{Multistage detector for  asynchronous
CDMA systems. Here,
$\pmb{\hslash}(1:K,n)=[\Mat{\Phi}_1\vecs_{1}^{(n)},\Mat{\Phi}_2\vecs_{2}^{(n)}, \ldots \Mat{\Phi}_K\vecs_{K}^{(n)}
]$}\label{Fig:Multistage_async}
\end{figure}

It has been shown in \cite{cottatellucci:04g} that, given the
weight vector $\vecw_{k}^{(m)}$ the detection of the symbol $b_{k}^{(m)}$
by the multistage detector of rank $L$ in
(\ref{gen_multistage_detector}) can be performed with finite delay
$L$ using the implementation scheme in Figure
\ref{Fig:Multistage_async}. Although infinite length vectors and
infinite dimension matrices appear in
(\ref{gen_multistage_detector}), the multistage detector in Figure
\ref{Fig:Multistage_async} implements exactly
(\ref{gen_multistage_detector})  and does not suffer from
truncation effects. Equivalently, the multistage detector in
Figure \ref{Fig:Multistage_async} can be considered as a
multistage detector processing data over an observation window of size $2 L.$
The projection of the received vector $\stacky$ onto the subspaces
$\chi_{L,k}^{(m)}(\stackH)$, for $k=1 \ldots K,$ is performed jointly
for all users and requires only multiplications between vectors and matrices.
The size of those vectors and matrices does not depend on the observation
window.  For further details the interested reader is referred to
\cite{cottatellucci:04g}, \cite{cottatellucci:05}.

The class of multistage detectors includes many popular multiuser
detectors:
\begin{itemize}
\item the single-user matched filter for $L=1$,
\item the linear parallel interference canceller (PIC) \cite{brown:01,guo:99} for weight coefficients chosen irrespective of the properties of the transfer matrix $\stackH$,
\item the polynomial expansion detector \cite{moshavi:96b} and the conjugate gradient method \cite{golub:96}, if the weight coefficients are identical for all users and chosen to minimize the mean square error,
\item the (reduced rank) multistage Wiener filter \cite{goldstein:98} if the weight coefficients are chosen to minimize the mean square error, but are allowed to differ from user to user.
\end{itemize}
Throughout this work we refer to detectors that minimize the MSE
in the projection subspace of the user of interest as
\emph{optimum detectors in the MSE sense}. More specifically this
class of multistage detectors includes the linear MMSE detector
and the multistage Wiener filter but not the polynomial expansion
detector.

In the following we focus on the design of multistage Wiener filters implemented as in Figure
\ref{Fig:Multistage_async}. This reduces the problem to the design of the
filter coefficients $\vecw_{k}^{(m)}.$
The multistage Wiener filter for the
detection of the symbol $m$ transmitted by user $k$ reads
\begin{equation}
\stackM_{k}^{(m)}= \sum_{\ell=0}^{L-1} (\vecw_{k}^{(m)})_{\ell-1}
\stackh_{k}^{(m)H} \stackT^{\ell}.
\end{equation}
The weight vector $\vecw_{k}^{(m)}$ that minimizes the MSE
$\mathrm{E}\{ \| \stackM_{k}^{(m)} \stacky -b_{k}^{(m)} \|^2 \}$  is given
by
\begin{align}
\vecw_{k}^{(m)}& = \underset{\overline{\vecw}_{k}^{(m)}}{\text{argmin}}
\mathrm{E} \left \{ \left\| \sum_{\ell=0}^{L-1}
(\overline{\vecw}_{k}^{(m)})_{\ell} \stackh_{k}^{(m)H} \stackT^{\ell}
\stacky-b_{k}^{(m)} \right\|^2 \right\} \\
&= \underset{\overline{\vecw}_{k}^{(m)}}{\text{argmin}} \mathrm{E}
\left\{ \left\| \overline{\vecw}_{k}^{(m)H} \vecx_{k}^{(m)} -b_{k}^{(m)}  \right\|^2
\right\}
\end{align}
where $\vecx_{k}^{(m)}$ is an $L$-dimensional vector with
$j^{\text{th}}$ element $(\vecx_{k}^{(m)})_j= \stackh_{k}^{(m)H}
\stackT^{j-1} \stacky.$ This optimization problem is solved by the
Wiener-Hopf theorem \cite{Kaybook} and $\vecw_{k}^{(m)}$ is
given by
\begin{equation}\label{weight_Wiener}
\vecw_{k}^{(m)}= (\boldsymbol{\Xi}_{k}^{(m)})^{-1} \boldsymbol{\xi}_{k}^{(m)}
\end{equation}
where $\boldsymbol{\Xi}_{k}^{(m)}=\mathrm{E}\{ \vecx_{k}^{(m)}
\vecx_{k}^{(m)H}\}$ and $ \boldsymbol{\xi}= \mathrm{E}\{ b_{k}^{(m)*}
\vecx_{k}^{(m)} \}.$ It is straightforward to verify that in this case
\begin{align}
{\boldsymbol{\Xi}}_{k}^{(m)}&=  \left ( \begin{array}{rclcrcl}
    (\stackR^{2})_{k,m}&+& \sigma^{2} (\stackR)_{k,m}  & \cdots & (\stackR^{L+1})_{k,m}&+& \sigma^{2} (\stackR^{L})_{k,m} \\
    (\stackR^{3})_{k,m}&+& \sigma^{2} (\stackR^{2})_{k,m}  & \cdots & (\stackR^{L+2})_{k,m}&+& \sigma^{2} (\stackR^{L+1})_{k,m} \\
    &\vdots &  &\ddots && \vdots \\
    (\stackR^{L+1})_{k,m}&+& \sigma^{2} (\stackR^{L})_{k,m}  & \cdots & (\stackR^{2L})_{k,m}&+& \sigma^{2} (\stackR^{2L-1})_{k,m}
    \
  \end{array}
  \right ) \nonumber \\
  {\boldsymbol{\xi}}_k^{(m)} &= \left( (\stackR)_{k,m}, (\stackR^{2})_{k,m},
\ldots,
  (\stackR^{L})_{k,m}\right)^T.
\end{align}
where $(\stackR^{s})_{k,m}=\vech^{(m)H}_{k} \stackT^{s-1}
\vech_{k}^{(m)}$ is the diagonal element of the matrix $\stackR^s$
corresponding to the $m^{\text{th}}$ symbol transmitted by user
$k.$

\section{Universal Weight Design}

Consider the SINR of any linear detector that admits a
multistage representation.
Let $\overline{\vecw}_{k,m}$ be the weight
vector for the detection of the $ m^{\mathrm{th}}$ symbol
transmitted by user $k.$ Then, the SINR at the output of the
multistage detector is given by
\begin{equation}\label{SINR_asynch_gen}
\mathrm{SINR}_k= \frac{\overline{\vecw}_{k}^{(m)H}
\boldsymbol{{\xi}}_{k}^{(m)} \boldsymbol{{\xi}}_{k}^{(m)T}
\overline{\vecw}_{k}^{(m)}}{\overline{\vecw}_{k}^{(m)H}
(\boldsymbol{{\Xi}}_{k}^{(m)} - \boldsymbol{{\xi}}_{k}^{(m)}
\boldsymbol{{\xi}}_{k}^{(m)T}) \overline{\vecw}_{k^{(m)H}}}.
\end{equation}
The performance of multistage Wiener filters  simplifies to
\begin{equation}\label{SINR_asynch_multistage_Wiener}
\mathrm{SINR}_k= \frac{\boldsymbol{{\xi}}_{k}^{(m)T}
\boldsymbol{{\Xi}}_{k}^{(m) \;-1} \boldsymbol{{\xi}}_{k}^{(m)}}{1-
\boldsymbol{{\xi}}_{k}^{(m)T} \boldsymbol{{\Xi}}_{k}^{(m) \; -1}
\boldsymbol{{\xi}}_{k}^{(m)}}.
\end{equation}
From (\ref{weight_Wiener}), (\ref{SINR_asynch_gen}), and
(\ref{SINR_asynch_multistage_Wiener}) it is apparent that the
diagonal elements of the matrix $\stackR^s$ play a fundamental
role in the design and analysis of multistage detectors.

It has been shown in \cite{cottatellucci:04} that, if the
spreading sequences are random and the CDMA system is synchronous,
the diagonal elements of the matrix $\stackR^s,$ $s \in
\mathbb{Z}^{+},$  converge to deterministic values as $K, N
\rightarrow \infty$ with constant ratio. This asymptotic
convergence holds for some classes of random matrices and is a stronger property
than the convergence of the eigenvalue distribution.
The Stieltjes transform of
the asymptotic eigenvalue distribution of $\stackR$ is related to
the SINR at the output of the linear MMSE detector, as pointed out
first in \cite{tse:99b} for synchronous CDMA systems. The
asymptotic eigenvalue moments of $\stackR$ enable the asymptotic
performance analysis of reduced rank multistage Wiener filters
\cite{honig:01} and the design of multistage detectors with
quadratic complexity order per bit \cite{li:04}, \cite{hachem:04}.
The convergence of the diagonal elements of $\stackR^s$ has been
utilized in \cite{cottatellucci:04} for the design of multistage
detectors with linear complexity order per bit in synchronous CDMA
systems and for the asymptotic analysis of any multistage detector
not necessarily optimum in a MSE sense.
In the following we extend the results in \cite{cottatellucci:04}
to the case of asynchronous CDMA systems making use of the
asymptotic properties of the random matrix $\stackR$ for
asynchronous CDMA systems.

The design of low complexity multistage detectors is based on the
approximation of the weight vectors $\vecw_{k}^{(m)}$ by their asymptotic
limit when $K, N \rightarrow \infty$ with constant ratio
$\beta$
\begin{equation}
\vecw_{k}^\infty= \lim_{K=\beta N \rightarrow \infty}
\boldsymbol{\Xi}_{k}^{(m)\; -1} \boldsymbol{\xi}_{k}^{(m)}.
\end{equation}
Thanks to the fact that the diagonal elements of
$\stackR^s$ can be computed by a polynomial in few macroscopic
system parameters, the computation of the weight vectors becomes
independent of the size of $\stackR$ and independent of $m$.
Thus, the effort for the computation of the weights becomes negligible and the
complexity of the detector is dominated by the
joint projection of the received signal $\stacky$ onto the
subspaces $\chi_{k}^{(m)}(\stackH),$ $k=1 \ldots K$ and $m \in
\mathbb{Z}.$ This projection has linear complexity per bit
if the multistage detector in Figure \ref{Fig:Multistage_async} is
utilized.
\newenvironment{Ventry}[1]%
    {\begin{list}{}{\renewcommand{\makelabel}[1]{\emph{##1:}\hfil}%
        \settowidth{\labelwidth}{\emph{#1:}}%
        \setlength{\leftmargin}{\labelwidth}%
        \addtolength{\leftmargin}{\labelsep}%
        \setlength{\parsep}{-4pt}%
        \setlength{\topsep}{0pt}%
        \setlength{\partopsep}{4pt}
        }}%
    {\end{list}}
\label{IVb}

The convergence of the diagonal elements of $\stackR^{\ell}$  to deterministic values is
established in the following theorem. The definitions and the
assumptions in the statement of Theorem
\ref{theo:diagonal_elements_chip_asynch} summarize and formalize
the characteristics of system model (\ref{matrix_model}) for
${\tau}_k \in [0, T_s].$

\begin{theor}\label{theo:diagonal_elements_chip_asynch}
Let $K, N \in \mathbb{N} $ and $\matA \in \mathbb{C}^{\K \times \K}$ be a diagonal matrix
with $k^{\text{th}}$ diagonal element $a_{k}\in \mathbb{C}.$ $T_s$ and $T_c$ are positive reals with $T_s=N T_c$. Given $\{\tau_1, \tau_2, \ldots \tau_K \}$ a set of delays in $[0, T_s),$ we introduce the sets of delays in $[0, T_c)$ defined as  $\{\widetilde{\tau}_k: \widetilde{\tau}_k=\tau_k \mathrm{mod} T_c, \, k=1, \ldots K \}$ and the set of normalized delays $\left\{\overline{\tau}_k: \overline{\tau}_k= \left\lfloor \frac{\tau_k}{T_c} \right\rfloor \right\}.$  Given a function $\Phi(\omega): \mathbb{R} \rightarrow \mathbb{C}$, let $\phi(\Omega,\tau)$ be as in
(\ref{discrete_Fourier_phi}). Given a positive integer $r,$ let $\Mat{\Phi}_k, \, k=1, \ldots K, $ be $r$-block-wise circulant matrices of order $\N$ defined in
(\ref{circulant_C_phi_r}) and $\matS^{(m)}= \left( \Mat{\Phi}_1 \vecs_1^{(m)}, \Mat{\Phi}_2 \vecs_2^{(m)}, \ldots \Mat{\Phi}_K \vecs_K^{(m)},  \right)$  with $ \vecs_k^{(m)}$ $N$-dimensional random column vector. Let ${\matH}=(\matH_u^{(m)T}, \matH_d^{(m)T})^T= {\matS}\matA$ with $\matH_u^{(m)}, \matH_d^{(m)} \in \mathbb{C}^{rN \times K}$ and $\stackH$ the infinite block row and block column matrix of the same form as in (\ref{unlimited_stack_H}), $\stackT=\stackH \stackH^H,$ $\stackR=\stackH^H \stackH,$ and $\stackh_k^{(m)}$ the column of $\stackH$ corresponding to $\Mat{\Phi}_k \vecs_k^{(m)}.$

We assume that the function $\Phi(\omega)$ is upper bounded and has finite support. The receive filter is such that the sampled discrete time noise process is white. The vectors  $\vecs_{k}$ are
independent with i.i.d.\ zero-mean circularly symmetric Gaussian elements with variance $\mathrm{E}\{|s_{ij}|^2\}= N^{-1}$.
Furthermore, the elements $a_{k}$ of the matrix $\matA$ are
uniformly bounded for any $K.$  The sequence of the empirical
joint distributions
$F_{|\matA|^2,\widetilde{T}}^{(\K)}(\lambda,\widetilde{\tau})=\frac{1}{\K} \sum_{k=1}^{\K}
1(\lambda-|a_{k}|^2)1(\widetilde{\tau}-\widetilde{\tau}_k)$ converges almost surely, as $\K \rightarrow \infty$, to a non-random distribution
function $F_{|\matA|^2,\widetilde{T}}(\lambda,\widetilde{\tau}).$

Then, conditioned on $(|a_{k}|^2, \widetilde{\tau}_k)$, the
corresponding diagonal elements of the matrices
$\stackR^{\ell}$  converge almost surely to the deterministic value
\begin{equation}
    \lim_{\K= \beta \N \rightarrow \infty}
    (\stackR^{\ell})_{k,m}=\lim_{\K= \beta \N \rightarrow \infty}
    \stackh_k^{(m)H} \stackT  \stackh_k^{(m)} \overset{a.s.}{=}
    {R}_{\ell}(|a_{k}|^2, \widetilde{\tau}_k)
\end{equation}
with ${R}_{\ell}(|a_{k}|^2, \widetilde{\tau}_k)$ determined by
the following recursion
\begin{equation}\label{R_chip_asynch}
    {R}_{\ell}(\lambda , \tau)= \sum_{s=0}^{\ell-1} g({\matT}_{\ell-s-1}, \lambda, \tau
    ) {R}_s(\lambda, \tau)
\end{equation}
and
\begin{align}
    {\matT}_{\ell}(\Omega)& = \sum_{s=0}^{\ell-1} \mathbf{f}({R}_{\ell-s-1},
    \Omega) {\matT}_s(\Omega) & -\pi \leq \Omega \leq \pi & \label{T_chip_asynch} \\
 \mathbf{f}({R}_{\ell}, \Omega)& = {\beta} \int \lambda \boldsymbol{\Delta}_{\phi,r}(\Omega,{\tau})\boldsymbol{\Delta}_{\phi,r}^H (\Omega,{\tau})  {R}_{\ell}(\lambda, \tau) \mathrm{d}\,F_{|\matA|^2, T}(\lambda, \tau)  &
 -\pi \leq \Omega \leq \pi & \label{f_Rs_chip_async} \\
g({\matT}_{\ell}, \lambda, \tau) & =  \frac{\lambda}{2 \pi}
\int_{-\pi}^{\pi}\boldsymbol{\Delta}_{\phi,r}^H(\Omega,\tau)
{\matT}_{\ell}(\Omega) \boldsymbol{\Delta}_{\phi,r}(\Omega,{\tau})
\mathrm{d}\,\Omega &  & \label{g_Ts_chip_async} \end{align} with
\begin{equation}\label{delta_phi_r_def}\Mat{\Delta}_{\phi,r}(\Omega,{\tau})=\left(
\begin{array}{c}
  {\phi}(\Omega, \tau) \\
  {\phi}(\Omega,\tau-\frac{T_c}{r}) \\
  \vdots \\
  {\phi}(\Omega, \tau-\frac{T_c (r-1)}{r}) \\
\end{array}
\right).
\end{equation}
The recursion is initialized by setting ${\matT}_0(\Omega)=\Mat{I}_r$
and ${R}_0(\lambda, \tau)=1.$
\end{theor}
Theorem \ref{theo:diagonal_elements_chip_asynch} is proven in
Appendix \ref{section:proof_theo_diagonal_elements_chip_asynch}.

Note that the asymptotic diagonal elements of $\stackR^{\ell}$ depend on the delay $\tau_k$ only via the delay of a chip pulse waveform within a chip, i.e. via $\widetilde{\tau}_k,$ while any delay multiple of $T_c$ leaves  the diagonal elements unchanged.

From Theorem \ref{theo:diagonal_elements_chip_asynch} we can
obtain $m_{\stackR}^{(\ell)}$, the asymptotic eigenvalue
moment of the matrix $\stackR$ of order $\ell$ by using
the relation
\begin{equation*}
m_{\stackR}^{(\ell)}=\mathrm{E}\{ {R}_{\ell}(\lambda,
\tau )\}
\end{equation*}
where the expectation is taken over the limit distribution $F_{|\matA|^2,\widetilde{T}}(\lambda,\widetilde{\tau}).$
For $r=1$ and $F_{|\matA|^2,\widetilde{T}}(\lambda, \widetilde{\tau})=
F_{|\matA|^2}(\lambda) \delta(\widetilde{\tau})$, i.e. for synchronous systems
sampled at the chip rate, and  $\Phi(\omega)$ satisfying the Nyquist
criterion the recursive equations (\ref{T_chip_asynch}),
(\ref{f_Rs_chip_async}), and (\ref{g_Ts_chip_async})  reduce to
the recursion in \cite{cottatellucci:04} Theorem 1.

This theorem is very general and holds for all chip pulses of
practical interest. Furthermore, no constraint is imposed on the
time delay distribution. The choice of the
front end in this work is restricted only by the applicability of
(\ref{SINR_asynch_gen})  or (\ref{SINR_asynch_multistage_Wiener}),
which imply white noise at the front end. Then, since both Front-end A and Front -end B keep the sampled noise white, Theorem \ref{theo:diagonal_elements_chip_asynch} applies to both of them.

Now, we specialize Theorem \ref{theo:diagonal_elements_chip_asynch} to a case of theoretical and practical interest, where
sufficient statistics are utilized in the detection, the chip
pulse waveform $\phi(t)$ is band-limited, and the sequence of the empirical distribution functions of the time delays converges to a uniform distribution function as $K \rightarrow +\infty.$
{ The constraint to use sufficient statistics restricts the class of front-ends. The following results apply to Front-end A but, in general, not to Front-end B.  }
\begin{corollary}\label{cor_proposition_raised_cosine}
Let us adopt the same definitions as in Theorem
\ref{theo:diagonal_elements_chip_asynch} and let the same
assumptions of Theorem \ref{theo:diagonal_elements_chip_asynch} be
satisfied. Additionally, assume that the random variables
$\lambda$ and $\widetilde{\tau}$ in $F_{|\matA|^2, \widetilde{T}}(\lambda, \widetilde{\tau})$  are
statistically independent and the random variable $\widetilde{\tau}$ is
uniformly distributed. Furthermore, $\Phi(\Omega)$ is bounded in absolute value, and bandlimited with bandwidth
$B \leq \frac{r}{2 T_c}.$ Then,
given $(|a_{k}|^2, \widetilde{\tau}_k)$ and $m \in \mathbb{Z}$, the corresponding diagonal element
of the matrix $\stackR^{\ell}$  converges almost surely to a
deterministic value, conditionally on $|a_{k}|^2$,
\begin{equation*}
    \lim_{\K= \beta \N \rightarrow \infty}
    (\stackR^{\ell})_{k,m} = \lim_{\K= \beta \N \rightarrow \infty}
    \stackh_k^{(m)H} \stackT^{\ell-1} \stackh_{k}^{(m)} \overset{a.s.}{=}
    {R}_{\ell}(|a_{k}|^2)
\end{equation*}
with ${R}_{\ell}(\lambda)|_{\lambda=|a_{k}|^2}$ determined by the
following recursion:
\begin{equation*}
    {R}_{\ell}(\lambda )= \sum_{s=0}^{\ell-1} \lambda
    {R}_s(\lambda) \nu_{\ell-s-1}
\end{equation*}
and
\begin{align}
    {T}_{\ell}(\omega)& =\frac{r}{T_c} \sum_{s=0}^{\ell-1}   f({R}_{\ell-s-1}) \frac{1}{T_c} \left|\Phi\left(\omega\right)\right|^2  {T}_s(\omega)  & -2 \pi B \leq \omega \leq 2 \pi B & \nonumber \\
 f({R}_{\ell})& =\beta \int \lambda   {R}_{\ell}(\lambda) \mathrm{d}\,F_{|\matA|^2}(\lambda)  &
  & \nonumber \\
\nu_{\ell} & =  \frac{r}{2 \pi T_c} \int_{-2 \pi B} ^{2 \pi B}
\left|\Phi\left(\omega \right)\right|^2 T_{\ell}(\omega)
\mathrm{d}\,\omega. &  & \nonumber
\end{align}
The recursion is initialized by setting ${T}_0(\omega)=1$ and
${R}_0(\lambda)=1.$
\end{corollary}

Corollary \ref{cor_proposition_raised_cosine} is derived in
Appendix \ref{section:proof_cor_proposition_raised_cosine}.

The eigenvalue moments of $\stackR$  can be expressed in terms of the auxiliary quantities $f({R}_s)$ and $\nu_{s}$ in the recursion of Corollary
\ref{cor_proposition_raised_cosine} by the following expression:
\begin{equation*}
m_{\stackR}^{(\ell)} = \mathrm{E} \{{R}_{\ell}(\lambda)
\} = \sum_{s=0}^{\ell-1} f({R}_s) \nu_{\ell-s-1}.
\end{equation*}

Applying Corollary \ref{cor_proposition_raised_cosine}  we obtain
the following algorithm to compute the asymptotic limits of the
diagonal elements of $\stackR^{\ell}$ and
its eigenvalue moments.
\begin{algorithm}\label{alg:raised_cosine}\mbox{}\\
\begin{Ventry}{\textsc{Initialization}}
  \item[Initialization]\label{init_rc}  Let $\rho_0(z)=1$ and $ \mu_0(y)=1$.
  \item[$l^{\mathrm{th}}$ step]\label{gen_step_rc} \begin{itemize}
    \item Define $u_{\ell-1}(y)={r}y \mu_{\ell-1}(y)$  and write it as a polynomial in $y$.
    \item Define $v_{\ell-1}(z)=z \rho_{\ell-1}(z)$  and write it as a polynomial in $z$.
    \item  Define \begin{align}\mathcal{E}_s&=\frac{1}{2 \pi T_c^{s}} \int_{-2 \pi B}^{2 \pi B} T_c |\Phi( {\omega} )|^{2s}  \mathrm{d}\,\omega
        \label{epsilon_raised_cosine}\end{align} and  replace all monomials $y, y^2, \ldots, y^{\ell}$ in the polynomial $u_{\ell-1}(y)$ by $\mathcal{E}_1/T_c,$ $\mathcal{E}_2/T_c, \ldots,$ $ \mathcal{E}_{\ell}/T_c$,  respectively. Denote the result by $U_{\ell-1}$.
    \item  Define $m_{|\matA|^2}^s=\E\{ |a_{k}|^{2s}\}$ and replace all monomials $z, z^2, \ldots, z^{\ell}$ in the polynomial $v_{\ell-1}(z)$ by the moments $m_{{|\matA|^2}}^{(1)}$, $m_{{|\matA|^2}}^{(2)}$,\ldots, $m_{{|\matA|^2}}^{(\ell)}$, respectively. Denote the result by $V_{\ell-1}$.
    \item Calculate
    \begin{align}
    \rho_{\ell}(z)&=\sum_{s=0}^{\ell-1} z U_{\ell-s-1} \rho_s(z)
    \nonumber \\
    \mu_{\ell}(y) &=\frac{r}{T_c} \sum_{s=0}^{\ell-1} \beta y V_{\ell-s-1}
    \mu_s(y).
    \nonumber
    \end{align}
    \item Assign $\rho_{\ell}(\lambda) $ to ${R}^{\ell}(\lambda).$

Replace all monomials $z, z^2, \ldots, z^{\ell}$ in the polynomial
$\rho_{\ell}(z)$ by the moments $m_{{|\matA|^2}}^{(1)}$,
$m_{{|\matA|^2}}^{(2)}$,\ldots, $m_{|\matA|^2}^{(\ell)}$,
respectively, and assign the result to
$m_{\boldsymbol{\mathcal{R}}}^{(\ell)}$.
\end{itemize}
\end{Ventry}\end{algorithm}

Algorithm \ref{alg:raised_cosine} is derived in Appendix
\ref{section:algorithm_raised cosine}.

Interestingly, the recursive equations in Corollary
\ref{cor_proposition_raised_cosine} do not depend on the  time
delay $\tau_k$ of the signal of user $k$, i.e. the performance of
a CDMA system with multistage detection is independent of the
sampling instants and time delays if the assumptions of Corollary
\ref{cor_proposition_raised_cosine} on the chip waveforms and on
the time delays are satisfied.

Additionally, the dependence of ${R}^{\ell}(\lambda)$ on the chip
pulse waveforms becomes clear from Algorithm
\ref{alg:raised_cosine}: ${R}^{\ell}(\lambda)$ depends on $\Phi(\omega)$ through the quantities $\mathcal{E}_s$, $s=1,2, \ldots$,
defined in (\ref{epsilon_raised_cosine}).

By applying Algorithm \ref{alg:raised_cosine}  we compute the
first five asymptotic eigenvalue moments
\begin{eqnarray}
 m_{\boldsymbol{\mathcal{R}}}^{(1)}& =& \frac{r}{T_c} m_{|\matA|^2}^{(1)} \mathcal{E}_1  \nonumber \\
m_{\stackR}^{(2)} & = &\left(\frac{r}{T_c} \right)^2
[\beta (m_{|\matA|^2}^{(1)})^2
\mathcal{E}_2+ m_{{|\matA|^2}}^{(2)} \mathcal{E}_1^2] \nonumber \\
m_{\stackR}^{(3)} & =& \left(\frac{r}{T_c} \right)^3 [
\beta^2 \mathcal{E}_3 (m_{|\matA|^2}^{(1)})^3+3
m_{|\matA|^2}^{(2)} \mathcal{E}_2 \beta m_{|\matA|^2}^{(1)}
\mathcal{E}_1 + m_{|\matA|^2}^{(3)}
\mathcal{E}_1^3] \nonumber \\
m_{\stackR}^{(4)} & =&  \left(\frac{r}{T_c} \right)^4 [2 \beta^2 \mathcal{E}_2^2 m_{|\matA|^2}^{(2)}  (m_{|\matA|^2}^{(1)})^2 + 4 \beta \mathcal{E}_1^2 \mathcal{E}_2  m_{|\matA|^2}^{(3)}  m_{|\matA|^2}^{(1)}+4 \beta^2 \mathcal{E}_1 \mathcal{E}_3 m_{|\matA|^2}^{(2)}  (m_{|\matA|^2}^{(2)})^2 + \beta^3 \mathcal{E}_4  (m_{|\matA|^2}^{(1)})^4  \nonumber \\
&& + 2 \beta \mathcal{E}_1^2 \mathcal{E}_2  (m_{|\matA|^2}^{(2)})^2   + \mathcal{E}_1^4 m_{|\matA|^2}^{(4)}]  \nonumber \\
m_{\stackR}^{(5)} & =& \left(\frac{r}{T_c} \right)^5 [
m_{|\matA|^2}^{(5)} \mathcal{E}_5 \beta^4+ \mathcal{E}_1^5
(m_{|\matA|^2}^{(1)})^5  + 5 \beta^3 \mathcal{E}_1
  \mathcal{E}_4 m_{|\matA|^2}^{(2)} (m_{|\matA|^2}^{(1)})^3 +5  \beta^3  \mathcal{E}_3 \mathcal{E}_2 m_{|\matA|^2}^{(2)} (m_{|\matA|^2}^{(1)})^3    \nonumber \\
&& + 5  \beta^2 \mathcal{E}_3 \mathcal{E}_1^2 m_{|\matA|^{(2)}}^3
(m_{|\matA|^2}^{(1)})^2 + 5 \beta^2  \mathcal{E}_1^2 \mathcal{E}_3
(m_{|\matA|^2}^{(2)})^2 m_{|\matA|^2}^{(1)} + 5  \beta^2
\mathcal{E}_1 \mathcal{E}_2^2 (m_{|\matA|^2}^{(2)})^2
m_{|\matA|^2}^{(1)} \nonumber \\
&& + 5 \beta^2 \mathcal{E}_2^2
\mathcal{E}_1 m_{|\matA|^2}^{(3)} (m_{|\matA|^2}^{(1)})^2 + 5
\beta \mathcal{E}_2 \mathcal{E}_1^3 m_{|\matA|^2}^{(4)}
m_{|\matA|^2}^{(1)} + 5 \mathcal{E}_2 \mathcal{E}_1^3
m_{|\matA|^2}^{(3)} m_{|\matA|^2}^{(2)}]. \nonumber
\end{eqnarray}

In general, the eigenvalue moments of $\stackR$ depend only on the system load $\beta$, the
sampling rate $\frac{r}{T_c}$, the eigenvalue distribution of the
matrix $\matA^H \matA$, and $\mathcal{E}_s$, $ s \in
\mathbb{Z}^{+}$. The latter coefficients take into account the
effects of the shape of the chip pulse or, equivalently, of the
frequency spectrum of the function ${\phi}(t)$. The asymptotic
limits of the diagonal elements of the matrix
$\stackR^{\ell}$ corresponding to user $k$ depends also
on $|a_{k}|^2$ but not on the time delay $\tau_k.$

In the special case of chip pulse waveforms $\psi(t)$  having
bandwidth not greater than the half of the chip rate, i.e. $B \leq
\frac{1}{2T_c} $ the result of Corollary
\ref{cor_proposition_raised_cosine} holds  for any sets of time
delays included synchronous systems.

In Theorem
\ref{theo:diagonal_elements_small_bandwidth}, chip pulse waveforms
with bandwidth $B \leq \frac{1}{2 T_c}$ are considered and the
diagonal elements of $\stackR^s$  are shown to be independent of the time
delays of the active users.
\begin{theor}\label{theo:diagonal_elements_small_bandwidth}
Let the definitions of Theorem
\ref{theo:diagonal_elements_chip_asynch} hold.

We assume that the function $\Phi(\omega)$ is bounded in
absolute value and has support $\mathcal{S} \subseteq \left[-\frac{\pi}{T_c}, \frac{\pi}{T_c} \right]$. The vectors $\vecs_{k}$ are
independent with i.i.d. Gaussian elements $s_{nk} \in \mathbb{C}$
such that $\E\{s_{nk}\}=0$ and  $\E\{|s_{nk}|^2
  \}=\frac{1}{\N}.$
Furthermore, the elements $a_{k}$ of the matrix $\matA$ are
uniformly bounded for any $K.$  The sequence of the empirical
distributions $F_{|\matA|^2}^{(\K)}(\lambda)=\frac{1}{\K}
\sum_{k=1}^{\K} 1(\lambda-|a_{k}|^2)$ converges in law almost
surely, as $\K \rightarrow \infty$, to a non-random distribution
function $F_{|\matA|^2}(\lambda).$

Then, given $|a_{k}|^2$, the $n$-th diagonal element of
the matrix $\stackR^{\ell},$ with $n \! \mod \! K\!=\!k,$  converges almost surely to a
deterministic value, conditionally on $|a_{k}|^2$,
\begin{equation*}
    \lim_{\K= \beta \N \rightarrow \infty}
    (\stackR^{\ell})_{k,m} = \lim_{\K= \beta \N \rightarrow \infty}
    \stackh_k^{(m)H} \stackT^{\ell-1} \stackh_k^{(m)} \overset{a.s.}{=}
    {R}_{\ell}(|a_{k}|^2)
\end{equation*}
with ${R}_{\ell}(|a_{k}|^2)$ determined by the following
recursion
\begin{equation}\label{R_small_bandwidth}
    {R}_{\ell}(\lambda)= \sum_{s=0}^{\ell-1} \lambda {R}_s(\lambda) \nu_{\ell-s-1}
\end{equation}
and
\begin{align}
    {T}_{\ell}(\omega)& = \frac{r}{T_c}  \sum_{s=0}^{\ell-1} \beta {f}({R}_{\ell-s-1}) \frac{1}{T_c} |\Phi( \omega)|^2 {T}_s(\omega) & \omega \in \mathcal{S} & \label{T_small_bandwidth} \\
{f}({R}_{\ell})& = \int \lambda {R}_{\ell}(\lambda)
\mathrm{d}\,F_{|\matA|^2}(\lambda)   &
  & \label{f_Rs_small_bandwidth} \\
  \nu_{\ell}&=\frac{r^2}{2 \pi T_c}  \int_{\mathcal{S}}  |\Phi( \omega)|^2
{T}_{\ell}(\omega)  \mathrm{d}\,\omega .& & \label{g_Ts_small_bandwidth}
\end{align}
The recursion is initialized by setting ${T}_0(\omega)=\frac{T_c}{r}$
and ${R}_0(\lambda)=1.$
\end{theor}
Theorem \ref{theo:diagonal_elements_small_bandwidth} is shown in
Appendix
\ref{section:proof_theo_diagonal_elements_small_bandwidth}.
{It applies to Front-end A but, in general, not to Front-end B since Front-end B implies the use of root Nyquist pulses. }
It is
straightforward to verify that Algorithm \ref{alg:raised_cosine}
can be applied to determine ${R}_{\ell}(\lambda),$ the asymptotic
limit of the diagonal elements  and the eigenvalue moments of
matrices $\stackR$  satisfying the conditions of Theorem
\ref{theo:diagonal_elements_small_bandwidth}.

The mathematical results presented in this section have important
implications on the design and analysis of asynchronous CDMA
systems and linear detectors for asynchronous CDMA systems. We
elaborate on them in the following section.

\section{Effects of Asynchronism, Chip Pulse Waveforms, and Sets of Observables}\label{chap:async_sec:chip_async_symb_sync_subsec:pulse_shape}
The theoretical framework developed in Section \ref{IVb}
enables the analysis and design of linear multistage detectors for
CDMA systems using optimum and suboptimum statistics and possibly
non ideal chip pulse waveforms. In this section we focus on the
following aspects:
\begin{enumerate}
  \item Analysis of the effects of chip pulse waveforms and time
  delay distributions when the multistage detectors are fed by
  sufficient statistics.
  \item Impact of the use of sufficient and suboptimum statistics on the complexity
  and the performance of multistage detectors.
\end{enumerate}

\subsection{Sufficient Statistics}\label{chap:async_sec:chip_async_symb_sync_subsubsec:sufficient_statistics}
Sufficient statistics impaired by discrete additive Gaussian noise
are obtained as output of detector Type A. For chip pulse
waveforms with bandwidth $B \leq \frac{1}{2 T_c}$ and any set of
time delays, Theorem \ref{theo:diagonal_elements_small_bandwidth}
applies. For $B > \frac{1}{2T_c}$ and uniform time delay
distribution,  Corollary \ref{cor_proposition_raised_cosine} holds.
In both cases, as $K,N \rightarrow \infty $ with constant ratio
the diagonal elements of the matrix $\stackR^{\ell}$ and
the eigenvalue moments $m_{\stackR}^{(\ell)}$ can be
obtained from Algorithm \ref{alg:raised_cosine}. As a consequence of
(\ref{SINR_asynch_gen}), the performance of the large class of
multiuser detectors that admit a representation as multistage
detectors depends only on the diagonal elements
$\stackR^{\ell}$ and the variance of the
noise. In large CDMA systems, the SINR
depends on the system load $\beta,$ the sampling rate
$\frac{r}{T_c}$, the limit distribution of the received powers
$F_{|\matA|^2}(\lambda),$ the variance of the noise $\sigma^2,$
the coefficients $\mathcal{E}_{\ell},$ $\ell \in \mathbb{Z}^{+}$
and the received powers $|a_{k}|^2,$ but it is independent of the
time delay $\tau_k$, in general.
%
For $B \leq \frac{1}{2T_c},$ the SINR is also independent
of the time delay distribution. Therefore we can state the
following corollary.
\begin{corollary}\label{cor:independence_delay_distribution}
If the bandwidth of the chip pulse waveform satisfies the
constraint $B \leq \frac{1}{2 T_c },$ large synchronous and
asynchronous CDMA systems have the same performance in terms of
SINR when a linear detector that admits a representation as
multistage detector is used at the receiver.
\end{corollary}
If the time delays and the received amplitudes of the signals are
known at the receiver and the sampling rate satisfies the
conditions of the sampling theorem, synchronous and asynchronous
CDMA systems have the same performance. In
\cite{mantravadi:02} is established the equivalence between
synchronous and asynchronous CDMA systems using an ideal Nyquist
sinc waveform ($B=\frac{1}{2 T_c}$) and linear MMSE detector.
Corollary \ref{cor:independence_delay_distribution} generalizes
that equivalence to any kind of chip pulse waveforms with
bandwidth $B \leq \frac{1}{2 T_c}$ and any linear multiuser
detector with a multistage representation.

By inspection of Algorithm \ref{alg:raised_cosine} we can verify
that the dependence of ${R}_{\ell}(|a_{k}|^2)$ and
$m_{\stackR}^{(\ell)}$ on the sampling rate
$\frac{r}{T_c}$ can be expressed by the following relations
\begin{equation}
{R}_{\ell}(|a_{k}|^2)=\left(\frac{r}{T_c} \right)^{\ell}
{R}^{*}_{\ell}(|a_{k}|^2)
\end{equation}
and
\begin{equation}
m_{\stackR}^{(\ell)}=\left(\frac{r}{T_c} \right)^{\ell}
m_{\stackR}^{* \,(\ell)}
\end{equation}
where ${R}^{*}_{\ell}(|a_{k}|^2)$ and $ m_{\stackR}^{*
\,(\ell)}$ are independent of the sampling rate $\frac{r}{T_c}. $
Thanks to this particular dependence and the fact that
$\sigma^2=\frac{r}{T_c}N_0,$ the quadratic forms appearing in (\ref{SINR_asynch_gen}) $\Mat{\xi}_{k,m}^{H}\Mat{\Xi}_{k,m}^{-1}\Mat{\xi}_{k,m},$
$\Mat{\xi}_{k,m}^{H}\Mat{\Xi}^{-1}\Mat{\xi},$  and
$\Mat{\xi}^{H}\Mat{\Xi}^{-1}\Mat{\Xi}_{k,m}\Mat{\Xi}^{-1}\Mat{\xi},$  are independent of the sampling rate for large systems,  when specialized to multistage Wiener filters and to polynomial expansion detectors. Thus, the
large system performance of (i) linear multistage detectors
optimum in a mean square sense (see
(\ref{SINR_asynch_multistage_Wiener})), (ii) of the polynomial
expansion detectors and (iii) the matched filters is independent of
the sampling rate. This property is not general. Detectors that
are not designed to benefit at the best from the available
sufficient statistics may improve their performance using
different sets of sufficient statistics. Therefore, the large
system performance of other multistage detectors like PIC
detectors depends on the sampling rate and can eventually improve
by increasing the oversampling factor $r.$

Given a positive real $\gamma,$  let us consider the chip pulse
\begin{equation}\label{sinc_waveform}
\Phi(\omega) =
  \begin{cases}
    \sqrt{\frac{T_c}{\gamma}} & \text{for  } |\omega| \leq \frac{\pi \gamma}{T_c}, \\
    0 & \text{otherwise}
  \end{cases}
\end{equation}
corresponding to a sinc waveform with bandwidth
$B=\frac{\gamma}{2T_c}$ and unit energy. For waveform
(\ref{sinc_waveform}) with $\gamma=1,$ $T_c=1,$ and $r=1$
Algorithm \ref{alg:raised_cosine} reduces to Algorithm 1 in
\cite{cottatellucci:05} for synchronous systems. Let us denote by
${R}_{\ell}^{(\text{syn})}(|a_{k}|^2, \beta)$ and
$m_{\stackR^{\mathrm{(syn)}}}^{(\ell)}(\beta)$ the values
of ${R}_{\ell}(|a_{k}|^2)$ and $m_{\stackR}^{(\ell)}$
for such a synchronous case and system load $\beta.$ Then, in general, for
chip pulse waveform (\ref{sinc_waveform}) Algorithm
\ref{alg:raised_cosine} yields
\begin{equation}\label{relation_sinc_vs_syn_diag_elem}
{R}_{\ell}^{(\text{sinc})}(|a_{k}|^2)=\left(\frac{r}{T_c}
\right)^{\ell}
\mathcal{R}^{\mathrm{(syn)}}_{\ell}\left(|a_{k}|^2,
\frac{\beta}{\gamma} \right)
\end{equation}
and
\begin{equation}
m_{\stackR^{(\text{sinc})}}^{(\ell)}=\left(\frac{r}{T_c}
\right)^{\ell} m_{\stackR^{(\text{syn})}}^{(\ell)}\left(
\frac{\beta}{\gamma} \right).
\end{equation}
Therefore, the same property  pointed out in part I of this paper
\cite{cottatellucci:07b} for linear MMSE detectors holds for
several multistage detectors (namely, multistage Wiener filters,
polynomial expansion detectors, matched filters): In a large
asynchronous CDMA system using a sinc function with bandwidth
$\frac{\gamma}{2T_c}$ as chip pulse waveform and system load
$\beta$ any multistage detector whose performance is independent
of the sampling rate performs as well as in a large synchronous
CDMA system with modulation based on root Nyquist chip
pulses and system load $\beta^{\prime}= \frac{\beta}{\gamma}.$

The comparison of synchronous and asynchronous systems with equal
chip pulse waveforms enables us to analyze the effects on the
system performance of the chip pulse waveforms jointly with the
effects of the distribution of time delays. We elaborate on these
aspects focusing on root raised cosine chip-pulse waveforms
with roll-off $\vartheta \in [0,1]$ and on chip pulse waveforms
(\ref{sinc_waveform}) with $\gamma \in [1,2]$. To simplify the
notation, we assume $T_c=1$. Let
\begin{equation*}
S(\omega)=  \begin{cases} 1 & 0 \leq
|\omega| \leq \pi (1-\vartheta) \\
\frac{1}{2} \left(1-\sin \left(\frac{|x|-\pi}{2 \vartheta} \right)
 \right) & \pi (1-\theta) \leq
|\omega| \leq \pi( 1+\vartheta) \\
0 & |\omega| \geq \pi (1+\vartheta).
 \end{cases}
\end{equation*}
The energy frequency spectrum of a root raised cosine
waveform with unit energy is given by $|\Psi_{\mathrm{sqrc}}(\omega)|^2=S(\omega).$ The large system analysis of an asynchronous CDMA
system using root raised cosine chip pulse waveform is
obtained applying Algorithm \ref{alg:raised_cosine}.  The
corresponding coefficients  $\mathcal{E}_{\mathrm{sqrc},s},$
$s=\mathbb{Z}^{+}$, are given by
\begin{equation*}
\mathcal{E}_{\mathrm{sqrt},s} \!\!=\!\! 2^s (1-\gamma)+\frac{1}{\pi} \!
\int_{\pi(1\!-\!\gamma)}^{\pi (1+\gamma)}\!\! \sin^s
\left(\frac{1}{2\gamma}\left(\pi \!- \! \omega \right)\right)
\!\! \mathrm{d}\omega.   \nonumber
\end{equation*}
It is well known that in a synchronous CDMA system the performance
is maximized using root Nyquist waveforms. In this case the
performance is independent of the specific waveform and the
bandwidth. It equals the performance of a large synchronous system
using the sinc function with bandwidth $\frac{1}{2T_c}$ as chip
pulse. Since the root raised cosine pulses are root
Nyquist waveforms, they attain the maximum SINR in synchronous
systems. The large system performance of multistage Wiener filters
for synchronous CDMA systems with a root raised cosine
waveform is obtained making use of
(\ref{SINR_asynch_multistage_Wiener}) and Algorithm
\ref{alg:raised_cosine} with $r=1$ and $\mathcal{E}_s=1,$ $s\in
\mathbb{Z}^{+}.$

In general, chip pulse waveform $(\ref{sinc_waveform})$ is not a
root Nyquist waveform. For this reason the performance
analysis of linear multistage Wiener filters for synchronous CDMA
sytems \cite{li:04}, \cite{cottatellucci:05} is not applicable. In
this case characterized by interchip interference we can still
apply Theorem \ref{theo:diagonal_elements_chip_asynch}, sampling
at rate $\frac{2}{T_c}$  and assuming a Dirac function
$f_T(\tau)=\delta(\tau)$ as probability density function of the
time delays. For the chip pulse waveform (\ref{sinc_waveform}), the
matrix $\Mat{Q}(\Omega)=\Mat{\Delta}_{\Phi,2}(\Omega,0)
\Mat{\Delta}_{\Phi,2}^H(\Omega,0)$ used in the recursion of Theorem
\ref{theo:diagonal_elements_chip_asynch} is given by
\begin{equation*}
\Mat{Q}(\Omega)= \begin{cases} \frac{1}{\gamma}
\left(\begin{array}{cc}
  1 & \mathrm{e}^{-j \frac{\Omega}{2}}  \\
  \mathrm{e}^{j \frac{\Omega}{2}}  & 1 \\
\end{array} \right) &  |\Omega| \leq 2 \pi \left(1-\frac{\gamma}{2} \right) \\ \frac{1}{\gamma}
\left(\begin{array}{cc}
  4 & 0 \\
  0 & 0 \\
\end{array} \right) & 2 \pi \left( 1-\frac{\gamma}{2}\right) \leq |\Omega| \leq \pi.   \\ \end{cases}
\end{equation*}
The large system analysis in the asynchronous case with chip pulse
(\ref{sinc_waveform}) can be readily performed making use of
(\ref{SINR_asynch_multistage_Wiener}) and
(\ref{relation_sinc_vs_syn_diag_elem}).

\begin{figure*}
\noindent \begin{minipage}[t]{.49\linewidth} \begin{minipage}[t]{\linewidth} %
\centering
\epsfig{file=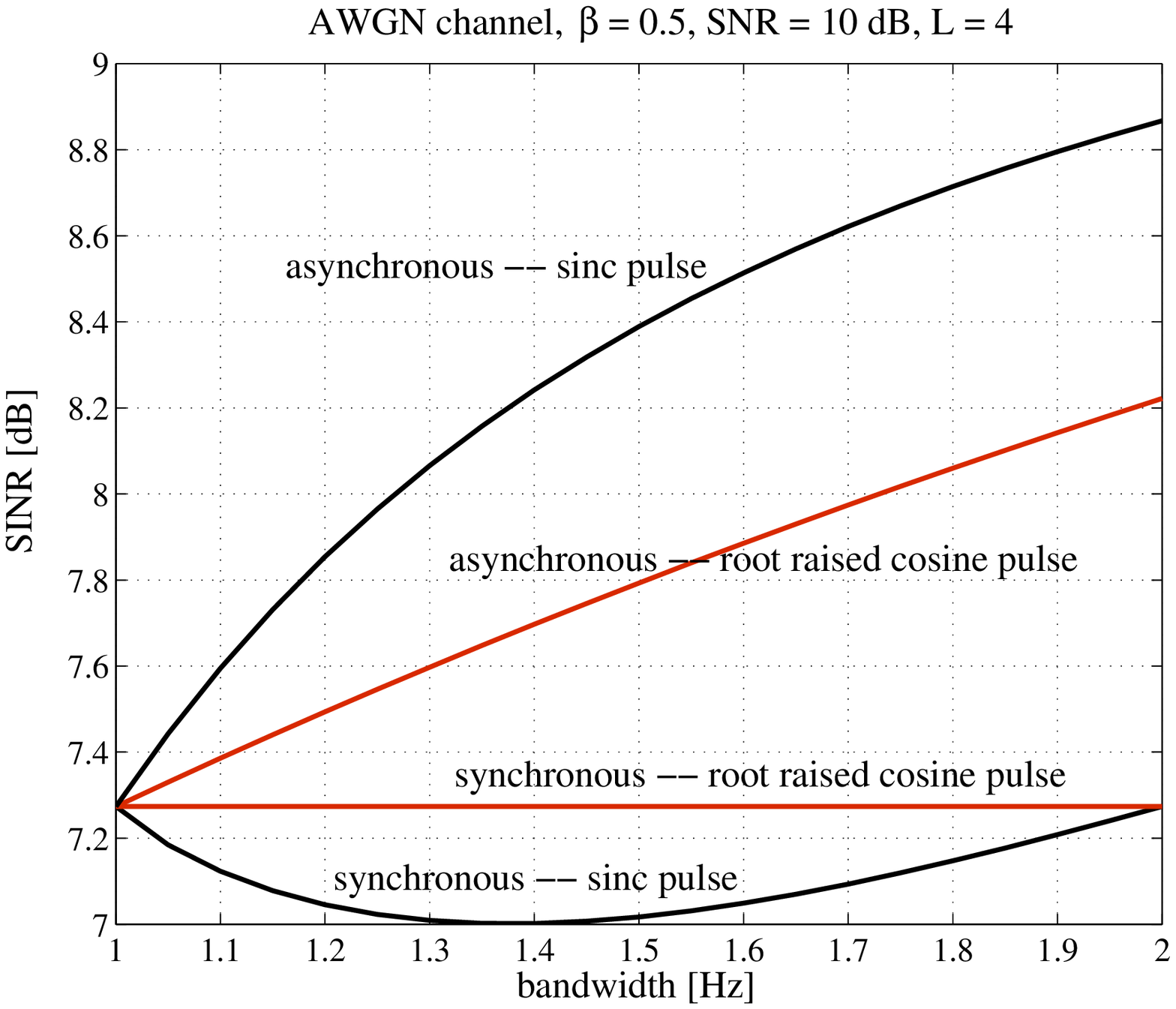,width=\linewidth}
\caption{Output SINR of a multistage Wiener filter with $L=4$
versus bandwidth. CDMA systems with equal received powers,
root raised cosine chip waveforms or sinc pulses, system
load $\beta=\frac{1}{2}$ and input $\mathrm{SNR}=10$ dB are
considered.
}\label{fig_square_root_raised_cosine_vs_sync_vs_bandwidth}
\end{minipage} \par\vspace*{0mm}\end{minipage}
\hfill \vspace*{0cm}
\begin{minipage}[t]{.49\linewidth}
\centering
\epsfig{file=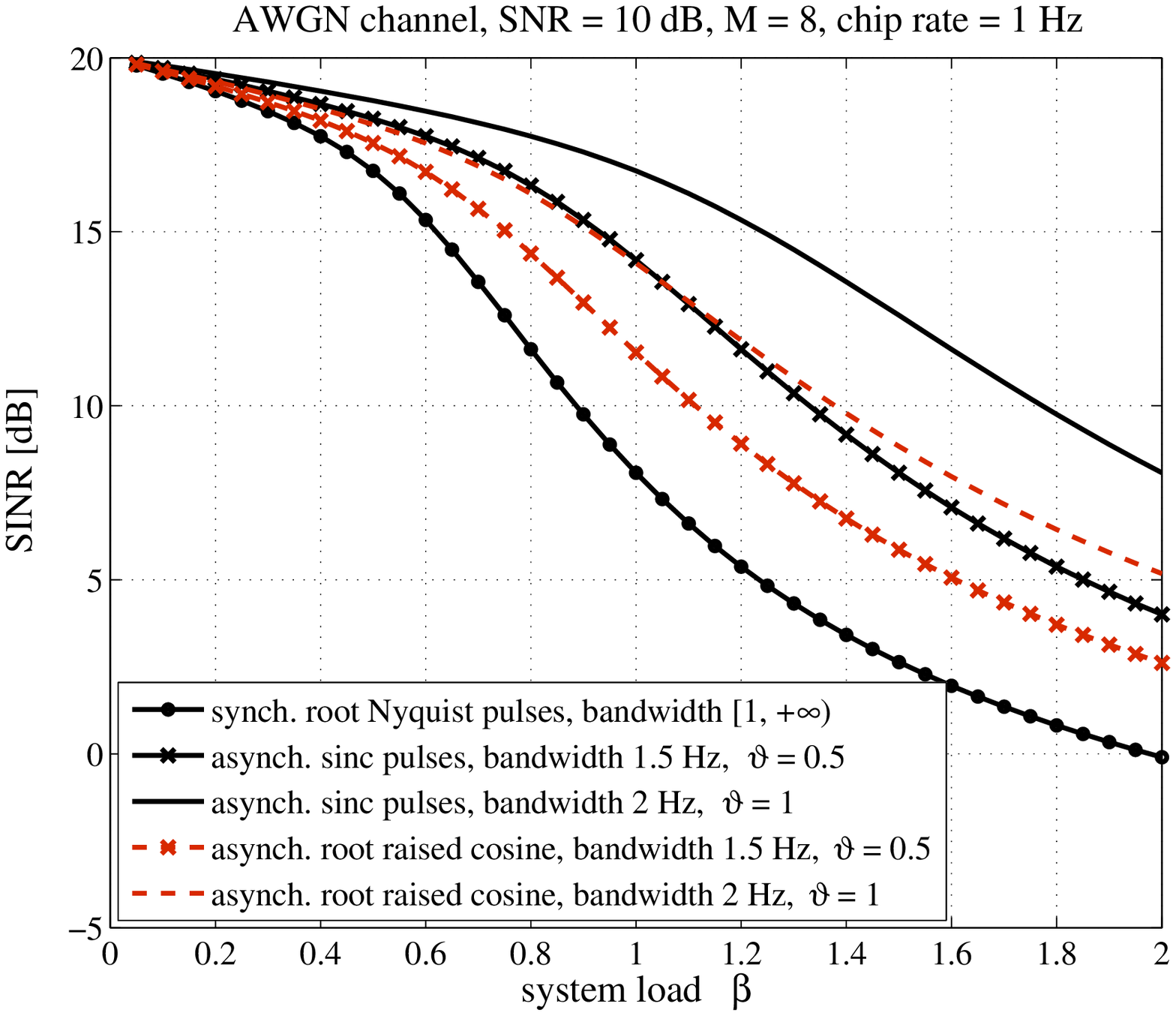,width=\linewidth}
\caption{Output SINR of a multistage Wiener filter with $L=8$
versus the system load. Asynchronous CDMA systems with equal
received powers, root raised cosine chip waveforms   or
sinc pulses with bandwidth $B=1.5, 2$ Hz, input $\mathrm{SNR}=10$
dB are compared to synchronous CDMA systems with root
Nyquist chip pulses. } \label{Fig_sync_vs_async_load}
\end{minipage}\end{figure*}

In Figure \ref{fig_square_root_raised_cosine_vs_sync_vs_bandwidth}
the large system  SINR at the output of a multistage Wiener filter
with $L=4$ is plotted as a function of the bandwidth for synchronous
and asynchronous CDMA systems based on modulation by  root
raised cosine or by pulse (\ref{sinc_waveform}). We assume perfect
power control, i.e. $\matA=\Mat{I},$ system load $\beta=0.5$, and
input $\mathrm{SNR}=10$ dB.

It is well known from the theory on synchronous CDMA that interchip interference colors the discrete-time spectrum of the signal and degrades performance.
Consistently with this effect, Figure \ref{fig_square_root_raised_cosine_vs_sync_vs_bandwidth}
shows that synchronous CDMA root raised cosine pulses outperform sinc pulses with non-integer ratios of bandwidth to chip rate, since the formers avoid interchip interference.
Asynchronous CDMA
systems with both chip pulse waveforms widely outperform the
corresponding synchronous systems. In contrast to the synchronous
case, sinc pulses exploit the additional degrees of freedom introduced by increasing the bandwidth better
than root raised cosine pulses, since they do not color the spectrum in continuous time domain. Thus, an asynchronous CDMA
system with sinc pulses considerably  outperforms
a system using root raised cosine pulses.
Note that for asynchronous systems, the spectral shape in continuous time is relevant, while for synchronous systems the spectral shape in discrete time matters.
In both cases the spectrum should be as white as possible to achieve high performance. For asynchronous systems, the spectrum is the less colored, the closer the delay distribution resembles an (eventually discrete) uniform distribution.

In Figure \ref{Fig_sync_vs_async_load} the SINR at the output of a
multistage Wiener filter with $L=8$ is plotted as a function of
the system load, parametric in the bandwidth, for
$\mathrm{SNR}=10$ dB. The improvement achievable by asynchronous
systems over synchronous systems increases as the the system load
increases.

\subsection{Chip Rate Sampling}\label{chap:async_sec:chip_async_symb_sync_subsubsec:square_root_Nyquist_waveforms}

Chip rate sampling is a widely used approach to generate
statistics for asynchronous CDMA systems. It implies the use of
root Nyquist chip pulses and makes use of front end Type B.
Hereafter, we refer to these CDMA systems as systems B, while we
refer to the systems that use sufficient statistics from a front
end Type A as systems A.

A bound on the performance of systems B with linear MMSE detectors
is in \cite{schramm:99}. The performance analysis of linear
multistage detectors as $K, N \rightarrow \infty$ with
$\frac{K}{N} \rightarrow \beta$ can be performed applying Theorem
\ref{theo:diagonal_elements_chip_asynch} to the chip pulse
waveform at the output of the chip matched filter $\Phi(\omega)=
\frac{1}{\sqrt{E_{\psi}}} |\Psi(\omega)|^2$ and assuming $r=1.$
In order to elaborate further on systems B we focus on the
root-raised cosine chip pulse with roll-off $\theta $
\cite{huber:92}
\begin{equation}
{\psi}(t)=\frac{4 \theta (\frac{t}{T_c}) \cos(\pi
(1+\theta)\frac{t}{T_c})+ \sin(\pi(1-\theta)\frac{t}{T_c})}{\pi t
(1-(4 \theta  \frac{t}{T_c})^2)} \qquad \theta \in [0,1].
 \end{equation}
In this case, the matrix function $\Mat{Q}(\Omega, \tau)=
\boldsymbol{\Delta}_{\phi,1}(\Omega,\tau)
\boldsymbol{\Delta}_{\phi,1}^H(\Omega,\tau)$ occurring in Theorem
\ref{theo:diagonal_elements_chip_asynch} reduces to the scalar
function
\begin{equation*}
\Mat{Q}(\Omega,{\tau})= \begin{cases} \frac{1}{2}+ \frac{1}{2}
\sin^2\left(\frac{1}{2\theta}(\Omega+\pi) \right) + \frac{\cos
2 \pi \tau}{2} \left(1-
\sin^2\left(\frac{1}{2 \theta}(\Omega+\pi) \right)\right) &
-\pi \leq \Omega \leq - \pi (1-\theta) \\
1 & - \pi (1-\theta) \leq \Omega \leq \pi (1-\theta) \\
\frac{1}{2}+ \frac{1}{2}
\sin^2\left(\frac{1}{2\theta}(\Omega-\pi) \right) + \frac{\cos
2 \pi \tau}{2} \left(1-
\sin^2\left(\frac{1}{2 \theta}(\Omega-\pi) \right) \right) &
\pi (1-\theta) \leq \Omega \leq \pi.
\end{cases}
\end{equation*}
due to the fact that $r=1$.  Equal received powers, system load
$\beta=\frac{1}{2}$, multistage Wiener filters with $L=3$ define
the scenario we consider for the asymptotic analysis.

The analysis shows a strong dependence of the performance on the
time delays. As expected, it is possible to verify that the best
SINR is obtained when the sampling instants coincide with the time
delays of the user of interest.

\begin{figure*}
\centering
\epsfig{file=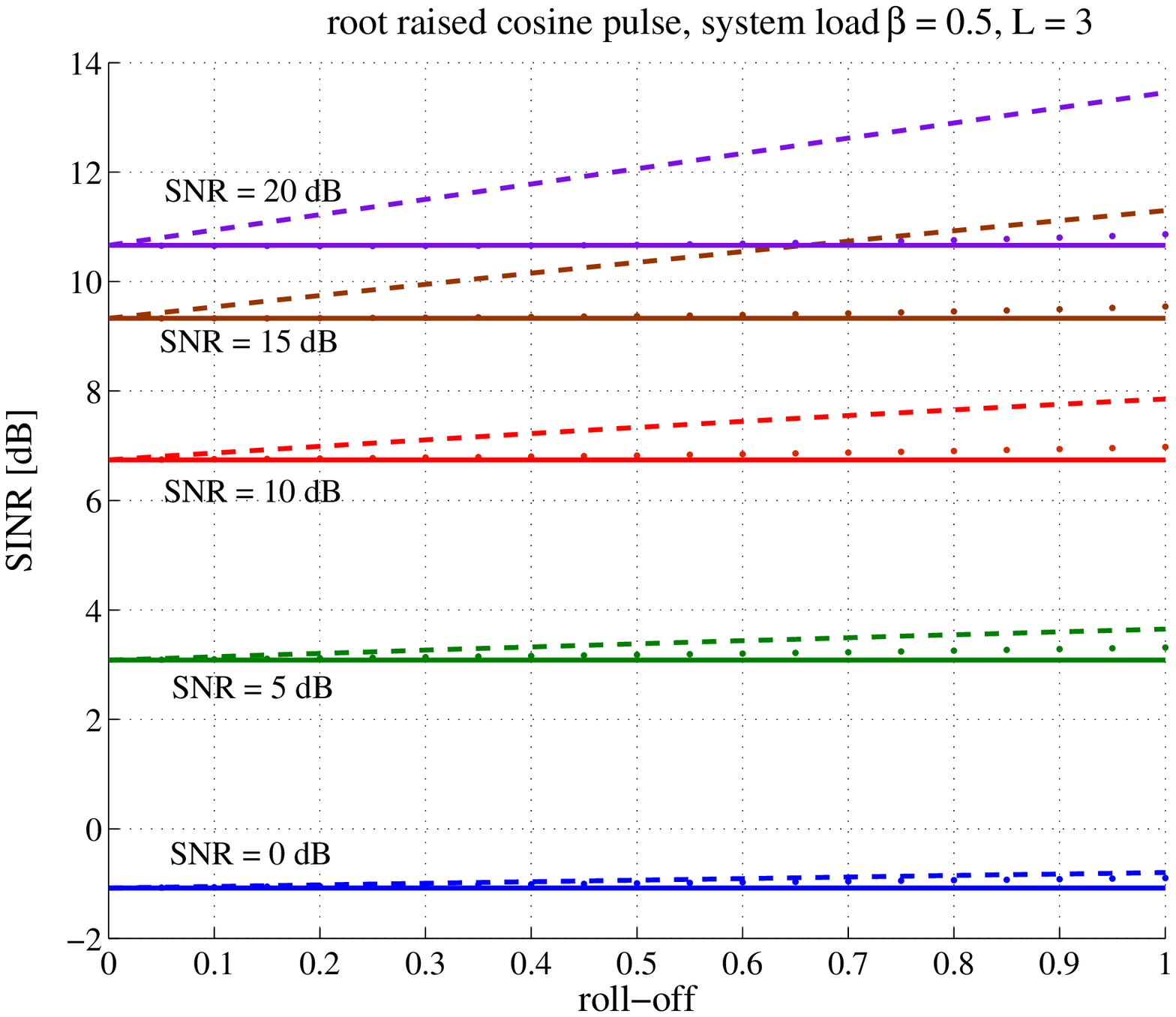,width=0.5\linewidth}
{\caption{\small Asymptotic output SINR of a multistage Wiener
filter with $L=3$ versus the roll-off $\theta $ as front-end A
(dashed lines) and front-end B (dots) are in use in an
asynchronous CDMA system. The solid lines show the reference
performance in synchronous CDMA systems. The curves are parametric
in the input SNR with SNR varying between $0$ dB and $20 $ dB in
steps of $5$ dB. }\label{Fig_fe_A_vs_fe_B}}
\end{figure*}

In Figure \ref{Fig_fe_A_vs_fe_B} we compare the performance of
system B with root raised cosine chip pulse to the SINR of
a system A with the same modulating pulse. In the comparison we
consider the best SINR for system B obtained when the sampling
times coincide with the time delays of the user of interest. The
curves represent the output SINR as a function of the roll-off
$\theta$ parameterized with respect to SNR. The parameter (SNR)
varies from 0 dB to 20 dB in steps of 5 dB. As reference we also
plot the performance of synchronous CDMA systems. As expected,
multistage detectors with front-end A outperform the corresponding
multistage detectors with front-end B.

Interestingly, while linear multistage detectors and asynchronism
in system A can compensate to some extent for the loss in spectral
efficiency caused by the increasing roll-off and typical of
synchronous CDMA systems such a compensation is not possible in
systems B. Systems B behave similarly to synchronous CDMA systems.
In fact, the SINR for system B is very close to the performance of
synchronous systems for any SNR level.

{A thorough explanation of these properties based on general analytical results is in Part I Section V \cite{cottatellucci:07b}. We recapitulate the main idea briefly here. The performance of a large asynchronous CDMA system is governed by an $r \times r$ matrix function in the frequency domain (eq. (24) in \cite{cottatellucci:07b})\footnote{Note that the matrices $\matT_{\ell}(\Omega)$ in Theorem \ref{theo:diagonal_elements_chip_asynch} can be interpreted as expansion coefficients of this matrix.}. To give an intuition, the system is then equivalent to a MIMO system with $r$ transmit and $r$ receive antennas. The structure of this matrix is such that the matrix is necessarily rank one for synchronous CDMA systems. Thus, only one dimension of the signal space is spanned. On the contrary, for arbitrary delay distributions, i.e. in general for asynchronous systems, the rank of the MIMO system can be higher, eventually, up to $r.$ This implies that asynchronous systems span more of the available dimensions of the signal space resulting in better exploitation of it. When the received signal is sampled at the chip rate, as in the case of Front-end B, and $r=1$ the processed signal for an asynchronous system only spans a single dimension, just like in synchronous systems, and the performances of synchronous and asynchronous systems are very similar.    }

Since the SINR in system B heavily depends on the
sampling instants with respect to $\tau_k$, different statistics
are needed for the detection of different users in order to obtain
good performance. As consequence, joint detection is not feasible
and each user has to be detected independently. This is a
significant drawback when several or all users have to be
detected (e.g. uplink) and has a relevant impact on the complexity
of the system.  For example, the complexity order per bit of a
multistage Wiener filter or polynomial expansion detector is linear in $rK$ in system A while the complexity order per bit of the
same detectors is quadratic in $K$ in system B. A similar increase
in complexity can be noticed also for other detectors (e.g. linear
MMSE detectors, or any multistage detector).

\section{Conclusions}\label{chap:async_sec:conclusions}

In Part II of this work we provided guidelines for the design of
asynchronous CDMA systems via the analysis of the effects of chip
pulse waveforms, time delay distributions, sufficient and
suboptimum observables on the complexity and performance of the
broad class of multiuser detectors with multistage representation.

Similarly to the results obtained in part I of this article
\cite{cottatellucci:07b}, i.e.\ the chip-pulse constrained spectral efficiency
and the performance of linear MMSE detectors, multistage detectors
show performance independent of the time delays of the active
users if the bandwidth of the chip pulse waveform is not greater
than half of the chip rate, i.e. $B \leq \frac{1}{2T_c}.$ Above
that threshold the performances of linear multistage detectors
depend on the time delay distributions and asynchronous CDMA systems
outperform synchronous CDMA systems.

The framework presented here enabled the analysis of
optimum and suboptimum multistage detectors based on  front ends whose sampled noise outputs are white. We focused on multistage detectors using statistics
(\ref{sufficient_statistics}), which are sufficient,  or
observables  (\ref{suboptimal_statistics}), which are suboptimum.
In the two cases of (i) chip pulses with bandwidth $B \leq
\frac{1}{2T_c}$ and (ii) chip pulses with bandwidth $B >
\frac{1}{2 T_c}$, sufficient statistics, and uniform distribution,
the effects of the chip pulse waveforms on the detector
performance are described by the coefficients
$\mathcal{E}_s=\frac{1}{2 \pi T_c^{s-1}} \int_{-2\pi B}^{2 \pi B} |\Psi(\omega)|^{2s} \mathrm{d}\omega.$ The output SINR of linear MMSE detectors,
multistage Wiener filters, polynomial expansion detectors, and
matched filters is independent of the sampling rate. In contrast,
the output SINR of other multistage detectors like PIC detectors
depends on the sampling rate and increases with it.

Comparing the performance of synchronous and asynchronous CDMA
systems with modulation based on root Nyquist pulses,
namely root raised cosine waveforms, and modulation based
on sinc functions with increasing bandwidth, it becomes apparent
that the chip pulse design for synchronous CDMA systems follows
the same guidelines as the chip pulse design for single user
systems. In contrast, chip pulse design for asynchronous CDMA
systems is governed by entirely different rules. In fact, for example, we found that  CDMA
systems with uniform delay distributions perform well if the spectrum of the received
signal is as white as possible.

The asymptotic analysis of  asynchronous CDMA systems using
statistics (\ref{suboptimal_statistics}) shows that the
performance of multistage Wiener filters is close to the SINR of
the corresponding synchronous CDMA systems for any bandwidth and
level of SNR. Therefore, this kind of front-end is not capable of exploiting the benefits of asynchronous CDMA.

The universal weights proposed for the design of low complexity
detectors account for the effects of asynchronism,
sub-optimality of the statistics, and non-ideality of
pulse-shapers. They depend on the sampling rate although the large
system performance of some multistage detectors, namely multistage Wiener filters, polynomial expansion detectors, and matched filters, does not.

From the asymptotic analysis and design performed in this work we
can draw the following conclusions:
\begin{itemize}
\item Multistage detectors with front end Type \ref{suboptimal_statistics} and universal weights are asymptotically suboptimal and
have the same complexity order per bit $\mathcal{O}(K^2)$ in
uplink as the linear MMSE detector.
\item Multistage Wiener filters and polynomial expansion detectors with statistics \ref{sufficient_statistics} and universal weights
are asymptotically optimum and have the same complexity order per
bit as the matched filter, i.e. $\mathcal{O}(rK)$ with $r \ll K$.
\item If only a user has to be detected, multistage detectors using statistics (\ref{suboptimal_statistics}) have slightly lower complexity than multistage
detectors with statistics (\ref{sufficient_statistics}), namely
they have a complexity per bit $\mathcal{O}(K^2)$ while in the
later case the complexity per bit is $\mathcal{O}(r K^2)$.
However, they perform almost as the multistage detectors for
synchronous systems at any SNR and do not provide the gain in
performance due to asynchronism in contrast to statistics
(\ref{sufficient_statistics}).
\end{itemize}

\section*{Acknowledgment}

The authors thank Dirk Slock for useful discussions.

\appendices
\section{Proof of Theorem \ref{theo:diagonal_elements_chip_asynch}}\label{section:proof_theo_diagonal_elements_chip_asynch}

Before going into the details of the proof we introduce some properties
of the convergence in probability and the almost sure convergence or convergence with probability one.

\emph{\textbf{Property A:}} Let us consider a finite number $q$ of
random sequences $\{a_n^{(1)}\}, \ldots, \{a_n^{(q)}\}$ that
converge in probability to deterministic limits $a_1, \ldots,
a_q,$ respectively. Then, any linear combination of such sequences
converges in probability to the linear combination of the limits.
Furthermore, if $|a_n^{(s)}-a_s|
\overset{\mathcal{P}}{\rightarrow} o(N^{-i_s}),$ with $i_s \in
\mathbb{R}^{+}, $  and $s=1, \ldots q,$ then any linear
combination of the random sequences converges as
$o(N^{-\min_{s=1,...q}(i_s)}),$ at worst.

\emph{\textbf{Property B:}} Let $\{a_n\}$ and $\{b_n\}$ be two random
sequences that converge in probability to $a$ and $b,$
respectively. Then, the sequence $\{a_n b_n\}$ converges in
probability to $ab.$

\emph{\textbf{Property C:}} If for large $n,$
$\mathrm{Pr}\{|a_n-a|>\varepsilon\}\leq o(n^{-s})$   and
$\mathrm{Pr}\{|b_n-b|>\varepsilon\} \leq o(n^{-t}),$ with $s,t \in
\mathbb{R}^{+}, $ then also
$\mathrm{Pr}\{|(a_n-a)(b_n-b)|>\varepsilon\} \leq
o(n^{-\min(s,t)}),$ at worst.

The convergence with probability one or almost sure convergence implies the convergence in probability. In general, the converse is not true. However, if a random sequence $a_k$ converge in probability to a constant $a$ with a convergence rate $o(n^{-s})$ and $s>1,$
i.e. $\mathrm{Pr}\{|a_n-a|>\varepsilon\}\leq o(n^{-s}),$ then, also the convergence with  probability one holds. This is a straightforward consequence of the Borel Cantelli lemma (see e.g. \cite{billingsley:95}).

In part I Theorem 3 of this work \cite{cottatellucci:07b} we have shown that, when $K, N \rightarrow + \infty$ with constant ratio $\beta,$ the eigenvalue distribution of the infinite matrix $\stackR$  is the same as the eigenvalue distribution of the matrix $\widetilde{\matR}=\matA^H \widetilde{\matS}^H  \widetilde{\matS} \matA= \widetilde{\matH}^H \widetilde{\matH}$ where $\widetilde{\matS}=(\widetilde{\Mat{\Phi}}_1 \vecs_1, \widetilde{\Mat{\Phi}}_2 \vecs_2, \ldots \widetilde{\Mat{\Phi}}_K \vecs_K)$ and $\widetilde{\Mat{\Phi}}_k$ is the $r$-block-wise circulant matrix of order $N$ defined in (\ref{circulant_C_phi_r}) with $\widetilde{\tau}_k= \tau_k \! \mod \! T_c.$

Let us consider the block diagonal matrix $\Mat{\Delta}_{\phi, r}(\widetilde{\tau}_k)$ with $r \times 1$ blocks
\begin{equation}\label{delta_phi_r__disc_def}\left(\Mat{\Delta}_{\phi,r}(\widetilde{\tau}_k)\right)_{\ell, \ell }=\left(
\begin{array}{c}
 {\phi}\left(2\pi \frac{\ell-1}{N}, \widetilde{\tau}_k \right) \\
 {\phi}\left(2\pi \frac{\ell-1}{N}, \widetilde{\tau}_k-\frac{T_c}{r} \right) \\
 \vdots \\
 {\phi}\left(2\pi \frac{\ell-1}{N}, \widetilde{\tau}_k-\frac{r-1}{r} T_c \right) \\
\end{array}
\right).
\end{equation}
and introduce the matrices
\begin{equation}\label{equivalent_matrix}
    \widehat{\matS}=(\Mat{\Delta}_{\phi, r}(\widetilde{\tau}_1) \vecs_1, \Mat{\Delta}_{\phi, r}(\widetilde{\tau}_2) \vecs_2, \ldots \Mat{\Delta}_{\phi, r}(\widetilde{\tau}_K) \vecs_K)
\end{equation}
and $\widehat{\matR}=\matA^H \widehat{\matS}^H \widehat{\matS} \matA.$

By applying the same approach as in part I Theorem 1 of this work
\cite{cottatellucci:07b} it can be shown that the eigenvalue distribution of the matrices $\widetilde{\matR}$ and $\widehat{\matR}$ coincide. Then, also the eigenvalue
moments of the two matrices coincide. The same
property holds for the diagonal elements of the matrices
$\widetilde{\Mat{R}}^{\ell}$ and $\widehat{\Mat{R}}^{\ell}$ with
$\ell \in \mathbb{Z}^{+}. $

In the following we focus on the asymptotic analysis of the
diagonal elements of the matrices $\widehat{\matR}^{\ell}.$

Throughout this proof we adopt the following notation. For $k=1,
\ldots, \K$ and $n= 1, \ldots, \N$
\begin{itemize}
\item $\widehat{\vech}_k$ is the $k^{\mathrm{th}}$ column of the
matrix $\widehat{\matH}$;
\item $\widehat{\vech}_{nk}$ is the $n^{\mathrm{th}}$ $r \times 1$ block of the vector $\widehat{\vech}_k$ and $\widehat{\vech}_{nk}=a_{k} ({\boldsymbol{\Delta}}_{\phi, r}(\widetilde{\tau}_k))_{nn} s_{nk}$;
\item $\widehat{\boldsymbol{\delta}}_n$
is the $n^{\mathrm{th}}$ block row of $\widehat{\matH}$ of
dimensions $r \times \K$; \item $\widehat{\matH}_{\vDash n}$ is
the matrix obtained from $\widehat{\matH}$ by suppressing
$\widehat{\boldsymbol{\delta}}_n;$ \item $\widehat{\matH}_{\sim
k}$ is the matrix obtained from $\widehat{\matH}$ by suppressing
$\widehat{\vech}_k$; \item $\widehat{\matT}=\widehat{\matH} \widehat{\matH}^H$ and $\widehat{\matT}_{\sim k}=\widehat{\matH}_{\sim k} \widehat{\matH}_{\sim k}^H$;
\item $\widehat{\matR}_{\vDash n}=\widehat{\matH}_{\vDash n}^H
\widehat{\matH}_{\vDash n};$
\item $\widehat{\boldsymbol{\sigma}}_n=({s}_{n1},{s}_{n2}, \ldots,
{s}_{n \K} );$ \item $\pmb{\nabla}_{n,t}$, for $t=1, \ldots, r$
and $n=1, \ldots, \N $, is a $\K \times \K$ diagonal matrix with
the $k^{\mathrm{th}}$ element equal to ${\phi}\left(2 \pi \frac{
n-1}{\N}, \widetilde{\tau}_k-\frac{(t-1)T_c}{r} \right).$ Note
that $\widehat{\boldsymbol{\sigma}}_n \pmb{\nabla}_{n,t} \matA $
coincides with the $(t+(n-1)r)^{\mathrm{th}}$ row of the matrix
$\widehat{\matH}$.
\item $\widehat{\matT}^s_{[nn]}$ is the $n^{\mathrm{th}}$
diagonal block of $\widehat{\matT}^s$ of dimensions $r  \times r
$.
\end{itemize}
Furthermore, since the channel gains $a_{k}$ are bounded, we
denote by $a_{\mathrm{MAX}}$ their upper bound, i.e.
$|a_{k}|<a_{\mathrm{MAX}}, \, \forall k.$ Finally, thanks to the
assumption that $\Phi(\omega)$ is bounded in absolute value
with finite support also $\phi(\Omega, {\tau})$ is upper bounded for
any $\Omega$ and $\tau$. We denote by $\Phi_{\mathrm{MAX}}$ its bound.

 Let us observe first that the eigenvalue moments of
the matrix $\widehat{\matR}$ (or equivalently of
$\widehat{\matT}$) are almost surely upper bounded by a finite
positive values $C^{(s)}$, i.e.
\begin{equation}\label{almost_sure_bound_finite_moments}
\exists C^{(s)}<+\infty : \qquad \qquad
\mathrm{Pr}\left\{\frac{1}{N} \mathrm{tr} \widehat{\matR}^s <
C^{(s)}\right\}=1 \qquad \text{as } K, N \rightarrow +\infty,
\frac{K}{N} \rightarrow \beta.
\end{equation}

In fact,
\begin{align}
\frac{1}{N} \mathrm{tr} \widehat{\matR}^s &= \frac{1}{N}\sum_{k_1,
\ldots k_s=1 }^{K}\sum_{n_1, \ldots n_s=1 }^{N}
\widehat{\vech}_{n_1,k_1}^H
\widehat{\vech}_{n_1,k_2}\widehat{\vech}_{n_2,k_2}^H
\widehat{\vech}_{n_2,k_3} \ldots \widehat{\vech}_{n_s,k_s}^H
\widehat{\vech}_{n_s,k_1} \nonumber \\
& = \frac{1}{N} \sum_{k_1, \ldots k_s=1 }^{K} |a_{k_1}|^2
\ldots |a_{k_s}|^2 \sum_{n_1, \ldots n_s=1 }^{N}
{\boldsymbol{\Delta}}_{\phi,r}(\widetilde{\tau_1})_{n_1n_1}^{H} {\boldsymbol{\Delta}}_{\phi,r}(\widetilde{\tau_2})_{n_1n_1}
\ldots{\boldsymbol{\Delta}}_{\phi,r}(\widetilde{\tau_s})_{n_sn_s}^{H}
{\boldsymbol{\Delta}}_{\phi,r}(\widetilde{\tau_1})_{n_sn_s}
\times \nonumber \\ & \times  {s}_{n_1,k_1}^{*}
{s}_{n_1,k_2}{s}_{n_2,k_2}^{*} {s}_{n_2,k_3} \ldots
{s}_{n_s,k_s}^{*} {s}_{n_s,k_1} \nonumber
\end{align}

Applying the approach of non-crossing partitions
\cite{speicher:98,yin:86},  it is possible to recognize that the
factors ${s}_{n_1,k_1}^{*} {s}_{n_1,k_2}{s}_{n_2,k_2}^{*}
{s}_{n_2,k_3} \ldots {s}_{n_s,k_s}^{*} {s}_{n_s,k_1}$ which do not
vanish asymptotically, correspond to the ones having nonzero
non-crossing partitions. Correspondingly, also the remaining
factors \begin{equation*}
{\boldsymbol{\Delta}}_{\phi,r}(\widetilde{\tau_1})_{n_1n_1}^{H}{\boldsymbol{\Delta}}_{\phi,r}(\widetilde{\tau_2})_{n_1n_1}
\ldots{\boldsymbol{\Delta}}_{\phi,r}(\widetilde{\tau_s})_{n_sn_s}^{H}{\boldsymbol{\Delta}}_{\phi,r}(\widetilde{\tau_1})_{n_sn_s}\end{equation*}
are positive and bounded by
\begin{equation*}
|{\boldsymbol{\Delta}}_{\phi,r}(\widetilde{\tau_1})_{n_1n_1}^{H}{\boldsymbol{\Delta}}_{\phi,r}(\widetilde{\tau_2})_{n_1n_1}
\ldots{\boldsymbol{\Delta}}_{\phi,r}(\widetilde{\tau_s})_{n_sn_s}^{H}{\boldsymbol{\Delta}}_{\phi,r}(\widetilde{\tau_1})_{n_sn_s}|\leq\frac{r^{2s}
\Delta_{\mathrm{MAX}}^{2s}}{T_c^{2s}}.
\end{equation*}
Therefore,
\begin{equation}\label{moment_bound}
  \frac{1}{N} \mathrm{Tr}\widehat{\matR}^s \leq \frac{r^{2s}
\Delta_{\mathrm{MAX}} a_{\mathrm{MAX}}^{2s}}{T_c^{2s}} \left(
\frac{1}{N} \sum_{k_1, \ldots k_s=1 }^{K}  \sum_{n_1, \ldots n_s=1
}^{N} {s}_{n_1,k_1}^{*} {s}_{n_1,k_2}{s}_{n_2,k_2}^{*}
{s}_{n_2,k_3} \ldots {s}_{n_s,k_s}^{*} {s}_{n_s,k_1}\right).
\end{equation}
The last factor in (\ref{moment_bound}) is the $s$-th eigenvalue
moment of a central Wishart matrix with zeromean i.i.d Gaussian
entries having variance $\frac{1}{N}.$ Well established results of
random matrix theory \cite{wachter:78,yin:86,mueller:01a} show
that the eigenvalue moments of such a matrix converge almost
surely to finite values. More specifically,
\begin{equation}\label{standard_wishart_moment}
\frac{1}{N}\sum_{n_1, \ldots n_s=1 }^{N} {s}_{n_1,k_1}^{*}
{s}_{n_1,k_2}{s}_{n_2,k_2}^{*} {s}_{n_2,k_3} \ldots
{s}_{n_s,k_s}^{*} {s}_{n_s,k_1}
\overset{a.s.}{\rightarrow}\sum_{i=0}^{s-1}\left(\begin{array}{c}
  s \\
  i
\end{array}\right)\left(\begin{array}{c}
  s \\
  i+1
\end{array}\right)\frac{\beta^i}{s}.
\end{equation}
Then, appealing to (\ref{moment_bound}) and
(\ref{standard_wishart_moment}), the eigenvalue moments of the
matrices $\widehat{\matR}$ and $\widehat{\matT}$ are upper bounded
almost surely by
\begin{equation}\label{upper_bound_eig_C_s}
C^{(s)}=\frac{r^{2s} \Delta_{\mathrm{MAX}
a_{\mathrm{MAX}}}^{2s}}{T_c^{2s}}
\sum_{i=0}^{s-1}\left(\begin{array}{c}
  s \\
  i
\end{array}\right)\left(\begin{array}{c}
  s \\
  i+1
\end{array}\right)\frac{\beta^i}{s}.
\end{equation}

The proof of Theorem \ref{theo:diagonal_elements_chip_asynch} is
based on strong induction. In the first step we prove the
following facts:
\begin{enumerate}
\item The diagonal elements of the matrix $\widehat{\matR}$
converge almost surely, as $\N \rightarrow \infty$, to
deterministic values ${R}_{1}(|a_{k}|^2, \widetilde{\tau}_k)$,
conditionally on $(|a_{k}|^2, \widetilde{\tau}_k).$ Furthermore,
$\forall \varepsilon >0$ and large $K=\beta N$
\begin{equation*}\mathrm{Pr}\{|\widehat{\matR}_{kk}-R_{1}(|a_{k}|^2,
\widetilde{\tau}_{k})|> \varepsilon \}\leq
o\left(\N^{-2}\right).\end{equation*}
\item $\widehat{\matT}_{[nn]}$, the $r \times r $ block diagonal
elements of the matrix $\widehat{\matT}=\widehat{\matH}
\widehat{\matH}^H$, converge almost surely to deterministic
blocks ${\matT}_1(\Omega)$, with $\Omega=\lim_{\N \rightarrow \infty}
2 \pi \frac{n}{\N}.$ Additionally,  $\forall \varepsilon >0,$ large
$K=\beta N$ and $u,v=1, \ldots r, $
\begin{equation*}\mathrm{Pr}\{|(\widehat{\matT}_{[nn]})_{uv}-(\matT_{1}(\Omega))_{uv}|>
\varepsilon \} \leq o\left(\N^{-2}\right).\end{equation*}
\end{enumerate}
Then, in the recursion step, we use the following induction
assumptions:
\begin{enumerate}
\item For $s=1, \ldots, \ell-1,$ the diagonal elements of the matrix $\widehat{\matR}^s$,
 converge almost surely, as $K =\beta \N \rightarrow \infty$, to
deterministic values ${R}_s(|a_{k}|^2, \widetilde{\tau}_k)$,
conditionally on $(|a_{k}|^2, \widetilde{\tau}_k).$ Additionally,
$\forall \varepsilon
>0$ and large $K=\beta N,$
$\mathrm{Pr}\{|(\widehat{\matR}^s)_{kk}-R_{s}(|a_{k}|^2,
\widetilde{\tau}_{k})|> \varepsilon \}\leq o\left(\N^{-2}\right).$
\item For $s=1, \ldots, \ell-1$, $\widehat{\matT}^s_{[nn]}$, the $r \times r $ block
diagonal elements of the matrix $\widehat{\matT}^s$ converge almost surely to deterministic blocks $\matT_s(\Omega)$,  with\footnote{Note that $n=n(N)$ is also a function of the matrix size $N.$}
$\Omega=\lim_{\N \rightarrow \infty} 2 \pi \frac{n}{\N}.$ Additionally,
$\forall \varepsilon >0,$ large $K=\beta N,$ and $u,v=1, \ldots r,
$
$\mathrm{Pr}\{|(\widehat{\matT}_{[nn]}^{s})_{uv}-(\matT_{s}(\Omega))_{uv}|>
\varepsilon \}\leq o\left(\N^{-2}\right).$
\end{enumerate}
We prove:
\begin{enumerate} \item The diagonal elements of the
matrix $\widehat{\matR}^{\ell}$,
 converge almost surely, as $\K= \beta \N \rightarrow \infty$, to
deterministic values $R^{\ell}(|a_{k}|^2, \widetilde{\tau}_k)$,
conditionally on $(|a_{k}|^2, \widetilde{\tau}_k).$ Furthermore,
$\forall \varepsilon >0$ and large $K=\beta N$
\begin{equation}\label{bound_infinitesimo_R_s}\mathrm{Pr}\{|(\widehat{\matR}^{\ell})_{kk}-R_{\ell}(|a_{k}|^2,
\widetilde{\tau}_{k})|> \varepsilon \} \leq
o\left(\N^{-2}\right).\end{equation}
\item The blocks $\widehat{\matT}^{\ell}_{[nn]}$, converge almost surely to deterministic blocks ${\matT}^{\ell}(\Omega)$ with
$\lim_{N \rightarrow \infty} 2 \pi \frac{n}{N}.$ Additionally, $\forall
\varepsilon
>0,$ large $N$ and $u,v=1, \ldots r, $
\begin{equation}\label{bound_infinitesimo_T_s}  \mathrm{Pr}\{|(\widehat{\matT}_{[nn]}^{\ell})_{uv}-(\matT_{\ell}(\Omega))_{uv}|>
\varepsilon \} \leq o\left(\N^{-2}\right).\end{equation}
\end{enumerate}

\emph{First step:} Consider $\widehat{\matR}_{kk}=
\widehat{\vech}_k^H \widehat{\vech}_k =|a_{k}|^2 {\vecs}_{k}^H
{\boldsymbol{\Delta}}_{\phi, r}^H(\widetilde{\tau}_k)
{\boldsymbol{\Delta}}_{\phi,r}(\widetilde{\tau}_k){\vecs}_{k}.$
Thanks to the bound $|\phi(\Omega, {\tau})|<\Phi_{\mathrm{MAX}}$ which
holds for any $\Omega$ and $\tau,$ also the eigenvalues of the matrix
${\boldsymbol{\Delta}}_{\phi,r}^H(\widetilde{\tau})
{\boldsymbol{\Delta}}_{\phi,r}(\widetilde{\tau})$ are upper
bounded. In fact, they are given by $\sum_{t=1}^{r}
\left|{\phi}\left(2 \pi \frac{n-1}{N},{\widetilde{\tau}_k-\frac{(t-1)T_c}{r}}\right)\right|^2$
for $n=1, \ldots, N$.  Therefore, the limit eigenvalue
distribution of the matrix
${\boldsymbol{\Delta}}_{\phi,r}^H(\widetilde{\tau})
{\boldsymbol{\Delta}}_{\phi,r}(\widetilde{\tau})$ has upper
bounded support $\Delta_{\mathrm{MAX}}$. Then, by appealing to
Lemma 9 in part I \cite{cottatellucci:07b} with $p=4$ and by
making use of the bound for any Hermitian matrix $\Mat{C} \in
\mathbb{C}^{\N \times \N},$ $(\mathrm{tr}\Mat{C})^2 \leq N
\mathrm{tr}(\Mat{C}^2)$ we obtain
\begin{align}
\zeta_1 &=\E\left| |a_{k}|^2 \vecs^H_k
{\boldsymbol{\Delta}}_{\phi,r}^H(\widetilde{\tau}_k){\boldsymbol{\Delta}}_{\phi,r}(\widetilde{\tau}_k)
\vecs_k - \frac{|a_{k}|^2}{N}
\mathrm{tr}({\boldsymbol{\Delta}}_{\phi,r}^H(\widetilde{\tau}_k)
{\boldsymbol{\Delta}}_{\phi,r}(\widetilde{\tau}_k)) \right|^4
\nonumber \\
& \leq  \frac{K_4 |a_{k}|^4}{N^3} \mathrm{tr}
({\boldsymbol{\Delta}}_{\phi,r}^H(\widetilde{\tau}_k)
{\boldsymbol{\Delta}}_{\phi,r}(\widetilde{\tau}_k))^4 \nonumber \\
& \leq \frac{K_4 |a_{k}|^4}{N^2}  \Delta_{\mathrm{MAX}}^4.
\nonumber
\end{align}
Since $|a_{k}|\leq a_{\mathrm{MAX}} < + \infty,$  the Bienaym\'e
inequality yields $\forall \varepsilon >0$

\begin{align}
\mathrm{Pr} \left\{\left| \widehat{\matR}_{kk}-
\frac{|a_{k}|^2}{N}
\mathrm{tr}({\boldsymbol{\Delta}}_{\phi,r}^H(\widetilde{\tau}_k)
{\boldsymbol{\Delta}}_{\phi,r}(\widetilde{\tau}_k)) \right| \geq
\varepsilon \right\}  & \leq \frac{\E \left| \widehat{\matR}_{kk}-
\frac{|a_{k}|^2}{N}
\mathrm{tr}({\boldsymbol{\Delta}}_{\phi,r}^H(\widetilde{\tau}_k)
{\boldsymbol{\Delta}}_{\phi,r}(\widetilde{\tau}_k)) \right|^4
}{\varepsilon^4} \nonumber \\
&\leq \frac{K_4 |a_{k}|^4 \Delta_{\mathrm{MAX}}^4}{N^2
\varepsilon^4} \label{bound_R_1}
\end{align}

Thanks to the bound (\ref{bound_R_1}) $\forall
\varepsilon>0$ \begin{equation*}\mathrm{Pr} \left\{\left|
\widehat{\matR}_{kk}- R_1(|a_{k}|^2, \widetilde{\tau}_k) \right|
\geq \varepsilon \right\}\leq o(N^{-2}). \end{equation*}

Furthermore, appealing to the Borel Cantelli lemma (see e.g. \cite{billingsley:95}), this bound implies the following almost sure convergence.

\begin{align}
    {R}_1(\lambda, \tau)|_{(\lambda, \tau)=(|a_{k}|^2, \tau_k )}&= \lim_{\K= \beta \N \rightarrow \infty} \widehat{\matR}_{kk} \nonumber \\ &= \lim_{\K= \beta \N \rightarrow \infty} \frac{|a_{k}|^2}{\N}
    \mathrm{tr}({\boldsymbol{\Delta}}_{\phi,r}^H(\widetilde{\tau}_k)
{\boldsymbol{\Delta}}_{\phi,r}(\widetilde{\tau}_k))     \nonumber \\
&= \lim_{\K= \beta \N \rightarrow \infty} \frac{|a_{k}|^2}{\N} \sum_{\ell=1}^{\N} ({\boldsymbol{\Delta}}_{\phi,r}^{H}(\widetilde{\tau}_k))_{\ell, \ell} ({\boldsymbol{\Delta}}_{\phi,r}(\widetilde{\tau}_k))_{\ell, \ell}   \nonumber \\
&= \left. \frac{\lambda}{2 \pi}  \int_{0}^{2 \pi} {\boldsymbol{\Delta}}_{\phi,r}^H
(\Omega,{\tau}) {\boldsymbol{\Delta}}_{\phi,r}(x,{\tau}) \mathrm{d} \,
x \right|_{(\lambda, \tau )=(|a_{k}|^2, \widetilde{\tau}_k)}.
\end{align}

 Let us now consider the block matrix $\widehat{\matT}_{[nn]}$ whose $(u,v)$ element $(\widehat{\matT}_{[nn]})_{uv}$ is given by
\begin{equation*}
(\widehat{\matT}_{[nn]})_{uv}=\widehat{\boldsymbol{\sigma}}_n
\matA \pmb{\nabla}_{n,u} \pmb{\nabla}_{n,v}^H \matA^H
\widehat{\boldsymbol{\sigma}}_n^H.
\end{equation*}
Thanks to the assumption  of Theorem
\ref{theo:diagonal_elements_chip_asynch} that the support of
$F_{|\matA|^2,T}(\lambda,\tau)$  is bounded  and $\phi(\Omega, {\tau})$
is bounded in absolute value, the diagonal elements of the
diagonal matrix $\matA \pmb{\nabla}_{n,u} \pmb{\nabla}_{n,v}^H
\matA^H$ are upper bounded in absolute value by a positive
constant $T_{\mathrm{MAX}}.$ Then, by appealing to Lemma 9 in part
I \cite{cottatellucci:07b} we obtain
\begin{align}
\E\left( \left|(\widehat{\matT}_{[nn]})_{u,v}
-\frac{1}{N}\mathrm{tr} \matA \pmb{\nabla}_{n,u}
\pmb{\nabla}_{n,v}^H \matA^H \right|^4 \right)& \leq
\frac{K_4}{N^3} \mathrm{tr} (\matA
\pmb{\nabla}_{n,u} \pmb{\nabla}_{n,v}^H \matA^H)^4 \nonumber \\
& \leq \frac{ K_4}{N^2} T_{\mathrm{MAX}}^4.
\label{bound_on_T_nn_quadratic}
\end{align}
By appealing again to the Bienaym\'e inequality and by making use
of the bound (\ref{bound_on_T_nn_quadratic}) we obtain $ \forall
\varepsilon > 0$
\begin{align}
\mathrm{Pr} \left\{ \left|(\widehat{\matT}_{[nn]})_{u,v}-
\frac{1}{N} \mathrm{tr} (\matA \pmb{\nabla}_{n,u}
\pmb{\nabla}_{n,v}^H \matA^H) \right| > \varepsilon \right\}& \leq
\frac{1}{\varepsilon^4}
\E\left(\left|(\widehat{\matT}_{[nn]})_{u,v}- \frac{1}{N}
\mathrm{tr} (\matA \pmb{\nabla}_{n,u} \pmb{\nabla}_{n,v}^H
\matA^H) \right|^4 \right) \nonumber \\ & \leq \frac{K_4
T_{\mathrm{MAX}}^4}{\varepsilon^4 N^2}. \label{bound_Tnn}
\end{align}
Thus, the following convergence in probability holds
\begin{align} \lim_{\K = \beta \N \rightarrow \infty } (\widehat{\matT}_{[nn]})_{u,v} & = \lim_{\K = \beta \N \rightarrow \infty } \frac{1}{\N}
\mathrm{tr}
\matA \pmb{\nabla}_{n,u} \pmb{\nabla}_{n,v}^H \matA^H \nonumber \\
& = \lim_{\K \! = \! \beta \! \N \! \rightarrow \! \infty }
\frac{\beta}{\K} \sum_{k=1}^{\K} |a_{k}|^2 \phi\left(2 \pi \frac{ n\!-
\!1}{\N}, \widetilde{\tau}_k \!- \!\frac{u\!-\!1}{r}T_c \right)
\phi^{*}\left(2 \pi \frac{n\!- \!1}{\N}, \widetilde{\tau}_k \!-
\!\frac{v\!-\!1}{r}T_c \right) \nonumber \\ &=  \beta \int \lambda
\phi \left(\Omega, \tau \!- \!\frac{u\!-\!1}{r}T_c\right) \phi\left( \Omega,
\tau \!- \!\frac{v\!-\!1}{r}T_c \right) \mathrm{d}\,
F_{|\Mat{A}|^2,T}(\lambda, \tau),
\end{align} with $\Omega= \lim_{N \rightarrow \infty} 2 \pi\frac{n}{N}$ and
$0 \leq \Omega \leq 2\pi.$ Therefore, the block matrix
$\widehat{\matT}_{[nn]}$ converges in probability and in mean
square sense to the $r \times r $ matrix
\begin{align}
{\matT}_1(\Omega) & = \lim_{\K= \beta \N \rightarrow \infty}
\widehat{\matT}_{[nn]} \nonumber\\
& = \beta \int\lambda \boldsymbol{\Delta}_{\phi,r}(\Omega,\tau)
\boldsymbol{\Delta}_{\phi,r}^H(\Omega,\tau) \mathrm{d} \,
F_{|\matA|^2,T}(\lambda, \tau) \nonumber
\end{align} with $0 \leq \Omega \leq 2 \pi$.
Thanks to the bound (\ref{bound_on_T_nn_quadratic}) for large $K=
\beta N $ and $\forall \varepsilon >0$ the bound \begin{equation*}
\mathrm{Pr} \left\{ \left|(\widehat{\matT}_{[nn]})_{u,v}-
(\matT(\Omega))_{u,v} \right|< \varepsilon \right\}\leq o(N^{-2})
\end{equation*} holds. Making use of this bound and applying the Borel Cantelli lemma the almost sure convergence is also proven.  This concludes the proof of the first step.

\emph{Step $\ell$:}

By appealing to the induction assumptions, i.e. the almost sure convergence of the diagonal elements of
$\widehat{\matR}^s$ and of the diagonal $r \times r$ blocks of
$\widehat{\matT}^s,$ for $s=1, \ldots, \ell-1$, we prove that the following
almost sure convergence holds:
\begin{align}
\lim_{\K = \beta \N \! \rightarrow \! \infty} \!\!\frac{\mathrm{tr} \matA
\pmb{\nabla}_{n,u} \widehat{\matR}_{\vDash n}^s
\pmb{\nabla}_{n,v}^H \matA^H}{N}  & =  \lim_{\K = \beta \N
\rightarrow \infty} \!\! \sum_{k=1}^{K}\! \frac{|a_{k}|^2}{N}
\phi\!\!\left(\!2 \pi \frac{n\!-\!1}{\N}, \widetilde{\tau}_k \!-
\!\frac{u\!-\!1}{r}T_c \right)\! \phi^{*}\!\!\left(\!
2 \pi \frac{n\!-\!1}{\N}, \widetilde{\tau}_k \!\!-\! \!
\frac{v\!-\!1}{r}T_c \right) \!(\widehat{\matR}_{\vDash n}^s)_{kk}
\nonumber
\\ &= \beta \int \lambda \phi\left(\Omega, \tau -\frac{u-1}{r}T_c
\right)\phi^{*}\left(\Omega, \tau -\frac{v-1}{r}T_c \right)
R_s(\lambda, \tau) \mathrm{d}F_{|\matA|^2, T}(\lambda, \tau)
\label{convergence_A_Nabla_R^s_NablaH_AH}
\end{align}
with $\Omega= \lim_{N \rightarrow \infty} 2 \pi \frac{n-1}{N},$ $s= 1, \ldots
\ell-1$ and
\begin{equation}\label{limit_Rs}
R_s(\lambda, \tau)|_{(\lambda, \tau)=(|a_{k}|^2,
\widetilde{\tau}_k)}=\lim_{K= \beta N \rightarrow \infty}
(\widehat{\matR}^s)_{kk}+ o(N^{-2})
\end{equation}
as from the recursion assumptions. Furthermore, we prove the following almost sure convergence
\begin{align}
\lim_{\K = \beta \N \rightarrow \infty} \frac{|a_{k}|^2}{N}
\mathrm{tr}\boldsymbol{\Delta}_{\phi,r}^H(\widetilde{\tau}_k)
\widehat{\matT}_{\sim k}^s
\boldsymbol{\Delta}_{\phi,r}(\widetilde{\tau}_k) &= \lim_{\K =
\beta \N \rightarrow \infty} \frac{|a_{k}|^2}{N}
\sum_{n=1}^{N}(\boldsymbol{\Delta}_{\phi,r}^H(\widetilde{\tau}_k))_{nn}
(\widehat{\matT}^s)_{nn}
(\boldsymbol{\Delta}_{\phi,r}(\widetilde{\tau}_k))_{nn} \nonumber
\\ &= \left. \frac{\lambda}{2 \pi} \int_{0}^{2 \pi} \boldsymbol{\Delta}_{\phi,r}^H(\Omega, \tau)\matT_s(\Omega) \boldsymbol{\Delta}_{\phi,r}(\Omega, \tau) \mathrm{d}\,\Omega
\right|_{(\lambda, \tau)=(|a_{k}|^2, \widetilde{\tau}_k )}
\label{convergence_AH_DeltaH_T^s_Delta_A}
\end{align}
with $s=1, \ldots \ell-1$ and
\begin{equation}\label{T_s_def}
\matT_s(\Omega)= \lim_{K=\beta N \rightarrow \infty}
(\widehat{\matT}^s)_{nn}.
\end{equation}

In fact, for (\ref{convergence_A_Nabla_R^s_NablaH_AH}) we can
write
 \begin{align}
 \zeta_2 &= \mathrm{Pr} \left\{\left| \frac{1}{\N}
\mathrm{tr} \matA \pmb{\nabla}_{n,u} \widehat{\matR}_{\vDash n}^s
\pmb{\nabla}_{n,v}^H \matA^H \right. \right. \nonumber \\ & \left.
\left. -\frac{1}{\N} \sum_{k=1}^{\K} |a_{k}|^2 \phi\left(\!\!
2 \pi \frac{n\!-\!1}{N}, \widetilde{\tau}_k \!\! - \!\!
\frac{u\!-\!1}{r} T_c \right) \phi^{*}\left(
\!2 \pi \!\frac{n\!-\!1}{N}, \widetilde{\tau}_k\!-\!\frac{v\!-\!1}{r}
T_c \!\! \right)
R_s(|a_{k}|^2, \widetilde{\tau}_k ) \right|> \varepsilon \right\} \nonumber \\
& \leq \zeta_{2a} + \zeta_{2b} \nonumber \end{align} where
\begin{equation}
\zeta_{2a}=\mathrm{Pr} \left\{ \left| \frac{1}{N} \mathrm{tr}
\matA \pmb{\nabla}_{n,u} (\widehat{\matR}^s -
\widehat{\matR}_{\vDash n}^s) \pmb{\nabla}_{n,v}^H   \matA^H
\right|> \frac{\varepsilon}{2} \right\} \nonumber
\end{equation}
and
\begin{equation}
\zeta_{2b}= \mathrm{Pr} \left\{\left|\frac{1}{\N} \sum_{k=1}^{K}
|a_{k}|^2 \phi\left(\!2 \pi \! \frac{n\!-\!1}{N}, \widetilde{\tau}_k
\!\! - \!\! \frac{u\!-\!1}{r} T_c \right)  \phi^{*}\left(
\!2 \pi\!\frac{n\!-\!1}{N}, \widetilde{\tau}_k\!-\!\frac{v\!-\!1}{r}
T_c \!\! \right) \left( (\widehat{\matR}^s)_{kk} - R_s(|a_{k}|^2,
\widetilde{\tau}_k ) \right) \right|>\frac{\varepsilon}{2}
\right\}. \nonumber
\end{equation}

Note that
\begin{equation}
\zeta_{2a}\leq\mathrm{Pr} \left\{ \left| \frac{1}{K} \mathrm{tr}
(\widehat{\matR}^s - \widehat{\matR}_{\vDash n}^s) \right|>
\frac{\varepsilon}{2 \beta a_{\mathrm{MAX}}^2
\phi_{\mathrm{MAX}}^2} \right\}. \nonumber
\end{equation}
The expansion of the matrix
$\widehat{\matR}^s=(\widehat{\matR}_{\vDash n
}+\widehat{\boldsymbol{\delta}}_n^H\widehat{\boldsymbol{\delta}}_n)^s$
yields
\begin{equation}
\mathrm{tr}\widehat{\matR}^s=\mathrm{tr}\widehat{\matR}^s_{\vDash
n } + \sum_{\begin{subarray}{c}
  (i_0, i_1, \ldots i_{s-1}) \\ i_0+\sum_{j=1}^{s-1} (j+1)i_j=s_0
\end{subarray}}\varphi(i_0, i_1, \ldots i_{s-1}) \prod_{u=0}^{s-1}
\left( \widehat{\boldsymbol{\delta}}_n^H \widehat{\matR}_{\vDash
n}^{u} \widehat{\boldsymbol{\delta}}_n\right)^{i_u} \nonumber
\end{equation}
where $\varphi(i_0, i_1, \ldots i_{s-1}) \leq 2^s$ is the number
of the terms of the expansion of $\widehat{\matR}^s$ whose trace
equals $\prod_{u=0}^{s-1} \left( \widehat{\boldsymbol{\delta}}_n^H
\widehat{\matR}_{\vDash n}^{u}
\widehat{\boldsymbol{\delta}}_n\right)^{i_u}.$ Then,
\begin{align}
\zeta_{2a} & \leq 2^s \sum_{\begin{subarray}{c}
  (i_0, i_1, \ldots i_{s-1}) \\ i_0+\sum_{j=1}^{s-1} (j+1)i_j=s_0
\end{subarray}} \mathrm{Pr} \left\{\frac{1}{N} \prod_{u=0}^{s-1}
\left( \widehat{\boldsymbol{\delta}}_n^H \widehat{\matR}_{\vDash
n}^u \widehat{\boldsymbol{\delta}}_n\right)^{i_u} >
\frac{\varepsilon}{ \beta a_{\mathrm{MAX}}^4 \phi_{\mathrm{MAX}}^4
2^{s+1} } \right\} \nonumber
\end{align}

Thanks to Property B on the convergence in probability,
$\zeta_{2a}$ converges in probability with rate
$o(N^{-2-\frac{4}{s}})$ at worst, i.e.   $\forall \varepsilon
>0,$ 
\begin{equation}\label{convergence_intermediate}
\lim_{K=\beta N \rightarrow \infty} \mathrm{Pr} \left\{
\frac{\prod_{u=0}^{s-1}\widehat{\boldsymbol{\delta}}_n^H
\widehat{\matR}_{\vDash n}^u \widehat{\boldsymbol{\delta}}_n}{{N}}
> \sqrt[s]{\frac{ \varepsilon}{\beta 2^{s+1} a_{\mathrm{MAX}}^{4}
\phi_{\mathrm{MAX}}^{4}}} \right\}
\leq o\left(\frac{1}{N^{2+\frac{4}{s}}} \right).
\end{equation}
In fact, for $\varepsilon^{\prime}= \frac{\varepsilon }{\beta
2^{s+1} a_{\mathrm{MAX}}^{4} \phi_{\mathrm{MAX}}^{4}}$
\begin{align}
\mathrm{Pr} \left\{
\frac{\prod_{u=0}^{s-1}(\widehat{\boldsymbol{\delta}}_n^H
\widehat{\matR}_{\vDash n}^u
\widehat{\boldsymbol{\delta}}_n)^{i_u}}{{N}} >
\varepsilon^{\prime} \right\} &\leq \sum_{u=0}^{s-1}
\mathrm{Pr}\left\{ \widehat{\boldsymbol{\delta}}_n^H
\widehat{\matR}_{\vDash n}^u \widehat{\boldsymbol{\delta}}_n>
\sqrt[s]{ \varepsilon^{\prime} N }
\right\}\nonumber \\
& \overset{(a)}{\leq}\sum_{u=0}^{s-1} \mathrm{Pr}\left\{
\left|\widehat{\boldsymbol{\delta}}_n^H \widehat{\matR}_{\vDash
n}^u \widehat{\boldsymbol{\delta}}_n - \frac{\mathrm{tr}
\widehat{\matR}_{\vDash n}^u }{N} \right|> \sqrt[s]{
\varepsilon^{\prime} N }- \frac{\mathrm{tr}
\widehat{\matR}_{\vDash n}^u }{N}
\right\}\nonumber \\
& \overset{(b)}{\leq}\sum_{u=0}^{s-1} \frac{\mathrm{E} \left\{
\left|\widehat{\boldsymbol{\delta}}_n^H \widehat{\matR}_{\vDash
n}^u \widehat{\boldsymbol{\delta}}_n - \frac{\mathrm{tr}
\widehat{\matR}_{\vDash n}^u }{N}
\right|^4\right\}}{\sqrt[s]{ (\varepsilon^{\prime} N)^4 }}  \nonumber \\
& \overset{(c)}{\leq} \frac{K_4 C^{(u)}}{N^2 ((N
\varepsilon^{\prime})^{\frac{1}{s}}- C^{(u)})^4}
\end{align}
where inequality (a) holds for $N$ sufficiently large, inequality
(b) follows from the Bienaym\'{e} inequality, and inequality (c)
is a consequence of Lemma 9 in part I \cite{cottatellucci:07b} and
the bound on the eigenvalues moments of the matrix
$\widehat{\matR}$.



Let us consider now the probability $\zeta_{2b},$
\begin{align}
\zeta_{2b} & \leq
\mathrm{Pr}\left\{\frac{1}{N}\sum_{k=1}^{K}|(\widehat{\matR}^s)_{kk}
- R_s(|a_{k}|^2, \widetilde{\tau}_k )| > \frac{\varepsilon}{
a^2_{\mathrm{MAX}} \phi^2_{\mathrm{MAX}}} \right\} \nonumber \\
& \leq \mathrm{Pr}\left\{ \max_k |(\widehat{\matR}^s)_{kk} -
R_s(|a_{k}|^2, \widetilde{\tau}_k )|> \frac{\varepsilon}{\beta
a^2_{\mathrm{MAX}} \phi^2_{\mathrm{MAX}}}  \right\}
\end{align}
for $s=1, \ldots \ell-1.$ Thanks to the assumption of the
recursive step that $\forall \varepsilon^{'} >0$ and large
$K=\beta N,$ $\mathrm{Pr}\{|(\widehat{\matR}^s)_{kk} -
R_s(|a_{k}|^2, \widetilde{\tau}_k )| > \varepsilon^{'} \} \leq
o(N^{-2}),$ $\zeta_{2b} \rightarrow o(N^{-2}),$ i.e. it vanishes
asymptotically as $N, K \rightarrow \infty$ with constant ratio
with the same converge rate as $o(N^{-2})$ at worst. Therefore,
(\ref{convergence_A_Nabla_R^s_NablaH_AH}) converges in probability with a rate
 as $o(N^{-2})$ for $N \rightarrow +\infty,$ at worst. This convergence rate enables the application of the Borel-Cantelli lemma to prove that (\ref{convergence_A_Nabla_R^s_NablaH_AH}) converges almost surely.

The proof of the convergence
(\ref{convergence_AH_DeltaH_T^s_Delta_A}) with probability one follows
along similar lines.

Following the same approach as in the proof of Theorem 1 in
\cite{cottatellucci:04}, we can expand
$(\widehat{\matR}^{\ell})_{kk}$ and
$\widehat{\matT}^{\ell}_{[nn]}$ as follows:
\begin{align}
(\widehat{\matR}^{\ell})_{kk} & = \sum_{s=0}^{\ell-1} \widehat{\vech}_k^H \widehat{\matT}_{\sim k}^{\ell-s-1} \widehat{\vech}_k (\widehat{\matR}^s)_{kk}  & \ell = 1,2, \ldots \label{inter_1} \\
\widehat{\matT}^{\ell}_{[nn]} & = \sum_{s=0}^{\ell-1}
\widehat{\boldsymbol{\delta}}_n \widehat{\matR}_{\vDash
n}^{\ell-s-1} \widehat{\boldsymbol{\delta}}_n^H
\widehat{\matT}^s_{[nn]}. & \ell = 1,2, \ldots  \label{inter_2}
\end{align}
being $\widehat{\matT}^0$ and $\widehat{\matR}^0$ the identity
matrices of dimensions $rN \times rN$ and $K \times K,$
respectively.

Thanks to Property A and Property B of the convergence in
probability of random sequences and the induction assumptions, the
 convergence in probability one of the sequences
$\{(\widehat{\matR}^{\ell})_{kk}\}$ and
$\{\widehat{\matT}^{\ell}_{[nn]}\}$ reduces to the following two steps. First we show  the
convergence in probability of $\widehat{\vech}_k^H
\widehat{\matT}_{\sim k}^{s}\widehat{\vech}_k$ and
$\widehat{\boldsymbol{\delta}}_n \widehat{\matR}_{\vDash n}^{s}
\widehat{\boldsymbol{\delta}}_n^H$ to a deterministic limit,
respectively. Then, we show that the convergence holds with an appropriate convergence rate which enables the application of the Borel Cantelli lemma. Let us define
\begin{equation*}
  \zeta_3= \widehat{\vech}_k^H
\widehat{\matT}_{\sim k}^{s}\widehat{\vech}_k -
\frac{|a_{k}|^2}{N} \mathrm{tr}
\Mat{\Delta}_{\phi,r}^H(\widetilde{\tau}_k) \widehat{\matT}_{\sim
 k}^{s} \Mat{\Delta}_{\phi,r}(\widetilde{\tau}_k).
\end{equation*}
Lemma 9 in part I \cite{cottatellucci:07b} applied to the
quadratic form $\widehat{\vech}_k^H \widehat{\matT}_{\sim
k}^{s}\widehat{\vech}_k$  with $p=4$ yields
\begin{align}
\mathrm{E} \left| \zeta_3 \right|^4 &<\frac{ K_4 |a_{k}|^4}{N^3}
\E\left(\mathrm{tr}(\Mat{\Delta}_{\phi,r}^H(\widetilde{\tau}_k)
\widehat{\matT}_{\sim k}^{s}
\Mat{\Delta}_{\phi,r}(\widetilde{\tau}_k))^4\right) \nonumber \\
& \leq \frac{K_4}{N^3} a_{\mathrm{MAX}}^8 \phi_{\mathrm{MAX}}^8
\mathrm{tr} (\widehat{\matT}_{\sim
k}^{4s}).\label{convergence_hhatH_Tsimk^s_hhat}
\end{align}

Thanks to the bound on the eigenvalues moments of the matrix
$\widehat{\matT},$  $\lim_{K= \beta N \rightarrow \infty}
\frac{1}{N} \E(\mathrm{tr} \widehat{\matT}^{4s}_{\sim k}) $ is
almost sure upper bounded $\forall s$ as $N=\beta K \rightarrow
+\infty.$ Therefore, $\mathrm{E}|\zeta_3|^4 \rightarrow 0 $ as
$K,N \rightarrow \infty$ with $\frac{K}{N} \rightarrow \beta$ and
$\widehat{\vech}_k^H \widehat{\matT}_{\sim
k}^{s}\widehat{\vech}_k$ converges in mean square sense, and thus
in probability. Furthermore, the Bienaym\'{e} inequality implies
that $\mathrm{Pr}\{|\zeta_3|>\varepsilon\} \leq o(N^{-2})$ as $N
\rightarrow +\infty.$ Thanks to
(\ref{convergence_AH_DeltaH_T^s_Delta_A})
\begin{align}
\lim_{\N = \beta \K \rightarrow \infty} \frac{|a_{k}|^2}{N}
\mathrm{tr} \Mat{\Delta}_{\phi,r}^H(\widetilde{\tau}_k)
\widehat{\matT}_{\sim k}^s
\Mat{\Delta}_{\phi,r}(\widetilde{\tau}_k) & = \left. \frac{\lambda}{2 \pi}
\int_0^{2 \pi} \boldsymbol{\Delta}_{\phi,r}^{H}(\Omega,\tau) \matT_s(\Omega)
\boldsymbol{\Delta}_{\phi,r}(\Omega,\tau) \mathrm{d} \, \Omega
\right|_{(\lambda, \tau)=(|a_{k}|^2,
\widetilde{\tau}_k) } +o(N^{-2})  \nonumber \\
&=g({\matT}_s, \lambda, \tau) +o(N^{-2}). \label{g_def}
\end{align}
then
\begin{equation}\label{convergence_rate_g}
  \mathrm{Pr}\{ |\widehat{\vech}_k^H
\widehat{\matT}_{\sim k}^{s}\widehat{\vech}_k - g({\matT}_s,
\lambda, \tau) |> \varepsilon \} \rightarrow o(N^{-2})
\end{equation}
thanks to property A. Thanks to the convergence rate in (\ref{convergence_rate_g}) and the Borel Cantelli lemma, the almost sure convergence (\ref{limit_Rs}) follows.

 The convergence with probability one of the diagonal blocks
$\widehat{\matT}^{\ell}_{[nn]}$ can be proven in a similar way.
More specifically, it can be shown that the $r \times r $ block
$\widehat{\boldsymbol{\delta}}_n \widehat{\matR}_{\vDash
n}^{s}\widehat{\boldsymbol{\delta}}_n^H$ converges to the $r
\times r$ deterministic matrix

\begin{align}\label{f_def}
    \mathbf{f}(R_s , \Omega) &  = \beta  \int \lambda
    \mathbf{\Delta}_{\phi,r}(\Omega,\tau)
\mathbf{\Delta}_{\phi, r}(\Omega, \tau)^H {R}_s(\lambda, \tau)
\mathrm{d} \, F_{|\matA|^2,T}(\lambda, \tau).
\end{align}
such that $\mathrm{Pr}\left\{\left| (\widehat{\boldsymbol{\delta}}_n)_u
\widehat{\matR}_{\vDash
n}^{s}(\widehat{\boldsymbol{\delta}}_n^H)_v -(\mathbf{f}(R_s ,
\Omega))_{u,v} \right| > \varepsilon \right\} \rightarrow o(N^{-2}).$

Finally, by making use of equations (\ref{inter_1}) and
(\ref{inter_2}) and  the definitions (\ref{limit_Rs}),
(\ref{T_s_def}), (\ref{f_def}),
 and (\ref{g_def}) we obtain
\begin{align}
{R}_{\ell}(\lambda, \tau) &  = \sum_{s=0}^{\ell-1}
g(\matT_{\ell-s-1}, \lambda, \tau) R_{s}(\lambda, \tau) \qquad
\ell=1,2, \ldots \label{R_l_recursion_asymptotic}
\end{align}
and
\begin{align}
\matT_{\ell}(\Omega) &  = \sum_{s=0}^{\ell-1}
\mathbf{f}(R_{\ell-s-1},\Omega) \matT_s(\Omega) \qquad \ell=1,2, \ldots
\label{T_l_recursion_asymptotic}.
\end{align}
with $g(\matT_s, \lambda, \tau)$ and $\mathbf{f}(R_s, \Omega)$ given in
(\ref{g_def}) and (\ref{f_def}), respectively. Consistently to the
definitions of $\widehat{\matT}^0$ and $\widehat{\matR}^0,$
$\matT_0(\Omega)=\Mat{I}_{r},$ being $\Mat{I}_{r}$ the $r \times r $
identity matrix and $R_{0}(\lambda)=1.$

Then, \mbox{$g(R_0, \lambda, \tau) = \frac{\lambda}{2 \pi}
\int_{-\pi}^{\pi}
\boldsymbol{\Delta}_{\phi,r}^H(\Omega, \tau)
\boldsymbol{\Delta}_{\phi,r}(\Omega, \tau) \mathrm{d}\Omega$} and
\mbox{$\Mat{f}(\matT_0,\!\Omega) \!= \!\beta \! \int \! \lambda
\boldsymbol{\Delta}_{\phi,r}(\Omega, \! \tau)
\boldsymbol{\Delta}_{\phi,r}^H(\Omega, \! \tau) \mathrm{d}F_{|\matA|^2,
T}(\lambda, \! \tau)$} and (\ref{R_l_recursion_asymptotic}) and
(\ref{T_l_recursion_asymptotic}) reduce to the asymptotic limits
$R_1(\lambda, \tau)$ and $\matT_1(\Omega)$ already derived in
\textit{step 1}. Therefore, we can begin the recursion with
$\ell=0,$  $R_0(\lambda, \tau)=1$ and $\matT_0(\Omega)=\Mat{I}_r.$

Properties A, B, and C, the induction assumptions, relations
(\ref{inter_1}) and (\ref{R_l_recursion_asymptotic}), the
convergence rates $\zeta_2 \rightarrow o(N^{-2}),$
$\mathrm{Pr}\{ \zeta_3 > \varepsilon \}\leq \rightarrow o(N^{-2}),$ and the Borel Cantelli lemma yield (\ref{bound_infinitesimo_R_s}). The proof of
(\ref{bound_infinitesimo_T_s}) follows immediately along similar
lines.


This concludes the proof of Theorem
\ref{theo:diagonal_elements_chip_asynch}.

\section{Proof of Corollary \ref{cor_proposition_raised_cosine}}\label{section:proof_cor_proposition_raised_cosine}
Corollary \ref{cor_proposition_raised_cosine} is derived by
specializing Theorem \ref{theo:diagonal_elements_chip_asynch} to a
unitary Fourier transform $\Phi(\omega)$ with bandwidth $B \leq
\frac{r}{2 T_c}$. Let us recall here that the unitary Fourier
transform in the discrete time domain is given by
\begin{equation}\label{phi_form_bandlimited}
{\phi}(\Omega,\tau)= \frac{1}{T_c} \mathrm{e}^{  j \frac{
\tau}{T_c}\Omega} \sum_{s=-\mathrm{sign}(\Omega)\left\lfloor \frac{r-1}{2}
\right\rfloor
}^{\mathrm{sign}(\Omega)\left\lfloor\frac{r}{2}\right\rfloor}
\mathrm{e}^{  j 2 \pi \frac{ \tau}{T_c}s} \Phi^{*} \left(\frac{\Omega+2 \pi s }{T_c} \right) \quad \text{for } \quad |\Omega|\leq \pi.
\end{equation} The matrix $\Mat{Q}(\Omega, \tau)=\Mat{\Delta}_{\phi,r}(\Omega,\tau)
\Mat{\Delta}_{\phi,r}(\Omega,\tau)^H,$  with
$\Mat{\Delta}_{\phi,r}(\Omega,\tau)$ defined in
(\ref{delta_phi_r_def}),  can be decomposed as $\Mat{Q}(\Omega,
\tau)=\Mat{Q}(\Omega)+\overline{\Mat{Q}}(\Omega, \tau)$ with the elements of
$\Mat{Q}(\Omega)$ and  $\overline{\Mat{Q}}(\Omega, \tau)$ defined by
\begin{equation}\label{elem_Q(x)}
(\Mat{Q}(\Omega))_{k,\ell}=\frac{1}{T_c^2}
\sum_{s=-\mathrm{sign}(\Omega)\left\lfloor \frac{r-1}{2}
\right\rfloor}^{\mathrm{sign}(\Omega)\left\lfloor\frac{r}{2}\right\rfloor}
\left|\Phi\left(\frac{\Omega+ 2 \pi s}{T_c} \right)\right|^2
\mathrm{e}^{-j\frac{ k-\ell}{r}(\Omega+2 \pi s)}  \quad \text{for }
\quad |\Omega|\leq \pi,
\end{equation}
and
\begin{multline}\label{elem_Q(x_tau)}
(\overline{\Mat{Q}}(\Omega,\tau))_{k,\ell}=\frac{1}{T_c^2}
\sum_{\begin{subarray}{c}
  s,u=-\mathrm{sign}(\Omega)\left\lfloor \frac{r-1}{2} \right\rfloor \\
  s \neq u
\end{subarray}}^{\mathrm{sign}(\Omega)\left\lfloor\frac{r}{2}\right\rfloor}
\Phi\left(\frac{\Omega + 2 \pi u}{T_c} \right)\Phi^{*}\left(\frac{\Omega+ 2
\pi s}{T_c} \right) \mathrm{e}^{-j 2 \pi \frac{
\tau}{T_c}(s-u)} \mathrm{e}^{-j  \left(\frac{
k-1}{r}(\Omega-2 \pi s)-\frac{ \ell-1}{r}(\Omega-2 \pi u) \right)} \\ \text{for } \quad
|\Omega|\leq \pi,
\end{multline}
respectively.

Equations (\ref{f_Rs_chip_async}) and (\ref{g_Ts_chip_async}) can
be rewritten as
\begin{align}
\Mat{f}(R_s,\Omega) & = \beta \Mat{Q}(\Omega) \int \lambda R_s(\lambda,
\tau) \mathrm{d}F_{|\matA|^2, T}(\lambda, \tau) \nonumber \\ & +
\beta \int \lambda R_s(\lambda, \tau) \overline{\Mat{Q}}(\Omega, \tau)
\mathrm{d}F_{|\matA|^2, T}(\lambda, \tau), & -\pi \leq \Omega
\leq \pi \label{intermidiate_1} \\
g(\matT_s,\lambda, \tau) &= \frac{\lambda}{2 \pi}
\int_{-\pi}^{\pi} \mathrm{tr} (\matT_s(\Omega)
\Mat{Q}(\Omega)) \mathrm{d}\Omega + \frac{\lambda}{2 \pi}
\int_{-\pi}^{\pi} \mathrm{tr} (\matT_s(\Omega)
\overline{\Mat{Q}}(\Omega,\tau)) \mathrm{d} \Omega, \label{intermidiate_2}
\end{align}
respectively. If the conditions of Corollary
\ref{cor_proposition_raised_cosine} are satisfied, i.e. if $B \leq
\frac{r}{2 T_c} $ and $\tau$ is uniformly distributed in
$[0,T_c],$ it can be shown that
\begin{itemize}
\item $R_{\ell}(\lambda, \tau)$, $\ell \in
\mathbb{Z}^{+}$, are independent of $\tau$ and
\item $\matT_{\ell}(\Omega)$ is a matrix of the form (\ref{matrixBform})
\begin{equation}\label{matrixBform}
\Mat{B}= \Mat{B}(\Omega)=\left[ \begin{array}{ccccc}
  b_0 & b_1 \mathrm{e}^{j \frac{\Omega}{r} } & \ldots & \ldots & b_{r-1} \mathrm{e}^{j \frac{(r-1)}{r} \Omega} \\
  b_{r-1} \mathrm{e}^{-j \frac{\Omega}{r} } & b_0 & b_1 \mathrm{e}^{j \frac{\Omega}{r} } & \ldots & b_{r-2} \mathrm{e}^{j \frac{(r-2)}{r} \Omega} \\
  \ldots & \ddots & \ddots & \ddots & \ddots \\
  b_{1} \mathrm{e}^{-j \frac{ (r-1)}{r} \Omega} & \ddots & \ddots & b_{r-1} \mathrm{e}^{-j \frac{\Omega}{r} } & b_0
\end{array} \right],
\end{equation}
being $b_0=b_0(\Omega),b_1=b_1(\Omega), \ldots b_{r-1}=b_{r-1}(\Omega),$
eventually functions of $\Omega.$
\end{itemize}

These properties can be proven by strong induction. It is
straightforward to verify that they are satisfied for $s=0$. In
fact, $R_{0}(\lambda, \tau)=1$ is independent of $\tau$ and
$\matT_0(\Omega)=\Mat{I}$ is of the form (\ref{matrixBform}) with
$b_{0}=1$ and $b_{i}(\Omega)=0$ with $i=1, \ldots r-1$. By appealing to
Lemma 1 in part I \cite{cottatellucci:07b} Appendix I
$\mathrm{tr}(\overline{\Mat{Q}}(\Omega,\tau))=0 $ and $g(\matT_0,
\lambda, \tau)= \frac{\lambda}{2 \pi} \int_{-\pi}^{\pi}
\mathrm{tr} (\Mat{Q}(\Omega)) \mathrm{d}\Omega.$ Hence, $g(\matT_0, \lambda,
\tau)$ is independent of $\tau.$

The induction step is proven using the following induction
assumptions: \begin{itemize} \item For $s=0,1, \ldots \ell-1, $
$R_s(\lambda, \tau)$ is independent of $\tau$;
\item For $s=0,1,\ldots \ell-1$, $\matT_s(\Omega)$ is of the
form (\ref{matrixBform}).
\end{itemize}

Thanks to the form (\ref{matrixBform}) of $\matT_s(\Omega),$ $s=1,
\ldots \ell-1,$ given by the induction assumptions and by applying
Lemma I in part I Appendix I we have $\mathrm{tr}(\matT_s(\Omega)
\overline{\Mat{Q}}(\Omega,\tau))=0$, for $s=0,1,\ldots,\ell-1$. Then,
(\ref{intermidiate_2}) reduces to $g(\matT_s, \lambda, \tau)=
\frac{\lambda}{2 \pi} \int_{-\pi}^{\pi} \mathrm{tr}
\left(\matT_s(\Omega) \Mat{Q}(\Omega) \right) \mathrm{d}\Omega $ and  $g(\matT_s,
\lambda, \tau)$ is independent of $\tau$ for $s=0,1, \ldots,
\ell-1.$ Therefore, all quantities that appear in the right hand
side of (\ref{R_chip_asynch}) are independent of $\tau$ and
$R_{\ell}(\lambda,\tau)$ is also independent of $\tau$. In the
following we will shortly write $R_{\ell}(\lambda)$ and
$g(\matT_s, \lambda)$ instead of $R_{\ell}(\lambda, \tau)$ and
$g(\matT_s, \lambda, \tau).$ Thanks to the fact that (i) $R_s(\lambda,
\tau )$ is independent of $\tau$ and (ii) $\lambda $ and $\tau$ are
statistically independent with $\tau$ uniformly distributed,
(\ref{intermidiate_1}) can be rewritten as
\begin{equation}\label{intermidiate_1_bis}
f(R_s,\Omega)= \beta \int \lambda R_s(\lambda) \mathrm{d} F_{|\matA|^2}
\left(\Mat{Q}(\Omega) + \frac{1}{T_c} \int_{0}^{T_c}
\overline{\Mat{Q}}(\Omega, \tau) \mathrm{d} \tau \right).
\end{equation}
It is straightforward to verify that $\int_{0}^{T_c}
\overline{\Mat{Q}}(\Omega, \tau) \mathrm{d} \tau=0$ from the definition
of $\overline{\Mat{Q}}(\Omega, \tau)$ in (\ref{elem_Q(x_tau)}). Then,
\begin{align}
\Mat{f}(R_s,\Omega) &= \beta \Mat{Q}(\Omega) \int \lambda R_s(\lambda)
\mathrm{d} F_{|\matA|^2}(\lambda) \nonumber \\ & = f(R_s)
\Mat{Q}(\Omega) \label{intermidiate_3}
\end{align}
with $f(R_s)= \beta \int \lambda R_s(\lambda) \mathrm{d}
F_{|\matA|^2}(\lambda).$ Substituting (\ref{intermidiate_3}) in
(\ref{T_chip_asynch}) yields
\begin{equation} \matT_{\ell}(\Omega)=\sum_{s=0}^{\ell-1}
f(R_{\ell-s-1}) \Mat{Q}(\Omega) \matT_s(\Omega), \qquad \qquad -\pi
\leq \Omega \leq  \pi. \label{matT_raised_cosine}
\end{equation}
Since $\matT_s(\Omega)$ is of form (\ref{matrixBform}), the conditions
of Lemma 2 in part I Appendix I are satisfied for
$\Mat{B}=\matT_s(\Omega).$ This implies that $\Mat{Q}(\Omega) \matT_s(\Omega)$ is
also of the form (\ref{matrixBform}). Since ${\matT}_{\ell}(\Omega)$ is
a linear combination of matrices of the form (\ref{matrixBform}),
$\matT_{\ell}(\Omega)$ is also a matrix of the form
(\ref{matrixBform}). Then, the statement of the strong induction
is proven.

Thanks to the properties shown by strong induction, the recursive
equations in Theorem (\ref{theo:diagonal_elements_chip_asynch})
reduce to the following set of recursive equations:
\begin{align}
R_{\ell}(\lambda)& =\sum_{s=0}^{\ell-1} g(\matT_{\ell-s-1},
\lambda )
R_s(\lambda)& \label{R_raised_cosine_final}\\[1mm] \matT_{\ell}(\Omega) & =  \sum_{s=0}^{\ell-1}
{f}(R_{\ell-s-1})\Mat{Q}(\Omega) \matT_s(\Omega) & -\pi \leq \Omega
\leq \pi \label{T_raised_cosine_final}\\[1mm]
 {f}(R_{s})& = \beta  \int \lambda
R_s(\lambda) \mathrm{d}\,F_{|\matA|^2}(\lambda), & \label{f_raised_cosine_final}\\[1mm]
g(\matT_s, \lambda) & = \frac{\lambda}{2 \pi} \int_{-\pi}^{\pi}
\mathrm{tr}( \matT_s(\Omega) \Mat{Q}(\Omega))  \mathrm{d}\,\Omega &
\label{g_raised_cosine_final}
\end{align}
with $\matT_0(\Omega)=\Mat{I}_r$ and $R_{0}(\lambda)=1.$

Then, applying again Theorem
\ref{theo:diagonal_elements_chip_asynch} we obtain the following
convergence with probability one
\begin{equation*}
    \lim_{\K= \beta \N \rightarrow \infty}
    (\widehat{\matR}^{\ell})_{kk} =
    R_{\ell}(\lambda)|_{\lambda=|a_{k}|^2}.
\end{equation*}

From (\ref{T_raised_cosine_final}) and $\matT_0(\Omega)=\Mat{I}_r$ it
is apparent that $\matT_{\ell}(\Omega)$ is a polynomial in
$\Mat{Q}^s(\Omega),$ for $s=0, 1, \ldots \ell.$ Then, $\matT_{\ell}(\Omega)$
has the same eigenvectors as $\Mat{Q}(\Omega)$ and it can written as
$\matT_{\ell}(\Omega)= \Mat{U}(\Omega) \Mat{\Lambda}_{\ell}(\Omega) \Mat{U}^H(\Omega)$
where $\Mat{\Lambda}_{\ell}(\Omega)$ is a diagonal matrix with diagonal
elements $t_{\ell,1},t_{\ell,2}, \ldots t_{\ell,r}$ and
\begin{equation}\label{U_definition}
\Mat{U}(\Omega)=\left( \Vec{e}\left( \Omega-\mathrm{sign}(\Omega)2 \pi \left \lfloor
\frac{r-1}{2} \right \rfloor \right), \ldots \Vec{e}\left( \Omega
\right) \ldots \Vec{e}\left( \Omega+\mathrm{sign}(\Omega) 2 \pi \left \lfloor
\frac{r}{2} \right \rfloor \right) \right)
\end{equation}
with $\Vec{e}\left( \Omega \right)$  r-dimensional column vector
defined by
\begin{equation*}
\Vec{e}\left( \Omega \right)= \frac{1}{\sqrt{r}}\left( 1,
\mathrm{e}^{-j \frac{\Omega}{r}}, \ldots \mathrm{e}^{-j\frac{r-1}{r} \Omega} \right)^T.
\end{equation*}  By making use
of the eigenvalue decomposition of the matrix $\Mat{Q}(\Omega)$ in part
I Appendix I  Lemma 3 the matrix equation
(\ref{T_raised_cosine_final}) reduces to $r$ scalar equations
{\small \begin{equation*} t_{\ell,u}(\Omega)= \sum_{s=0}^{\ell-1}
f(R_{\ell-s-1}) \frac{r}{T_c^2}\left|\Phi\left( \frac{\Omega}{T_c}- \mathrm{sign}(\Omega) \frac{2 \pi}{T_c} \left(\left\lfloor
\frac{r-1}{2} \right\rfloor -u+1\right) \right) \right|^2
t_{s,u}(\Omega) \qquad u=1,\ldots r \quad \text{and} \quad |\Omega|\leq
\pi.
\end{equation*}}
By substituting $y=\Omega- \mathrm{sign}(\Omega) 2 \pi \left(\left\lfloor
\frac{r-1}{2} \right\rfloor -u+1\right) $ for $|\Omega|\leq
\pi$ we obtain
\begin{equation}\label{aux_1}
t_{\ell, u}\left(y+ 2 \pi \left( \left\lfloor \frac{r-1}{2} \right\rfloor -u+1
\right)\right) = \sum_{s=0}^{\ell-1}
f(R_{\ell-s-1})\frac{r}{T_c^2}\left|\Phi\left( \frac{y}{T_c}
\right) \right|^2 t_{s,u}\left(y+2 \pi \left( \left\lfloor \frac{r-1}{2}
\right\rfloor -u+1 \right) \right)
\end{equation}
for $0 \leq y+2 \pi \left(\left\lfloor \frac{r-1}{2} \right\rfloor -u+1 \right) \leq
\pi$ and
\begin{equation}\label{aux_2}
t_{\ell, u}\left(y- 2 \pi \left( \left\lfloor \frac{r-1}{2} \right\rfloor
-u+1 \right) \right) =\sum_{s=0}^{\ell-1}
f(R_{\ell-s-1})\frac{r}{T_c^2}\left|\Phi\left(  \frac{y}{T_c}
\right) \right|^2 t_{s,u}\left(y-2 \pi \left(\left\lfloor \frac{r-1}{2}
\right\rfloor -u+1\right) \right)
\end{equation}
for $ -\pi \leq y-2 \pi \left(\left\lfloor \frac{r-1}{2} \right\rfloor
-u+1 \right) \leq 0.$ Then, for $u=1, \ldots r,$ the $r$ functions
(\ref{aux_1}) and (\ref{aux_2}) defined in not overlapping
intervals in $\left[-2 \pi r, 2 \pi r \right]$ can be combined in a unique
scalar functions $T_{\ell}^{\prime}(y)$ in the interval $|y|\leq 2 \pi r$
satisfying the recursive equation
\begin{equation*}
T_{\ell}^{\prime}(y)= \sum_{s=0}^{\ell-1} \frac{r}{T_c^2} f(R_{\ell-s-1})
\left|\Phi\left(  \frac{y}{T_c}  \right) \right|^2 T_{s}^{\prime}(y).
\end{equation*}
Similar arguments applied to (\ref{g_raised_cosine_final}) yield
\begin{equation*}
g(T_s, \lambda)=\frac{\lambda}{2 \pi} \int_{-r\pi}^{r\pi} \frac{r}{T_c^2} T_s^{\prime}(y)
\left|\Phi\left(  \frac{y}{T_c} \right) \right|^2
\mathrm{d}y.
\end{equation*}
The substitutions $\omega= \frac{y}{T_c}$ and $T_{\ell}^{\prime}(\omega T_c)= T_{\ell}(\omega)$ yield to the recursive equations in Corollary \ref{cor_proposition_raised_cosine}.

This concludes the derivation of Corollary
\ref{cor_proposition_raised_cosine} from Theorem
\ref{theo:diagonal_elements_chip_asynch}.

\section{Derivation of Algorithm  \ref{alg:raised_cosine}}\label{section:algorithm_raised cosine}
Algorithm \ref{alg:raised_cosine} can be derived from the
recursive equations of Corollary
\ref{cor_proposition_raised_cosine} by using the following
substitutions\footnote{Note that the substitution of $\lambda$
with $z$ is redundant. It is used to obtain polynomials in the
commonly used variable $z$.}:
\begin{align}\nonumber
& \lambda & \! & \rightarrow &\!& z \nonumber \\
& R_s(\lambda) & \! & \rightarrow & \! & \rho_s(z) \nonumber\\
& \lambda R_s(\lambda) & \! & \rightarrow & \! & v_s(z)
\nonumber\\
& \mathrm{E}(\lambda R_s(\lambda))= \frac{1}{\beta} f(R_s) & \! & \rightarrow & \!
& V_s
\nonumber \\
& \frac{1}{T_c} \left| \Phi\left( \omega \right)
\right|^2 & \! & \rightarrow & \! & y \nonumber\\
& T_s(\cdot) & \! & \rightarrow & \! & \mu_s(y) \nonumber \\
& \frac{r}{T_c} \left| \Phi\left(\omega \right)
\right|^2 T_s(\omega)& \! & \rightarrow & \! & u_s(y) \nonumber\\
&  \frac{r}{2 \pi T_c}\int_{-2 \pi B}^{2 \pi B} \left|
\Phi\left(\omega \right) \right|^2 T_s(\omega)
\mathrm{d}\omega & \! & \rightarrow & \! & U_s. \nonumber
\end{align}

Then, the initial step is obtained by defining $\mu_0(y)=1$ and
$\rho_0(z)=1.$ The recursive equations in step $\ell$ are obtained
by using the previous substitutions. In order to derive $U_s$ let
us observe that $\frac{1}{T_c}\left| \Phi\left(\omega \right) \right|^2 T_s(\omega)$ is a polynomial in $y=\frac{1}{T_c}\left|
\Phi\left(\omega \right) \right|^2$ of degree $s+1.$
Then, $U_s$ is a linear combination of $\frac{\mathcal{E}_n}{T_c}$ where
\begin{equation*}
\mathcal{E}_n=\frac{1}{2 \pi T_c^{n-1}}\int_{-2 \pi B}^{2 \pi B} \left|
\Phi\left(\omega \right) \right|^{2n}
\mathrm{d}\omega
\end{equation*}
The coefficients of the linear combination are obtained by
expanding $u_s(y)$ as a polynomial in $y.$

We conclude the derivation of Algorithm \ref{alg:raised_cosine} by
summarizing the previous considerations and substitutions:
\begin{itemize}
  \item \begin{align} \rho_{\ell}(z) &= \sum_{s=0}^{\ell-1}z
  U_{\ell-s-1}\rho_s(z) \nonumber \\
\mu_{\ell}(y) &= \frac{r}{T_c}\sum_{s=0}^{\ell-1} \beta y
  V_{\ell-s-1}\mu_s(y). \nonumber
  \end{align}
  \item $U_s$ and $V_s$ are obtained from $u_{s}(y)=y \mu_{s}(y)$
  and $v_{s}(z)=z \rho_{s}(z)$, respectively by
\begin{itemize}
  \item expanding $u_{s}(y)$ and $v_{s}(z)$ as polynomials in $y$
  and $z,$ respectively,
  \item replacing the monomials $y^n$ and $z^n$, $n\in \mathbb{Z}^{+}$
  with $\frac{\mathcal{E}_n}{T_c}$ and $m_{|\matA|^2}^{(s)}$, respectively.
\end{itemize}
\end{itemize}
Then, $R_{\ell}(\lambda)=\rho_{\ell}(\lambda)$ and the eigenvalue
moment $m_{\stackR}^{(\ell)}= \E\{R_{\ell}(\lambda)\}$ is
obtained by replacing all monomials $z, z^2, \ldots, z^{\ell}$ in
the polynomial $\rho_{\ell}(z)$ by the moments $m_{|\matA|^2}^1,
m_{|\matA|^2}^2, \ldots, m_{|\matA|^2}^{\ell}$, respectively.

\section{Proof of Theorem \ref{theo:diagonal_elements_small_bandwidth}}\label{section:proof_theo_diagonal_elements_small_bandwidth}
The proof of Theorem \ref{theo:diagonal_elements_small_bandwidth}
follows along the line of the proof of Theorem
\ref{theo:diagonal_elements_chip_asynch}. As in the proof of Theorem 1, we can focus on the spreading matrix $\overline{\matS}$ in (\ref{equivalent_matrix}) and the autocorrelation $\overline{\matR}.$

For a signal with bandwidth $B \leq \frac{1}{2T_c},$
\begin{equation*}\underline{\phi}(\Omega, \tau)= \frac{1}{T_c} \mathrm{e}^{j \frac{\tau \Omega}{T_c}} \Phi^{*}\left(\frac{\Omega}{T_c} \right)
\qquad \qquad |\Omega|\leq \pi \end{equation*} and $\phi(\Omega,
\tau)=\underline{\phi}(\Omega-2 \pi \left\lfloor
\frac{\Omega}{\pi} \right\rfloor, \tau)$ for any $\Omega.$
Correspondingly, we define
\begin{equation*}
\underline{\boldsymbol{\Delta}}_{\phi,r}(\Omega,\tau)=
\frac{1}{T_c}\Phi\left( \frac{\Omega}{T_c}\right) \mathrm{e}^{-\frac{j \tau \Omega}{T_c}} \Vec{e}(\Omega), \qquad \qquad |\Omega| \leq \pi
\end{equation*}
with $\Vec{e}(\Omega)=(1, \mathrm{e}^{j \frac{\Omega}{r}  }, \ldots
\mathrm{e}^{j  \frac{(r-1) }{r} \Omega })$ and
\begin{equation*}
\boldsymbol{\Delta}_{\phi,r}(\Omega,\tau)=\underline{\boldsymbol{\Delta}}_{\phi,r}\left(\Omega-2 \pi \left\lfloor
\frac{\Omega}{\pi} \right\rfloor ,\tau \right) \qquad \text{for any } \; \Omega.
\end{equation*}
We adopt here the same notation as in the proof of Theorem
\ref{theo:diagonal_elements_chip_asynch}. Then, the $K \times K$
diagonal matrix $\pmb{\nabla}_{nt},$ for $t=1, \ldots r$ and $n=1,
\ldots N $ is given by
\begin{equation*}
\pmb{\nabla}_{nt}= \frac{1}{T_c} \Phi^{*}\left(\frac{j 2
\pi}{T_c}\underline{n} \right) \mathrm{e}^{-\frac{j 2 \pi
\underline{n} (t-1)}{r}} \mathrm{diag}\left(\mathrm{e}^{\frac{j 2
\pi \underline{n} \widetilde{\tau}_1}{T_c}}, \mathrm{e}^{\frac{j 2
\pi \underline{n} \widetilde{\tau}_2}{T_c}}, \ldots
\mathrm{e}^{\frac{j 2 \pi \underline{n} \widetilde{\tau}_K}{T_c}}
\right)
\end{equation*}
with $\underline{n}= \frac{n-1}{N}- \left\lfloor 2\frac{n-1}{N}
\right \rfloor $ and
$\boldsymbol{\Delta}_{\phi,r}(\widetilde{\tau}_k)$ is the $rN
\times N$ block diagonal matrix with $n$ diagonal block
$\boldsymbol{\Delta}_{\phi,r}(\underline{n}, \widetilde{\tau}_k).$
We develop the proof by strong induction as in Theorem
\ref{theo:diagonal_elements_chip_asynch} with similar initial step
and similar induction step.

\emph{Step 1:} In this case
\begin{equation*}
\widehat{\matR}_{kk}=|a_{k}|^2 \vecs_k^H
\boldsymbol{\Delta}_{\phi,r}^H
(\widetilde{\tau}_k)\boldsymbol{\Delta}_{\phi,r}(\widetilde{\tau}_k)
\vecs_k = |a_{k}|^2 \vecs_k^H \boldsymbol{\Phi} \vecs_k
\end{equation*}
where $\boldsymbol{\Phi}$ is a matrix independent of
$\widetilde{\tau}_k$ and the $n^{\mathrm{th}}$ element is given by
$\boldsymbol{\Phi}_{nn}= \frac{r}{T_c}\left| \Phi\left( \frac{j 2
\pi \underline{n} }{T_c}\right) \right|^2.$

By following the same approach as in Theorem
\ref{theo:diagonal_elements_chip_asynch} it results $\forall
\varepsilon > 0$
\begin{equation*}
\mathrm{Pr} \left\{ \left|\widehat{\matR}_{kk}-\frac{r
|a_{k}|^2}{T_c N} \sum_{n=0}^{N-1}\left|\Phi \left( \frac{j 2 \pi
\underline{n} }{T_c} \right) \right|^2 \right|> \varepsilon
\right\} \leq \frac{K_4 |a_{k}|^4 \Delta_{\mathrm{MAX}}^4}{N^2
\varepsilon^4}
\end{equation*}
being $\Delta_{\mathrm{MAX}} = \max_{\Omega \in \left[-\pi,
\pi \right]} \left| \Phi \left( \frac{\Omega }{T_c}
\right) \right|^2$ and
\begin{align}
\left. R_1(\lambda)\right|_{\lambda=|a_{k}|^2} &= \lim_{K=\beta N
\rightarrow \infty } \frac{|a_{k}|^2}{N} \sum_{\ell=0}^{N-1}
\left|\Phi \left( \frac{ 2 \pi  }{T_c} \left(\frac{n}{N} - \left
\lfloor \frac{2n}{N} \right \rfloor \right) \right)\right|^2
\nonumber \\
&= \left. \frac{\lambda}{2\pi} \int_{-\pi}^{\pi} \left|\Phi
\left( \frac{\Omega }{T_c} \right)\right|^2 \mathrm{d}\Omega
\right|_{\lambda=|a_{k}|^2}.
\end{align}
Furthermore, as in Theorem \ref{theo:diagonal_elements_chip_asynch}, it can be shown that $\mathrm{Pr}\left\{|\widehat{\matR}_{kk}
-R_1(|a_{k}|^2)| > \varepsilon \right\} \leq o\left( N^{-2}
\right) $ with the consequent  convergence with probability one by the Borel Cantelli lemma
\begin{equation*}
    \lim_{K=\beta N \rightarrow + \infty} \widehat{\matR}_{kk}\overset{a.s.}{=}R_1(|a_{k}|^2)_{\lambda=|a_{k}|^2}.
\end{equation*}

Similarly, $(\widehat{\matT}_{[nn]})_{uv},$ the $(u,v)$-element of
the matrix $\widehat{\matT}_{[nn]}$ is given by
\begin{align}
\widehat{\matT}_{[nn]} &= \widehat{\boldsymbol{\sigma}}_n \matA
\pmb{\nabla}_{n,u} \pmb{\nabla}_{n,v}^H \matA^H
\widehat{\boldsymbol{\sigma}}_n^{H} \nonumber \\
& = \frac{1}{T_c} \left|\Phi\left(\frac{ 2 \pi
\underline{n}}{T_c} \right) \right| \mathrm{e}^{-j2 \pi
\underline{n}\frac{v-u}{r} } \widehat{\boldsymbol{\sigma}}_n\matA
\matA^H\widehat{\boldsymbol{\sigma}}_n^H.
\end{align}
As in Theorem \ref{theo:diagonal_elements_chip_asynch} it can been
shown that
\begin{equation*}
\mathrm{Pr}\left\{ \left| (\widehat{\matT}_{[nn]})_{uv}
-\frac{1}{N T_c}\left|\Phi\left(\frac{ 2 \pi \underline{n}}{T_c}
\right) \right|^2 \mathrm{e}^{-j2 \pi \underline{n}\frac{v-u}{r}
}\mathrm{tr}(\matA \matA^H)
  \right| > \varepsilon
\right\} \leq \frac{K_4 T_{\mathrm{MAX}}^4}{N^2 \varepsilon^4}
\end{equation*}
with $T_{\mathrm{MAX}} = \left(\max_{\Omega \in \left[-\pi,
\pi\right]} \left|\Phi\left(\frac{ 2 \pi
\underline{n}}{T_c} \right) \right|^2 \right) \left(  \sup_{K} \max_{k}
|a_{k}|^2 \right)$ and the following convergence in probability holds
\begin{align}
\lim_{K= \beta N \rightarrow \infty} (\widehat{\matT}_{[nn]})_{uv}
&= \lim_{K= \beta N \rightarrow\infty} \frac{\beta}{T_c K}
\left|\Phi\left(\frac{ 2 \pi \underline{n}}{T_c} \right)
\right|^2 \mathrm{e}^{-j2 \pi \underline{n}\frac{v-u}{r} }
\sum_{k=1}^{K} |a_{k}|^2
\nonumber \\
& \overset{\mathcal{P}}{=} \frac{\beta}{T_c} \left|\Phi\left(\frac{\Omega}{T_c}
\right) \right|^2 \mathrm{e}^{-j2 \pi \underline{n}\frac{v-u}{r} }
\int \lambda \mathrm{d}F_{|\matA|^2}(\lambda) \nonumber
\end{align}
with $\Omega=2 \pi \lim_{N \rightarrow \infty} \underline{n}$ and $|\Omega|\leq
\pi.$ Thus, the diagonal block converges in probability as
follows
\begin{align}
\matT_1(\Omega) & {=} \lim_{K= \beta \N \rightarrow \infty}
(\widehat{\matT}_{[nn]})_{uv} \nonumber \\
&\overset{\mathcal{P}}{=}\frac{\beta}{T_c} \left|\Phi\left(\frac{\Omega}{T_c} \right)
\right|^2 \int \lambda \mathrm{d}F_{|\matA|^2}(\lambda) \Vec{e}(\Omega)
\Vec{e}^{H}(\Omega) \label{conv_aux}
\end{align}
Furthermore,
\begin{equation*}
\mathrm{Pr}\left\{ \left| (\widehat{\matT}_{[nn]})_{uv}
-(\matT_1(\Omega))_{uv} \right|> \varepsilon \right\} \leq o(N^{-2}).
\end{equation*}
Then, the convergence in probability (\ref{conv_aux}) holds also with probability one by the Borel Cantelli lemma.
This concludes the first step of the induction.

\emph{Step $\ell$:} Let us observe that
\begin{align*}
\vartheta_1 &=\frac{1}{\N} \mathrm{tr}\matA
\pmb{\nabla}_{n,u}\widehat{\matR}^{s}_{\vDash n}
\pmb{\nabla}_{n,u}^H \matA^H \\
& = \frac{\mathrm{e}^{-j 2 \pi \underline{n}\frac{u-v}{r}}}{N}
\sum_{k=1}^{K} \frac{|a_{k}|^2}{T_c^2} \left|\Phi\left( \frac{ 2
\pi \underline{n} }{T_c} \right) \right|^2
(\widehat{\matR}^s_{\vDash n})_{kk}
\end{align*}
and
\begin{align*}
\vartheta_2 & =\frac{|a_{k}|^2}{\N}
\mathrm{tr}\Mat{\Delta}_{\Phi,r}^H(\widetilde{\tau}_k)\widehat{\matT}^{s}_{\sim
k}
\Mat{\Delta}_{\Phi,r}(\widetilde{\tau}_k) \\
& = \frac{|a_{k}|^2}{N} \sum_{n=1}^{N} \frac{1}{T_c^2}
\left|\Phi\left( \frac{2 \pi \underline{n} }{T_c} \right)
\right|^2 \Vec{e}^H(2 \pi \underline{n})(\widehat{\matT}^s_{\sim
k})_{nn} \Vec{e}(2 \pi \underline{n}).
\end{align*}

By following the same approach as in Theorem
\ref{theo:diagonal_elements_chip_asynch} it can be shown that
$\vartheta_1$ and $\vartheta_2$ converge almost surely to the
following limits
\begin{align}
\lim_{K = \beta N \rightarrow \infty} \vartheta_1 &=
\frac{\beta}{T_c^2}\mathrm{e}^{-j 2 \pi
\underline{n}\frac{u-v}{r}} \left|\Phi\left( \frac{\Omega}{T_c}
\right) \right|^2 \int \lambda R_s(\lambda)
\mathrm{d}F_{|\matA|^2}(\lambda) \nonumber
\end{align}
and
\begin{align}
\lim_{K = \beta N \rightarrow \infty} \vartheta_2 &= \left.
\frac{\lambda}{2 \pi T_c^2}  \int_{-\pi}^{\pi}
\left|\Phi\left( \frac{\Omega}{T_c} \right) \right|^2
\Vec{e}^H(\Omega) \matT_s(\Omega) \Vec{e}(\Omega) \mathrm{d}\Omega
\right|_{\lambda=|a_{k}|^2} \nonumber
\end{align}
with $\left. R_s(\lambda) \right|_{\lambda=|a_{k}|^2}= \lim_{K=
\beta N \rightarrow \infty} (\widehat{\matR}^s)_{kk} $ and $\left.
\matT_s(\Omega) \right|= \lim_{K= \beta N \rightarrow \infty}
\widehat{\matT}^s_{[nn]} $ given by the recursion assumptions.

Additionally, it can be shown that the following almost sure convergence  holds
\begin{align}
\left. g(\matT_s, \lambda) \right|_{\lambda=|a_{k}|^2} &=
\lim_{K= \beta N \rightarrow \infty} \widehat{\vech}_k^H
\widehat{\matT}^s_{\sim k} \widehat{\vech}_k
\nonumber \\
& = \left. \frac{\lambda}{2 \pi T_c} \int_{-\pi}^{\pi }
\left|\Phi\left( \frac{\Omega}{T_c} \right) \right|^2
\Vec{e}^H(\Omega) \matT_s(\Omega) \Vec{e}(\Omega) \mathrm{d}\Omega
\right|_{\lambda=|a_{k}|^2} \label{convergence_theta_1}
\end{align}
and
\begin{align}
\Mat{f}(R_s, \Omega) &= \lim_{K=\beta N \rightarrow \infty}
\widehat{\boldsymbol{\delta}}_n \widehat{\matR}^s_{\vDash n}
\widehat{\boldsymbol{\delta}}^H_n \nonumber \\
&= \frac{\beta}{T_c^2}
\left|\Phi\left( \frac{\Omega}{T_c} \right) \right|^2
\Vec{e}(\Omega)  \Vec{e}^H(\Omega) \int \lambda R_s(\lambda) \mathrm{d}F_{|\matA|^2} (\lambda)  \label{convergence_theta_2}
\end{align}
Furthermore, the convergence  satisfies the bounds
\begin{equation*}
\mathrm{Pr} \left\{ |\widehat{\vech}_k^H \widehat{\matT}^s_{\sim
k} \widehat{\vech}_k - g(\matT_s, |a_{k}|^2)| > \varepsilon
\right\} <o(N^{-2})
\end{equation*}
and
\begin{equation*}
\mathrm{Pr} \left\{ |(\widehat{\boldsymbol{\delta}}_n)_u
\widehat{\matR}^s_{\vDash n} (\widehat{\boldsymbol{\delta}}^H_n)_v
- (\Mat{f}(R_s, \Omega))_{u,v}| > \varepsilon \right\} <o(N^{-2})
\end{equation*}
for large $N$ and $\forall \varepsilon.$

The recursion assumptions and the limits
(\ref{convergence_theta_1}) and (\ref{convergence_theta_2}) in
(\ref{inter_1}) and (\ref{inter_2}) yield
\begin{align}
\left. R_{\ell}(\lambda)\right|_{\lambda=|a_{k}|^2} &=
\sum_{s=0}^{\ell-1} g(\matT_{\ell-s-1}, \lambda) R_s(\lambda)
\nonumber \\
&= \left. \sum_{s=0}^{\ell-1} R_s(\lambda) \frac{\lambda}{2 \pi T_c^2}
\int_{-\pi}^{\pi } \left|\Phi\left( \frac{\Omega}{T_c} \right) \right|^2  \mathrm{tr}\left(\matT_s(\Omega)
\Vec{e}(\Omega)\Vec{e}^H(\Omega)\right) \mathrm{d}\Omega
\right|_{\lambda=|a_{k}|^2} \label{recursion_R_small}
\end{align}
and
\begin{align}
\matT_{\ell}(\Omega)&= \sum_{s=0}^{\ell-1} \Mat{f}(R_{\ell-s-1}, \Omega)
\matT_{s}(\Omega) \nonumber\\
& \sum_{s=0}^{\ell-1} \frac{\beta}{T_c^2} \left|\Phi\left( \frac{\Omega}{T_c} \right) \right|^2 \int \lambda R_s(\lambda)
\mathrm{d}F_{|\matA|^2}(\lambda) \; \Vec{e}(\Omega)\Vec{e}^H(\Omega) \matT_s(\Omega)
\label{recursion_T_small}
\end{align}
where $R_0(\lambda)=1$ and $\matT_0(\Omega)=\Mat{I}_r.$  With a similar
approach as in Theorem \ref{theo:diagonal_elements_chip_asynch} it
can be proven that for large $N$ and $\forall \varepsilon>0$
\begin{equation*}
\mathrm{Pr}\left \{ \left|\widehat{\matR}_{kk}^{\ell} -
R_{\ell}(|a_{k}|^2) \right|> \varepsilon \right \} \leq o(N^{-2})
\end{equation*}
and
\begin{equation*}
\mathrm{Pr}\left \{ \left|(\widehat{\matT}_{[nn]}^{\ell})_{uv} -
(\matT_{\ell}(\Omega))_{uv} \right|> \varepsilon \right \} \leq
o(N^{-2}).
\end{equation*}

In contrast to Theorem \ref{theo:diagonal_elements_chip_asynch}
the recursive equations (\ref{recursion_R_small}),
(\ref{recursion_T_small}), (\ref{convergence_theta_1}), and
(\ref{convergence_theta_2}) are independent of the time delay
$\widetilde{\tau}_k.$

The recursive equations can be further simplified by observing
that  $(\Vec{e}(\Omega) \Vec{e}^H(\Omega))^m=r^{m-1} \Vec{e}(\Omega)
\Vec{e}^H(\Omega).$ Then, it is straightforward to verify by recursion
that the matrix $\matT_s(\Omega)$, $s=1,2, \ldots,\ell-1,$ is
proportional to the matrix $\Vec{e}(\Omega) \Vec{e}^H(\Omega)$ and we can
express it as $\matT_s(\Omega)=T_s(\Omega)\Vec{e}(\Omega) \Vec{e}^H(\Omega),$
$s=1,2,\ldots.$ Thus, the recursive equations can be rewritten as
\begin{align}\nonumber
R_{\ell}(\lambda) & =\sum_{s=0}^{\ell-1} g(\matT_{\ell-s-1},
\lambda )
R_s(\lambda) & \\[1mm]
\label{T_ell} T_{\ell}(\Omega) \Vec{e}(\Omega) \Vec{e}^H(\Omega) &
=\sum_{s=1}^{\ell-1} \Mat{f}(R_{\ell-s-1}, \Omega) {T}_s(\Omega)\Vec{e}(\Omega)
\Vec{e}^H(\Omega) + \Mat{f}(R_{\ell-1}, \Omega)
\matT_0(\Omega) & \ell=1,2, \ldots \\[1mm]
\label{matfRx} \Mat{f}(R_{s}, \Omega)&={f}(R_{s}, \Omega) \Vec{e}(\Omega) \Vec{e}^H(\Omega) & \\[1mm]
\nonumber {f}(R_{s},\Omega) & = \frac{\beta}{T_c^2}
\left|\Phi\left(\frac{\Omega}{T_c}\right)\right|^2 \int \lambda
{R}_s(\lambda) \mathrm{d}\,F_{|\matA|^2}(\lambda) & -\pi
\leq \Omega \leq \pi  \\[1mm] \nonumber
g(T_s, \lambda)&=
  \begin{cases}
   \frac{r^2 \lambda}{2 \pi T_c^2} \int_{-\pi}^{\pi}
\left|\Phi\left(\frac{\Omega}{T_c} \right)\right|^2 {T}^s(\Omega)  \mathrm{d}\,\Omega & s=1,2, \ldots \\
    \frac{r \lambda}{2 \pi T_c^2} \int_{-\pi}^{\pi}
\left|\Phi\left(\frac{\Omega}{T_c} \right)\right|^2
\mathrm{d}\,\Omega  & s=0.
  \end{cases}
 &
\end{align}
with $\matT_0(\Omega)=\Mat{I}_r$ and $R_{0}(\lambda)=1.$

Substituting (\ref{matfRx})  in (\ref{T_ell}) we obtain
\begin{align}
T_{\ell}(\Omega) \Vec{e}(\Omega) \Vec{e}^H(\Omega) &=\sum_{s=1}^{\ell-1}
f(R_{\ell-s-1}, \Omega) T_s(\Omega)\Vec{e}(\Omega) \Vec{e}^H(\Omega)  \Vec{e}(\Omega)
\Vec{e}^H(\Omega)+ {f}(R_{\ell-1}, \Omega)
\matT_0(\Omega)\Vec{e}(\Omega) \Vec{e}^H(\Omega) \nonumber \\
& = r \sum_{s=1}^{\ell-1} f(R_{\ell-s-1}, \Omega) T_s(\Omega)\Vec{e}(\Omega)
\Vec{e}^H(\Omega)+ {f}(R_{\ell-1}, \Omega) T_0^{'}(\Omega)\Vec{e}(\Omega) \Vec{e}^H(\Omega)
\label{matfRx_2}
\end{align}

Recalling that $\matT_0(\Omega)=\Mat{I}_r$ and defining $
{T}_{0}^{'}(\Omega)=\frac{1}{r},$  we obtain from (\ref{matfRx_2}) the
scalar ${T}_{\ell}(\Omega)$:
\begin{align}
{T}_{\ell}(\Omega) & = r \left(\sum_{s=1}^{\ell-1} f(R_{\ell-s-1},\Omega)
{T}_s(\Omega)+ f(R_{\ell-1},\Omega)  {T}_0^{'}(\Omega) \right) \label{matfRx_3}.
\end{align}

The following equations summarize the recursion in terms of only
scalar functions.
\begin{align}\nonumber
{R}_{\ell}(\lambda)&=\sum_{s=0}^{\ell-1} g(T_{\ell-s-1}, \lambda )
R_s(\lambda) & \\[1mm]
\nonumber T_{\ell}(\Omega) &= r \sum_{s=0}^{\ell-1}
{f}(R_{\ell-s-1}, \Omega) T_s(\Omega) & \\[1mm]
\nonumber {f}(R_{s},\Omega) & = \frac{\beta}{T_c^2}
\left|\Phi\left(\frac{\Omega}{ T_c}\right)\right|^2 \int
\lambda R_s(\lambda) \mathrm{d}\,F_{|\matA|^2}(\lambda) &
| x| \leq \pi  \\[1mm]
g(T_s, \lambda) & =\frac{r^2 \lambda}{2 \pi T_c^2}
\int_{-\pi}^{\pi} \left|\Phi\left(\frac{\Omega}{
T_c}\right)\right|^2 T_s(\Omega) \mathrm{d}\,\Omega  &  s=0, 1, \ldots
\nonumber
\end{align}
with  $T_0(\Omega)=\frac{T_c}{r}$ and $R_{0}(\lambda)=1.$ Let us
observe that  the different expressions of $g(T_s, \lambda)$ for
$s=0, 1, \ldots$ could be absorbed in a unified expression by
initialize the recursion with $T_0(\Omega)=\frac{T_c}{r}$ instead of
using $T_0^{'}(\Omega)=\frac{1}{r}.$

The recursion in the statement of Theorem
\ref{theo:diagonal_elements_small_bandwidth} is obtained by
defining
\begin{equation*}
f(R_s)=\int\lambda R_s(\lambda) \mathrm{d}F_{|\matA|^2} (\lambda)
\end{equation*}
and \begin{equation*}
\nu(T_s)=\frac{r^2}{2 \pi T_c}\int_{-\pi/T_c}^{\pi/T_c}
\left|\Phi\left(\omega\right)\right|^2 T_s(\omega)
\mathrm{d}\,\omega
\end{equation*}
and by expressing $R_{\ell}(\lambda)$ and $T_{\ell}(\omega)$ as
recursive functions of $f(R_s)$ and $\nu(T_s).$

\bibliographystyle{ieeetran}

\begin{biography}[]
{Laura Cottatellucci} is currently working as assistant professor at the department of Mobile Communications at Eurecom, France. She received the degree in Electrical Engineering  and the PhD from University "La Sapienza", Italy in 1995 and from Technical University of Vienna, Austria in 2006, respectively. She worked in Telecom Italia from 1995 until 2000. From April 2000 to September 2005 she was Senior Research at ftw., Vienna, Austria in the group of information processing for wireless communications. From October 2005 to December 2005 she was research fellow on  ad-hoc networks at INRIA, Sophia Antipolis, France and guest researcher at Eurecom,  Sophia Antipolis, France. From January 2006 to November 2006, Dr. Cottatellucci was appointed research fellow at the Institute for Telecommunications Research, University of South Australia, Adelaide, Australia working on information theory for networks with uncertain topology.  Her research interests lie in the field of network information theory, communication theory, and signal processing for wireless communications.
\end{biography}

\begin{biography}[]
{Ralf R. M\"uller} (S'96-M'03-SM'05) was born in Schwabach, Germany, 1970. He received the Dipl.- Ing. and Dr.-Ing. degree with distinction from University of Erlangen-Nuremberg in 1996 and 1999, respectively. From 2000 to 2004, he directed a research group at Vienna Telecommunications Research Center in Vienna, Austria and taught as an adjunct professor at Vienna University of Technology. Since 2005 he has been a full professor at the Department of Electronics and Telecommunications at the Norwegian University of Science and Technology (NTNU) in Trondheim, Norway. He held visiting appointments at Princeton University, US, Institute Eurecom, France, University of Melbourne, Australia, University of Oulu, Finland, National University of Singapore, Babes-Bolyai University, Cluj-Napoca, Romania, Kyoto University, Japan, and University of Erlangen-Nuremberg, Germany. Dr. M\"uller received the Leonard G. Abraham Prize (jointly with Sergio Verd\'u) for the paper "Design and analysis of low-complexity interference mitigation on vector channels" from the IEEE Communications Society. He was presented awards for his dissertation "Power and bandwidth efficiency of multiuser systems with random spreading" by the Vodafone Foundation for Mobile Communications and the German Information Technology Society (ITG). Moreover, he received the ITG award for the paper "A random matrix model for communication via antenna arrays," as well as the Philipp-Reis Award (jointly with Robert Fischer). Dr. M\"uller served as an associate editor for the IEEE TRANSACTIONS ON INFORMATION THEORY from 2003 to 2006.
\end{biography}

\begin{biography}[]{
M{\'e}rouane Debbah} was born in Madrid, Spain. He entered the Ecole Normale Suprieure de Cachan (France) in 1996 where he received his M.Sc and Ph.D. degrees respectively in 1999 and 2002. From 1999 to 2002, he worked for Motorola Labs on Wireless Local Area Networks and prospective fourth generation systems. From 2002 until 2003, he was appointed Senior Researcher at the Vienna Research Center for Telecommunications (FTW) (Vienna, Austria) working on MIMO wireless channel modeling issues. From 2003 until 2007, he joined the Mobile Communications department of Eurecom (Sophia Antipolis, France) as an Assistant Professor. He is presently a Professor at Supelec (Gif-sur-Yvette, France), holder of the Alcatel-Lucent Chair on Flexible Radio. His research interests are in information theory, signal processing and wireless communications. M{\'e}rouane Debbah is the recipient of the "Mario Boella" prize award in 2005, the 2007 General Symposium IEEE GLOBECOM best paper award, the Wi-Opt 2009 best paper award as well as the Valuetools 2007,Valuetools 2008 and CrownCom2009 best student paper awards. He is a WWRF fellow.
\end{biography}

\end{document}
